\documentclass{LMCS}

\def\dOi{11(1:2)2015}
\lmcsheading%
{\dOi}
{1--66}
{}
{}
{Oct.~17, 2012}
{Feb.~27, 2015}
{}

\ACMCCS{[{\bf Theory of computation}]:  Semantics and
  reasoning---Program reasoning---Program verification}


\newif\ifcolors
\colorsfalse
\usepackage{hyperref}\usepackage[anythingbreaks]{breakurl}
\usepackage{amsthm} 
\usepackage{amsmath} 
\usepackage{amsfonts} 
\usepackage{xspace}
\usepackage{stmaryrd} 
\usepackage{color,soul}
\usepackage{bussproofs}
\usepackage{nicefrac}
\usepackage{bbm} 
\usepackage{ifpdf}
\usepackage{graphicx} 
\usepackage[normalem]{ulem} 
\usepackage{comment}
\usepackage{caption}
\usepackage{subcaption}
\usepackage{todo}

  \makeatletter
  \renewcommand\paragraph{\@startsection{paragraph}{4}{\z@}%
			 {-4\p@ \@plus -2\p@ \@minus -2\p@}%
			 {-0.5em \@plus -0.22em \@minus -0.1em}%
			 {\normalfont\normalsize\itshape}}
  \makeatother 

\newcommand{\RULE}[1]{{\ifmmode\mbox{\rm(#1)}\else{\rm(#1)}\fi}}

\renewcommand{\iff}{\mbox{iff}}

\def\und{\leavevmode{\kern0.03em\vbox{\hrule width0.47em}\kern0.03em}}

\newcommand{\modifyboxdimen}[2]
  {
   \dimen0=#1%
   \advance\dimen0 by #2%
   #1=\dimen0 
  }

\makeatletter
\def\judg{\@ifnextchar [{\@judgAB}{\@judgB}}
\def\@judgAB[#1]#2{#1\vdash#2\@isC}
\def\@judgB#1{\emset\vdash#1\@isC}
\def\@isC{\@ifnextchar [{\@yesC}{}}
\def\@yesC[#1]{\mathbin{\ty}#1}
\makeatother

\newenvironment{smallmath}
   {\small \begin{center}$}
   {$\end{center}\normalsize\ignorespaces}

\newenvironment{restate-lemma}[1]%
  {\begin{trivlist}\item[]{\normalsize\bf Proof of #1.}\it}
  {\end{trivlist}}
\newenvironment{restate-proposition}[1]%
  {\begin{trivlist}\item[]{\normalsize\bf Proof of #1.}\it}
  {\end{trivlist}}
\newenvironment{restate-theorem}[1]%
  {\begin{trivlist}\item[]{\normalsize\bf Proof of #1.}\it}
  {\end{trivlist}}

  {\\[\smallskipamount]\begin{minipage}{\textwidth}\begin{tabular}[b]{r@{\ \ }l@{\quad }l}}
  {\\[\smallskipamount]\end{tabular}\end{minipage}}

\newcommand{\caseskip}{\indent}

\newcommand{\Case}[1]{%
  \paragraph{\caseskip {\bf Case \theCaseCtr,} {\it #1:}}}

\newcommand{\AnonCase}{%
  \paragraph{\caseskip {\bf Case \theCaseCtr:}}}

\newcommand{\Subcase}[1]{%
  \paragraph{\caseskip {\bf Case \theCaseCtr.\theSubcaseCtr,} {\it #1:}}}

\newcommand{\AnonSubcase}[1]{%
  \paragraph{\caseskip {\bf Case \theCaseCtr.\theSubcaseCtr:}}}

\newcommand{\Subsubcase}[1]{%
  \paragraph{\caseskip {\bf Case \theCaseCtr.\theSubcaseCtr.\theSubsubcaseCtr,} {\it #1:}}}

\newcommand{\Subsubsubcase}[1]{%
  \paragraph{\caseskip {\bf Case \theCaseCtr.\theSubcaseCtr.\theSubsubcaseCtr.\theSubsubsubcaseCtr,} {\it #1:}}}

\newcounter{CaseCtr}
\newcounter{SubcaseCtr}
\newcounter{SubsubcaseCtr} 
\newcounter{SubsubsubcaseCtr} 
\newcounter{InProofEnum}
\newcounter{InCaseEnum}
\newcounter{InSubcaseEnum}
\newcounter{InSubsubcaseEnum}
\newcounter{InSubsubsubcaseEnum}
\newcounter{Store}

\newcommand{\labelandtext}[2]{\refstepcounter{rule}\addToLabel{#1}\label{#2} & \text{#1}}

\renewcommand{\theInProofEnum}{%
  {\rm(\arabic{InProofEnum})}}

\renewcommand{\theInCaseEnum}{%
  {\rm(\theCaseCtr.\arabic{InCaseEnum})}}

\renewcommand{\theInSubcaseEnum}{%
   {\rm(\theCaseCtr.\theSubcaseCtr.\arabic{InSubcaseEnum})}}

\renewcommand{\theInSubsubcaseEnum}{%
  {\rm(\theCaseCtr.\theSubcaseCtr.\theSubsubcaseCtr.\arabic{InSubsubcaseEnum})}}

\renewcommand{\theInSubsubsubcaseEnum}{%
  {\rm(\theCaseCtr.\theSubcaseCtr.\theSubsubcaseCtr.\theSubsubsubcaseCtr.\arabic{InSubsubsubcaseEnum})}}

\newenvironment{penumerate}[2]
  {\begin{list}{#1}
     {\setcounter{Store}{\value{#2}}
      \usecounter{#2}
      \setcounter{#2}{\value{Store}}
      \setlength{\leftmargin}{0.5em}}}
  {\end{list}}

\renewcommand{\qed}{\hspace*{\fill}$\Box$}

  {\setcounter{InProofEnum}{0}
   \begin{trivlist}\item[]{\normalsize\bf Proof of #1.}\ }
  {\qed\end{trivlist}
   \setcounter{CaseCtr}{1}
   \setcounter{SubcaseCtr}{1}
   \setcounter{SubsubcaseCtr}{1}
   \setcounter{SubsubsubcaseCtr}{1}}

  {\setcounter{InCaseEnum}{0}
   \Case{#1}}
  {\refstepcounter{CaseCtr}
   \setcounter{SubcaseCtr}{1}
   \setcounter{SubsubcaseCtr}{1}
   \setcounter{SubsubsubcaseCtr}{1}}

  {\setcounter{InCaseEnum}{0}
   \AnonCase}
  {\refstepcounter{CaseCtr}
   \setcounter{SubcaseCtr}{1}
   \setcounter{SubsubcaseCtr}{1}
   \setcounter{SubsubsubcaseCtr}{1}}

  {\setcounter{InSubcaseEnum}{0}
   \AnonSubcase}
  {\refstepcounter{SubcaseCtr}
   \setcounter{SubsubcaseCtr}{1}
   \setcounter{SubsubsubcaseCtr}{1}}

  {\setcounter{InSubcaseEnum}{0}
   \Subcase{#1}}
  {\refstepcounter{SubcaseCtr}
   \setcounter{SubsubcaseCtr}{1}
   \setcounter{SubsubsubcaseCtr}{1}}

  {\setcounter{InSubsubcaseEnum}{0}
   \Subsubcase{#1}}
  {\refstepcounter{SubsubcaseCtr}
   \setcounter{SubsubsubcaseCtr}{1}}

  {\setcounter{InSubsubsubcaseEnum}{0}
   \Subsubsubcase{#1}}
  {\refstepcounter{SubsubsubcaseCtr}}

\makeatletter
  \newcommand{\addToLabel}[1]{%
    \protected@edef\@currentlabel{\@currentlabel#1}%
  }
\makeatother

\newcounter{rule}

\newcommand{\staterule}[4][]{%
  \refstepcounter{rule}%
  \addToLabel{#2}%
  $\begin{array}[b]{@{}l}%
   \mbox{#2#1}\\%
   \begin{array}{c}
   #3\\
   \hline
   \raisebox{0ex}[2.5ex]{\strut}#4%
   \end{array}
  \end{array}$}

\newcommand{\staterulelabel}[4][]{
  \refstepcounter{rule}%
  \addToLabel{#2}%
  $\begin{array}{c l}
   #3 & \raisebox{-6pt}[-6pt][-6pt]{~#2#1} \\
   \cline{1-1}
   \raisebox{0ex}[2.5ex]{\strut}#4%
   \end{array}$}

\newcommand{\staterulelabelbis}[4][]{
  \refstepcounter{rule}%
  \addToLabel{#2}%
  $\begin{array}{c l}
   #3 & \raisebox{-12pt}[-12pt][-12pt]{~#2#1} \\
   \cline{1-1}
   \raisebox{0ex}[2.5ex]{\strut}#4%
   \end{array}$}

\newcommand{\stateaxiom}[3][]{%
  \refstepcounter{rule}%
  \addToLabel{#2}%
   $\begin{array}[b]{@{}l}%
      \mbox{#2#1}\\%
      \quad #3 
   \end{array}$}

\newcommand{\stateaxiomcond}[4][]{%
  \refstepcounter{rule}%
  \addToLabel{#2}%
   $\begin{array}[b]{@{}l}%
      \mbox{#2#1} \quad \begin{array}{l} #3\end{array}\\%
      \quad #4 
   \end{array}$}

\newcommand{\stateaxiomshortlongcond}[5][]{%
  \refstepcounter{rule}%
  \addToLabel{#2}%
   $\begin{array}[b]{@{}l}%
      \mbox{#2#1} \quad \begin{array}{l} #3\end{array}\\%
      \begin{array}{l}
        \quad #4 \\ \quad #5
      \end{array}
   \end{array}$} 

\newcommand{\staterulecond}[5][]{
  \refstepcounter{rule}%
  \addToLabel{#2}%
  $\begin{array}[b]{@{}l}%
   \mbox{#2#1}  \begin{array}{l} \quad  #3\end{array}\\%
   \begin{array}{c}
   #4\\
   \hline
   \raisebox{0ex}[2.5ex]{\strut}#5%
   \end{array}
  \end{array}$}

\newcommand{\GAP}{2ex}

\newcommand{\recallmath}[2]{%
\begin{array}{c}
 #1\\ 
 \hline
 \raisebox{0ex}[2.5ex]{\strut}#2%
 \end{array}}

\newcommand{\hbra}{
\hbox to \textwidth{\vrule width0.3mm height 1.8mm depth-0.3mm
                    \leaders\hrule height1.8mm depth-1.5mm\hfill
                    \vrule width0.3mm height 1.8mm depth-0.3mm}}
\newcommand{\hket}{
\hbox to \textwidth{\vrule width0.3mm height1.5mm
                    \leaders\hrule height0.3mm\hfill
                    \vrule width0.3mm height1.5mm}}

\newcommand{\ratio}{.35}
\newenvironment{display}[1]{\small\begin{tabbing}
  \hspace{1.5em} \= \hspace{\ratio\linewidth-1.5em} \= \hspace{1.5em} \= \kill
  \textbf{#1}\\*[-.8ex]
  \hbra\\*[-.8ex]
  }{\\*[-.8ex]\hket\nopagebreak
  \end{tabbing}}

\newcommand{\clause}[2]{$#1$\>\>#2}

\newcommand{\kw}[1]{\mathsf{#1}} 

\newcommand{\fv}[1]{\kw{fv}(#1)}               

\newcommand{\deq}{\mathrel{\stackrel{\scriptscriptstyle\Delta}{=}}}

\newcommand{\Is}{ \ ::= \ }
\newcommand{\Or}{ \ | \ }
\newcommand{\Deq}{\ \ \deq\ \ }

\newcommand{\set}[1]{\{#1\}}
\newcommand{\setcomp}[2]{\{#1\, |\, #2\}}

\newcommand{\lvec}[1]{\bar{#1}} 

\newcommand{\xs}{\lvec x}

\newcommand{\Ts}{\lvec T}

\newcommand{\Us}{\lvec U}
\newcommand{\Vs}{\lvec V}
\newcommand{\Ws}{\lvec W}

\newcommand{\vs}{\lvec v}
\newcommand{\ws}{\lvec w}

\newcommand{\alphas}{\lvec \alpha}

\newcommand{\mvec}[1]{\bar {#1}}

\newcommand{\Es}{\lvec E}

\newcommand{\es}{\lvec e}

\newcommand{\Fd}{\mvec F}
\newcommand{\Gd}{\mvec G}
\newcommand{\Fs}{\mvec F}

\newcommand{\Ed}{\mvec E}

\newcommand{\fs}{\lvec f}

\newcommand{\initcmd}[1]{\kw{init}(#1)}

\newcommand{\calj}{{\mathcal{J}}}

\newcommand{\hoaresimpl}[3]{\set {#1} #2 \set {#3}}

\newcommand{\initOp}{\mathsf{initStore}}
\newcommand{\fieldsOp}{\mathsf{fld}}

\newcommand{\OwnedUnlocked}[2]{\GGet {#1} {\java{traversable}}\lpa #2 \rpa}

\newcommand{\WtClSpecValSet}{\kw{WellTypClSpecVal}}
\newcommand{\specvaleq}{\simeq}

\newcommand{\staterulelabelbisecart}[4][]{
  \refstepcounter{rule}%
  \addToLabel{#2}%
  $\begin{array}{c l}
   #3 & \raisebox{-15.5pt}[-15.5pt][-15.5pt]{~#2#1} \\
   \cline{1-1}
   \raisebox{0ex}[2.5ex]{\strut}#4%
   \end{array}$}

\newcommand{\writes}[1]{\kw{writes}(#1)}

\newcommand{\compatibleSym}{\#}
\newcommand{\locked}[2]{#1.\java{locked}\lpa #2 \rpa}
\newcommand{\unlocked}[2]{#1.\java{unlocked}\lpa #2 \rpa}

\newcommand{\MathPred}[3]{\TyApp {#1.#2} {#3}}

\newcommand{\PPPointsTo}[3]{\GGet {#1}{#2} \mapsto {#3}}

\newcommand{\FPPointsTo}[4]{\GGet {#1}{#2} \stackrel{#3}{\longmapsto} {#4}}

\newcommand{\optypeFun}{\kw{type}}
\newcommand{\optype}[1]{\optypeFun(#1)}

\newcommand{\cfv}[1]{\kw{cfv}(#1)}
\newcommand{\hastype}[2]{#1 : #2}

\newcommand{\goodin}[3]{\good {#1} {#2} {\;\kw{in}\;#3}}
\newcommand{\goodrsc}[2]{#1 \vdash \hastype {#2} \ok}

\newcommand{\hpsup}{\supseteq_{\kw{hp}}}

\newcommand{\main}{\kw{main}} 

\newcommand{\pe}{\mathcal{E}}

\newcommand{\OLDwtt}[6]{#1 \vdash \hastype{#2,#3,#5,#6} \ok}

\newcommand{\return}{\jkw{return}}
\newcommand{\returnand}[3]{\SetVar {#1} {\return\lpa{#2}\rpa} #3}

\newcommand{\PredSet}{\kw{Pred}}

\newcommand{\pred}{\jkw{pred}}

\newcommand{\semval}{\mu}

\newcommand{\ClSubstSet}[1]{\kw{ClosingSubst}(#1)}

\newcommand{\goodsubst}[2]{#1 \vdash \hastype {#2} \ok}

\newcommand{\pef}{\mathcal{F}}

\newcommand{\assures}[2]{#1\ \java{assures}\ #2}
\newcommand{\ispartof}[2]{#1\ \java{ispartof}\ #2}

\newcommand{\restrict}[2]{#1_{{|#2}}}

\newcommand{\Bind}[3]{\Seq {#1 \shortleftarrow #2} {#3}}

\newcommand{\axioms}[1]{\kw{axiom}(#1)}

\newcommand{\SemValSet}{\kw{SemVal}}

\newcommand{\compatible}[2]{#1 \#  #2}

\newcommand{\axiom}[3]{#2;#1 \vdash #3}

\newcommand{\SpecValSet}{\kw{SpecVal}}

\newcommand{\PermTy}{\jkw{perm}}
\newcommand{\LocksetTy}{\jkw{lockset}}
\newcommand{\LogVarSet}{\kw{LogVar}}
\newcommand{\specval}{\pi}
\newcommand{\specvals}{\mvec\specval}
\newcommand{\logvar}{\alpha}
\newcommand{\logvars}{\mvec\logvar}

\newcommand{\perm}{\specval} 

\newcommand{\one}{1}

\newcommand{\HdAssert}[1]{\tjkw{assert}\lpa #1 \rpa}
\newcommand{\Assert}[2]{\Seq {\HdAssert{#1}} {#2}}

\newcommand{\quant}{\mathit{qt}}
\newcommand{\Qt}[3]{\lpa \quant\;{#2}\,{#1} \rpa\lpa #3 \rpa}
\newcommand{\Ex}[3]{\lpa\jkw{ex}\; {#2}\, {#1}\rpa\lpa #3 \rpa}
\newcommand{\Fa}[3]{\lpa\jkw{fa}\; {#2}\, {#1}\rpa\lpa #3 \rpa}

\newcommand{\bop}{\mathit{lop}}

\newcommand{\full}{1}

\newcommand{\goodenv}[1]{#1 \vdash \ok}

\newcommand{\Sound}[3]{\TyApp {#1} {\TypedVar {#2} {#3}}\ \kw{sound}}

\newcommand{\declared}[1]{\kw{declared}(#1)}
\newcommand{\Extends}[4]{\TyApp {#1}{\TypedVar{#2}{#3}}\ \kw{ext}\ #4}
\newcommand{\TypeExtends}[4]{\TyApp {#1}{\TypedVar{#2}{#3}}\
       \kw{type\texttt{-}extends}\ #4} 
\newcommand{\Implements}[4]{\TyApp {#1}{\TypedVar{#2}{#3}}\ \kw{impl}\
       #4} 
\newcommand{\undef}{\kw{undef}}
\newcommand{\goodtype}[2]{ #1 \vdash \hastype {#2} \ok}
\newcommand{\good}[2]{ #1 \vdash \hastype {#2} \ok}

\newcommand{\ptype}[3]{\kw{ptype}(#1,\TyApp {#2}{#3})}
\newcommand{\pptype}[2]{\kw{ptype}(#1,#2)}
\newcommand{\pbody}[4]{\kw{pbody}(\TyApp{#1}{#2}, \TyApp {#3}{#4})}
\newcommand{\ppbody}[3]{\kw{pbody}(\TyApp{#1}{#2}, #3)}

\newcommand{\Pbody}[2]{#1\;\kw{ext}\;#2}

\newcommand{\mlkup}[1]{\kw{mlkup}(#1)}
\newcommand{\plkup}[1]{\kw{plkup}(#1)}

\newcommand{\alkup}[1]{\kw{axiom}(#1)}

\newcommand{\Pred}[3]{\TyApp {\GGet{#1}{#2}} {#3}}

\newcommand{\InterfaceSet}{\kw{Interface}}
\newcommand{\ClassSet}{\kw{Class}}

\newcommand{\initpred}{\jkw{init}}

\newcommand{\NewPD}[4]{\PT{#1}{#2}{#3} \jeq #4}
\newcommand{\PDNoArg}[3]{\PTNoArg {#1}{#2} \jeq #3}
\newcommand{\NewPDNoArg}[3]{\PTNoArg {#1}{#2} \jeq #3}
\newcommand{\pdty}{\mathit{pt}}
\newcommand{\pdtys}{\SEQ\pdty}
\newcommand{\PT}[3]{\jkw{pred}\;#1\ANGLE{\TypedVar{#2}{#3}}}
\newcommand{\PTNew}[3]{\pred\;#1\ANGLE{\TypedVar{#2}{#3}}}

\newcommand{\PTNoArg}[2]{#1\;\jkw{pred}\;#2}

\newcommand{\MD}[8]{\MMT{#1}{#2}{#3}{#4}{#5}{#6}{#7}\lbr #8 \rbr}

\newcommand{\fd}{\mathit{fd}}
\newcommand{\fds}{\SEQ\fd}
\newcommand{\md}{\mathit{md}}
\newcommand{\mds}{\SEQ\md}

\newcommand{\TypedVar}[2]{#1\,#2}

\newcommand{\PredIdSet}{\kw{PredId}}

\newcommand{\pd}{\mathit{pd}}
\newcommand{\pds}{\SEQ\pd}

\newcommand{\extends}{\jkw{ext}}
\newcommand{\implements}{\jkw{impl}}

\newcommand{\CClass}[6]{\jkw{class}\;\TyApp{#1}{#2\;#3}\;\extends\;#4\;
\implements \;#5\; \lbr #6 \rbr}  

\newcommand{\inter}{\mathit{int}}
\newcommand{\Interface}[5]{\jkw{interface}\;
\TyApp{#1}{#2\;#3}\,\extends\;#4\,\lbr #5 \rbr}

\newcommand{\ax}{\mathit{ax}}
\newcommand{\axs}{\SEQ\ax}
\newcommand{\AX}[1]{\jkw{axiom}\;#1}

\newcommand{\SMT}[6]{%
  \TyAbs {#1}\, \SMTNoLogic{#2}{#3}{#4}{#5}{#6}}

\newcommand{\SMTNoLogic}[5]{%
  \Mspec{#1}{#2} #3\,#4\lpa #5 \rpa}
\newcommand{\MMT}[7]{%
  \TyAbs {\TypedVar{#1}{#2}}\, #3\; #4\;#5 \lpa\TypedVar{#6}{#7}\rpa}
  
\newcommand{\MMMT}[9]{%
  \TyAbs {\TypedVar{#1}{#2}}\, #3\; #4\,#5 \lpa\TypedVar{#6}{#7};
  \TypedVar{#8}{#9}\rpa}

\newcommand{\mt}{\mathit{mt}}
\newcommand{\mts}{\SEQ\mt}

\newcommand{\IntIdSet}{\kw{IntId}}

\newcommand{\PAt}[2]{#1\tjkw{@}#2}

\newcommand{\Bag}[1]{\kw{Bag}(#1)}

\newcommand{\emptybag}{[\,]}

\newcommand{\pts}{\mathcal S}

\newcommand{\sentails}[4]{#1 \vdash #2;#3 \models #4}

\newcommand{\scalar}[2]{#1 \cdot #2}

\newcommand{\bit}{\mathit{bit}}
\newcommand{\bits}{\mathit{bits}}
\newcommand{\BitsSet}{\kw{Bits}}
\newcommand{\binfr}{\mathit{fr}}
\newcommand{\BinFrSet}{\kw{BinFrac}}
\newcommand{\All}{\kw{all}}
\newcommand{\BinFr}[1]{\kw{fr}(#1)}

\newcommand{\LLAnd}{\tjkw{*}}
\newcommand{\LAnd}{\,\LLAnd\,}
\newcommand{\CCAnd}{\tjkw{\&}}
\newcommand{\CAnd}{\tjkw{\,\CCAnd\;}}
\newcommand{\CCOr}{\tjkw{|}}
\newcommand{\COr}{\tjkw{\,|\,}}

\newcommand{\LImplies}{\tjkw{\,-\;\!\!*\,}}
\newcommand{\LLImplies}{\tjkw{-\;\!\!*}}
\newcommand{\LEquiv}{\tjkw{\,*\;\!\!-\;\!\!*\,}}
\newcommand{\MLI}{\tjkw{\,-\;\!\!*\,}}

\newcommand{\LLEquiv}{\tjkw{*\;\!\!-\;\!\!*}}

\newcommand{\MCA}{\tjkw{\,\CCAnd\;}}

\newcommand{\Split}[1]{\jkw{split}(#1)}
\newcommand{\Perm}[3]{\jkw{Perm}\lpa \GGet{#1}{#2},#3 \rpa}

\newcommand{\Contains}[4]{\jkw{PointsTo}\lpa \GGet {#1} {#2},#3,#4 \rpa}
\newcommand{\PointsTo}[4]{\jkw{PointsTo}\lpa \GGet {#1} {#2},#3,#4 \rpa}

\newcommand{\True}{\jkw{true}}
\newcommand{\False}{\jkw{false}}

\newcommand{\Fun}[2]{#1 \rightarrow #2}

\newcommand{\dfFun}{\kw{df}}
\newcommand{\df}[1]{\dfFun(#1)}

\newcommand{\free}{\kw{free}}

\newcommand{\hc}{\mathit{hc}} 
\newcommand{\scmd}{\mathit{sc}} 

\newcommand{\ParFun}[2]{#1 \rightharpoonup #2}   

\newcommand{\isclassof}[1]{#1\;\java{classof}}
\newcommand{\iisclassof}[2]{#1\;\java{classof}\;#2}

\newcommand{\iinstanceof}[1]{\java{instanceof}\;#1}

\newcommand{\BareThread}[2]{#2\;\kw{in}\;#1}
\newcommand{\IdThread}[2]{#1\;\kw{is}\;#2}
\newcommand{\Thread}[3]{\IdThread {#1} {(\BareThread {#2} {#3})}}
\newcommand{\State}[2]{\langle #2,\, #1\rangle}

\newcommand{\holds}[5]{#1 \vdash #2;#3;#4 \models {#5}}
 
\newcommand{\OLDholds}[6]{#1 \vdash #2;#3;#5 \models {#6}}

\newcommand{\pt}{\mathcal{P}}  

\newcommand{\goodstate}[1]{\hastype {#1} \diamond}

\newcommand{\goodpool}[2]{#1 \vdash \hastype {#2} \diamond}

\newcommand{\goodthread}[2]{#1 \vdash \hastype {#2} \diamond}

\newcommand{\wildcard}{\_}

\newcommand{\Rationals}{\mathbb{Q}}

\newcommand{\updtfun}[3]{#1[#2 \mapsto #3]}

\newcommand{\hpenv}[1]{#1_{\kw{hp}}}

\newcommand{\ignore}[1]{ }

\newcommand{\perms}{\mvec \perm}

\newcommand{\osem}[3]{\sem{#1}^{#2}(#3)}

\newcommand{\semo}[2]{[\!| #1 |\!]^{#2}}

\newcommand{\init}[1]{\initOp(#1)}

\newcommand{\rvar}{\imath}
\newcommand{\rvars}{\lvec{\rvar}}

\newcommand{\FinDcl}[4]{#1\;#2 \jeq #3;\, #4}

\newcommand{\esem}[3]{\sem{#1}^{{#2}}_{{#3}}}

\newcommand{\goodheap}[2]{#1\vdash \hastype {#2} \diamond}
\newcommand{\goodstore}[2]{#1\vdash \hastype {#2} \diamond}

\newcommand{\goodobj}[2]{#1\vdash \hastype {#2} \diamond}

\newcommand{\Fst}[1]{{#1}_{1}}
\newcommand{\Snd}[1]{{#1}_{2}}

\newcommand{\DynTy}[2]{#1(#2)_1}

\newcommand{\Seq}[2]{#1;\,#2}   

\newcommand{\HdSetVar}[2]{#1 \jeq #2}
\newcommand{\HdOp}[3]{#1 \jeq #2(#3)}
\newcommand{\HdSet}[3]{#1.#2 \jeq #3}
\newcommand{\HdGet}[3]{#1 \jeq #2.#3}

\newcommand{\HdCall}[4]{#1 \jeq #2.#3 \lpa #4 \rpa}
\newcommand{\HdNew}[3]{#1 \jeq \jkw{new}\;\TyApp {#2} {#3}}

\newcommand{\HdInvWhile}[3]{\jkw{invariant}\;#2;\;\jkw{while}\lpa #1 \rpa \lbr #3 \rbr} 
\newcommand{\HdWhile}[2]{\jkw{while}\, \lpa #1 \rpa \lbr #2 \rbr}
\newcommand{\HdCond}[3]{\jkw{if}\, \lpa #1 \rpa \lbr #2 \rbr \jkw{else} \lbr
#3 \rbr} 

\newcommand{\HdCmdSet}{\kw{HeadCmd}}
\newcommand{\SpecCmdSet}{\kw{SpecCmd}}

\newcommand{\HdFSetNoFin}[3]{#1.#2 \jeq #3}

\newcommand{\Cond}[4]{\Seq {\HdCond {#1} {#2} {#3}} {#4}}

\newcommand{\java}[1]{{\mathtt{#1}}} 
\newcommand{\jkw}[1]{{\mathtt{#1}}} 

\newcommand{\tjkw}[1]{\mbox{\tt #1}} 

\newcommand{\SEQ}[1]{#1\mbox{\bf *}}

\newcommand{\This}{\jkw{this}}

\newcommand{\lbr}{\texttt{\symbol{'173}}}
\newcommand{\rbr}{\texttt{\symbol{'175}}}
\newcommand{\lpar}{\texttt{(\,}}
\newcommand{\rpar}{\texttt{\,)}}
\newcommand{\lpa}{\texttt{\upshape(}}
\newcommand{\rpa}{\texttt{\upshape)}}
\newcommand{\Object}{\jkw{Object}}
\newcommand{\ThreadTy}{\jkw{Thread}}

\newcommand{\jeq}{\,\mbox{\tt =}\,}
\newcommand{\jneq}{\,\mbox{\tt !=}\,}
\newcommand{\Null}{\jkw{null}}
\newcommand{\Void}{\jkw{void}}
\newcommand{\fork}{\java{fork}}
\newcommand{\join}{\java{join}}

\newcommand{\postJoin}{\java{postJoin}}
\newcommand{\run}{\java{run}}

\newcommand{\result}{\jkw{result}}
\newcommand{\jplus}{\;\tjkw{+}\;}

\newcommand{\jor}{\;\tjkw{|}\;}
\newcommand{\jnot}{\tjkw{!}}
\newcommand{\jdeq}{\;\tjkw{==}\;}
\newcommand{\jjdeq}{\tjkw{==}}

\newcommand{\ClIdSet}{\kw{ClassId}}
\newcommand{\ObjIdSet}{\kw{ObjId}}

\newcommand{\MthIdSet}{\kw{MethId}}
\newcommand{\FldIdSet}{\kw{FieldId}}
\newcommand{\VarSet}{\kw{Var}}
\newcommand{\RdWrVarSet}{\kw{RdWrVar}}
\newcommand{\RdVarSet}{\kw{RdVar}}
\newcommand{\ValSet}{\kw{OpenVal}}

\newcommand{\ExpSet}{\kw{Exp}}
\newcommand{\CmdSet}{\kw{Cmd}}
\newcommand{\TySet}{\kw{Type}}

\newcommand{\PermFormSet}{\kw{Formula}}

\newcommand{\meta}[1]{\mathit{#1}}
\newcommand{\lvar}{\ell}                      

\newcommand{\cmd}{c}
\newcommand{\cl}{\mathit{cl}}                      
\newcommand{\cls}{\mathit{ct}}

\newcommand{\ty}{\mathit{ty}}         
\newcommand{\mspec}{\mathit{spec}}   

\newcommand{\Fld}[2]{\TypedVar{#1}{#2}}

\newcommand{\ANGLE}[1]{\mbox{\tt <}#1\mbox{\tt >}}

\newcommand{\TyApp}[2]{#1\ANGLE{#2}}
\newcommand{\TyAbs}[1]{\ANGLE{#1}}

\newcommand{\FSetNoFin}[4]{#1.#2 \jeq #3;\, #4}
\newcommand{\Get}[4]{{#1} \jeq #2.#3;\, #4}
\newcommand{\GGet}[2]{#1.#2}

\newcommand{\SetVar}[3]{{#1} \jeq #2;\, #3}

\newcommand{\FSetVarNoFin}[4]{{#1}\;#2 \jeq #3;\, #4}

\newcommand{\Dcl}[3]{{#1}\;{#2};\, #3}

\newcommand{\CCall}[3]{#1.{#2} \lpa #3 \rpa}
\newcommand{\ECCall}[4]{\TyApp{#1.{#2}}{#3} \lpa #4 \rpa}
\newcommand{\EHdCall}[5]{#1 \jeq \ECCall{#2}{#3}{#4}{#5}}
\newcommand{\Call}[5]{{#1}\jeq \CCall{#2}{#3}{#4};\, #5}

\newcommand{\New}[4]{{#1} \jeq \jkw{new}\;\TyApp {#2} {#3};\, #4}

\newcommand{\While}[3]{%
  \jkw{while}\, \lpa #1 \rpa \lbr #2 \rbr; #3}

\newcommand{\Mspec}[2]{\prett\,#1;\posttt\,#2;}

\newcommand{\IntSet}{\kw{Int}}
\newcommand{\IntTy}{\jkw{int}}
\newcommand{\BoolSet}{\kw{Bool}}
\newcommand{\BoolTy}{\jkw{bool}}

\newcommand{\op}{\mathit{op}}
\newcommand{\arityFun}{\kw{arity}}
\newcommand{\arity}[1]{\arityFun(#1)}

\newcommand{\Op}[4]{#1 \jeq #2(#3);\, #4}

\newcommand{\sem}[1]{[\![ #1 ]\!]}

\newcommand{\ok}{\diamond}

\newcommand{\subcl}{\preceq}
\newcommand{\subty}{<:}

\newcommand{\prett}{\mathtt{requires}}
\newcommand{\posttt}{\mathtt{ensures}}

\newcommand{\subst}[3]{#3[#1/#2]}

\newcommand{\fields}[1]{\fieldsOp(#1)}
\newcommand{\mtype}[2]{\kw{mtype}(#1,#2)}

\newcommand{\entails}[4]{#2;#1;#3 \vdash #4}

\newcommand{\hoare}[6]{#2;#1 \vdash \set{#3} \hastype {#4} {#5} \set{#6}}

\newcommand{\hhoare}[5]{#2;#1 \vdash \set{#3} {#4} \set{#5}}

\newcommand{\goodform}[2]{#1\vdash \hastype {#2} \diamond}

\newcommand{\Cpl}[2]{(#1,#2)} 
\newcommand{\Tpl}[3]{(#1,#2,#3)} 

\newcommand{\enn}{n}
\newcommand{\bee}{b}

\newcommand{\dom}[1]{\kw{dom}(#1)}

\newcommand{\StoreSet}{\kw{Stack}}
\newcommand{\StackSet}{\kw{Stack}}
\newcommand{\ThreadSet}{\kw{Thread}}
\newcommand{\ObjStoreSet}{\kw{ObjStore}}

\newcommand{\HeapSet}{\kw{Heap}}
\newcommand{\ClValSet}{\kw{Val}}

\newcommand{\StateSet}{\kw{State}}
\newcommand{\ThreadPoolSet}{\kw{ThreadPool}}

\newcommand{\tpool}{\mathit{ts}}  
\newcommand{\fr}{\mathit{fr}}     
\newcommand{\str}{s}              
\newcommand{\thr}{t}               
\newcommand{\ostr}{\mathit{os}}    
\newcommand{\obj}{\mathit{obj}}    
\newcommand{\hp}{\mathit{h}}     
\newcommand{\stt}{\mathit{st}}     

\newcommand{\parpop}{\;|\;}       

\newcommand{\updtstore}[3]{#1[#2 \mapsto #3]}
\newcommand{\updtheap}[4]{#1[\GGet{#2}{#3} \mapsto #4]}
\newcommand{\mbody}[2]{\kw{mbody}(#1,#2)}  
\newcommand{\Mbody}[3]{(#2).#3}            

\newcommand{\step}{\,\rightarrow\,}

\newcommand{\sstep}[1]{\,\rightarrow_{{#1}}\,}
\newcommand{\ssteps}[1]{\,\rightarrow_{#1}^*\,}

\newcommand{\eg}{e.g.,\xspace}
\newcommand{\ie}{i.e.,\xspace}
\newcommand{\etc}{etc.\xspace}
\newcommand{\etal}{et al.\xspace}
\newcommand{\cf}{cf.\xspace}
\newcommand{\wrt}{w.r.t.\xspace}

\newcommand{\un}{(1)\xspace} 
\newcommand{\deux}{(2)\xspace}
\newcommand{\trois}{(3)\xspace}

\newcommand{\HdLock}[1]{\GGet {#1} {\mathtt{lock}}\lpa\rpa}
\newcommand{\HdUnlock}[1]{\GGet {#1} {\mathtt{unlock}}\lpa\rpa}

\newcommand{\heldkwd}{\jkw{Held}}

\newcommand{\Held}[2]{#1.\heldkwd\lpa#2\rpa}

\newcommand{\Statelm}[3]{\langle #2,\,#3,\, #1\rangle}

\newcommand{\lm}{\mathit{l}}

\newcommand{\whatever}{\_}

\newcommand{\lt}{\mathcal{L}}

\newcommand{\LockTableSet}{\kw{LockTable}}

\newcommand{\monInvpred}{\jkw{inv}}
\newcommand{\MonInv}[1]{\GGet {#1} \monInvpred}

\newcommand{\readyFun}{\kw{ready}}
\newcommand{\ready}[1]{\readyFun(#1)}

\newcommand{\HdCommit}[1]{\GGet {#1} {\jkw{commit}}}

\newcommand{\conc}[1]{\kw{conc}(#1)}
\newcommand{\concapp}[2]{\kw{conc}(#1)(#2)}

\newcommand{\Initialized}[1]{\GGet {#1} {\tjkw{initialized}}}
\newcommand{\FreshFun}{\tjkw{fresh}}
\newcommand{\Fresh}[1]{\GGet {#1} \FreshFun}
\renewcommand{\Join}[2]{\tjkw{Join(}#1,#2\tjkw{)}}

\newcommand{\HdWait}[2]{#1 \jeq \GGet {#2} {\jkw{wait}\lpa\rpa}}
\newcommand{\HdNotify}[2]{#1 \jeq \GGet {#2} {\jkw{notify}\lpa\rpa}}
\newcommand{\HdNotifyAll}[2]{#1 \jeq \GGet {#2} {\jkw{notifyAll}\lpa\rpa}}
\newcommand{\Wait}[3]{\Seq {\HdWait {#1} {#2}} {#3}}
\newcommand{\Notify}[3]{\Seq {\HdNotify {#1} {#2}} {#3}}
\newcommand{\NotifyAll}[3]{\Seq {\HdNotifyAll {#1} {#2}} {#3}}
\newcommand{\HdWaiting}[2]{\GGet {#1} {\jkw{waiting}\lpa#2\rpa}}
\newcommand{\HdResume}[2]{\GGet {#1} {\jkw{resume}\lpa#2\rpa}}
\newcommand{\Waiting}[3]{\Seq {\HdWaiting {#1} {#2}} {#3}}
\newcommand{\Resume}[3]{\Seq {\HdResume {#1} {#2}} {#3}}

\newcommand{\Sex}[6]{(#1,#2,#3,#4,#5,#6)} 

\newcommand{\nil}{\java{nil}}
\newcommand{\MsCupFun}{\cdot}
\newcommand{\MsCup}[2]{#1\MsCupFun#2}
\newcommand{\MsCupN}[2]{{#1}^{#2}}

\newcommand{\N}{\mathbb{N}}

\newcommand{\Lockset}[1]{\java{Lockset}\lpa #1 \rpa}
\newcommand{\contains}[2]{#1\;\java{contains}\;#2}

\newcommand{\cali}{{\mathcal{ I}}}
\newcommand{\calf}{{\mathcal{ F}}}

\newcommand{\ahp}{\mathcal{H}}  
\newcommand{\ahph}{\ahp_{\kw{hp}}}

\newcommand{\ahpg}{\ahp_{\kw{join}}}
\newcommand{\ahplock}{\ahp_{\kw{lock}}}
\newcommand{\ahpinit}{\ahp_{\kw{init}}}

\newcommand{\AugHeapSet}{\kw{AugHeap}}

\ifcolors\newcommand{\hlspec}[1]{\textcolor{deepblue}{#1}}
\else\newcommand{\hlspec}[1]{#1}
\fi
\definecolor{mlightgray}{rgb}{0.840,0.840,0.840}
\newcommand{\changed}[1]{\colorbox{mlightgray}{\ensuremath{#1}}}

\newcommand{\spty}{\mathit{pt}}
\newcommand{\ptys}
{\SEQ\spty}

\definecolor{MyBlue}{rgb}{0,0.06,0.80}
\newcommand{\xy}{\tt\color{MyBlue}}

\begin{document}

\title[Permission-Based Separation Logic for Multithreaded Java Programs]{Permission-Based Separation Logic for \\Multithreaded Java Programs\rsuper*}

\author[A.~Amighi]{Afshin Amighi\rsuper a}
\address{{\lsuper{a,c}}University of Twente, The Netherlands}
\email{a.amighi,@utwente.nl, marieke.huisman@ewi.utwente.nl}
\thanks{{\lsuper{a,c}}Amighi and Huisman are supported  by ERC grant 258405 for
the VerCors project.}

\author[C.~Haack]{Christian Haack\rsuper b}
\address{{\lsuper b}aicas GmbH, Karslruhe, Germany}
\email{christian.haack@aicas.de}
\thanks{{\lsuper b}Part of the work done while the author was at Radboud University Nijmegen, Netherlands.}

\author[M.~Huisman]{Marieke Huisman\rsuper c}
\address{\vspace{-18 pt}}
\thanks{{\lsuper c}Part of the work done while the author was at INRIA Sophia Antipolis -- M\'editerran\'ee, France.}

\author[C.~Hurlin]{Cl\'ement Hurlin\rsuper d}
\address{{\lsuper d}Prove \& Run, Paris, France}
\email{clement.hurlin@provenrun.com}
\thanks{{\lsuper d}Part of the work done while the author was at INRIA Sophia Antipolis -- M\'editerran\'ee, France, and visiting the University of Twente, Netherlands; and then at INRIA -- Bordeaux Sud-Ouest, France, Microsoft R\&D, France and IRISA/Universit\'e de Rennes 1, France.}

\titlecomment{{\lsuper*}This work was funded in part by the 6th Framework programme of the EC under the MOBIUS project IST-FET-2005-015905 (Haack, Hurlin and Huisman) and ERC grant 258405 for the VerCors project (Huisman and Amighi).}

\begin{abstract}

This paper presents a program logic
for reasoning about multithreaded Java-like programs with dynamic thread
creation, thread joining and reentrant object monitors.
The logic is based on concurrent separation logic.
It is the first detailed adaptation of
concurrent separation logic to a multithreaded Java-like language.

The program logic associates a unique
static access permission with each heap location, ensuring
exclusive write accesses and ruling out data races. Concurrent
reads are supported through fractional permissions.
Permissions can be transferred between threads upon thread starting, thread
joining, initial monitor entrancies and final monitor exits. In order to
distinguish between initial monitor entrancies and  
monitor reentrancies,
auxiliary variables keep track of multisets of currently held monitors. 
Data abstraction and behavioral subtyping are facilitated 
through abstract predicates, which are also used to represent monitor
invariants, preconditions for thread starting and postconditions
for thread joining. Value-parametrized types allow to conveniently capture 
common strong global invariants, like static object ownership relations.  

The program logic is presented for a model language with Java-like classes and 
interfaces, the soundness of the program logic is proven, and a number of
illustrative examples are presented.

\end{abstract}

\maketitle
\vfill

\section{Introduction}\label{sec:introduction}

\subsection{Motivation and Context}

In the last decade, researchers have spent great efforts on developing 
advanced program analysis tools for popular object-oriented
programming languages, like Java or C\#. Such tools include 
software model-checkers~\cite{VisserHBPL03}, static analysis tools for data
race and deadlock detection~\cite{NaikAW06,NaikPSG09},
type-and-effect systems for atomicity~\cite{FlanaganQ03b,AbadiFF06}, and program
verification tools based on interactive 
theorem proving~\cite{Huisman01}. 

A particularly successful
line of research is concerned with static contract checking
tools, based on Hoare logic. Examples 
include ESC/Java~\cite{FlanaganLLNSS02} --- a highly automatic, but deliberately
unsound, tool based on a weakest precondition calculus and a SMT solver, 
the Key tool ~\cite{Beckert07} --- a sound verification tool for Java
programs based on dynamic logic and symbolic execution, and Spec\#~\cite{BarnettDFLS04}
--- a sound modular verification tool for C\# programs
 that achieves modular soundness by imposing a dynamic object ownership
 discipline. While still primarily used in academics,
these tools are mature and usable enough, so that programmers other than the
tool developers can employ them for constructing realistic, verified
programs. A restriction, however, is that support for 
concurrency in static contract checking tools is still limited. Because most real-world 
applications written in Java or C\# are multithreaded, this limitation is
a serious obstacle for bringing assertion-based verification to the real
world. Support for concurrency is therefore the most important
next step for this technique.

What makes verification of shared-variable concurrent programs difficult is
the possibility of thread interference. 
Any assertion that has been established by one thread can potentially be
invalidated by any other thread at any time. 
Some traditional program logics for shared-variable concurrency,  \emph{e.g.},
Owicki-Gries~\cite{OwickiG75} or Jones's rely-guarantee
method~\cite{Jones83}, account for thread interference in the most general 
way. Unfortunately, the generality of these logics makes them tedious to use,
perhaps even unsuitable as a practical foundation for verifying 
Java-like programs. In comparison to these logics, Hoare's logics for conditional 
critical regions \cite{Hoare72} and monitors \cite{Hoare74}  
are much simpler, because they rely on syntactically enforceable synchronization disciplines that limit
thread interference to a few synchronization points (see \cite{Andrews91} for a survey).

Because Java's main thread synchronization mechanism is based on
monitors, Hoare's logic for monitors is a good basis for the
verification of Java-like programs.  Unfortunately, however, a safe
monitor synchronization discipline cannot be enforced syntactically
for Java. This is so, because Java threads typically share heap memory
including possibly aliased variables.  Recently, O'Hearn
\cite{OHearn07} has generalized Hoare's logic to programming languages
with heap. To this end, he extended \emph{separation
  logic}~\cite{IshtiaqO01,Reynolds02}, a new program logic, which had
previously been used for reasoning about sequential pointer programs.
\emph{Concurrent separation logic (CSL)} \cite{OHearn07, Brookes04}
enforces correct synchronization of heap accesses \emph{logically},
rather than \emph{syntactically}.  Logical enforcement of correct
synchronization has the desirable consequence that all CSL-verified
programs are guaranteed to be data-race free.

CSL has since been extended in various directions to make it
more suitable to reason about more complex concurrent programs. For instance,
Bornat and others have combined separation logic with permission
accounting in order to support concurrent reads \cite{BornatOCP05},
while Gotsman \etal \cite{GotsmanBCRS07} and Hobor
\etal~\cite{HoborAZ08} have generalized concurrent separation logic to
cope with Posix-style threads and locks, that is they can reason about
dynamic allocation of locks and threads.

In this paper, we further adapt CSL and its extensions, to make it suitable to
reason about a Java-like language. This requires several challenges to
be addressed:

\begin{itemize}
\item Firstly, in Java locks are reentrant, dynamically allocated, and
  stored on the heap, and thus can be aliased. Reasoning about
  \emph{storable locks} has been addressed before by Gotsman
  \etal~\cite{GotsmanBCRS07} and Hobor \etal~\cite{HoborAZ08}, however
  these approaches do not generalise to reentrant locks. 
  Supporting reentrant locks has important advantages, as they can
  avoid deadlocks due to attempted reentrancy. Such deadlocks would,
  for instance, occur when synchronized methods call synchronized
  methods on the current self: a very common call-pattern in Java.
  Therefore, any practical reasoning method for concurrent Java
  programs needs to provide support to reason about lock reentrancy.
\item Secondly, Java threads are
based on thread identifiers (represented by thread objects) that are
dynamically allocated on the heap, can be stored on the heap and can
be aliased. Additionally, a join-operation that is parametrized by a
thread identifier allows threads to wait for the termination of other
threads.  A crucial difference with Posix threads is that Java threads
can be joined multiple times, and the logic has to cater for this
possibility. 
\item Finally, Java has a notifying mechanism to wake threads up
  while waiting for a lock. This is an efficient mechanism to allow
  threads to exchange information about the current shared state,
  without the need for continuous polling. A reasoning technique for
  Java thus should support this wait-notify mechanism.
\end{itemize}

\noindent The
resulting proof system supports Java's main concurrency primitives:
dynamically created threads and monitors that can be stored on the
heap, thread joining (possibly multiple times), monitor reentrancy,
and the wait-notify mechanism. Furthermore, the proof
system is carefully integrated into a Java-like type system, enriched
with value-parametrized types.  The resulting formal system allows
reasoning about multithreaded programs written in Java. Since the use
of Java is widespread (\emph{e.g.},~internet applications, mobile
phones and smart cards), this is an important step towards reasoning
about realistic multi-threaded software.

\subsection{Separation Logic Informally}
\label{subsec:sl:informally}

Before discussing our contribution in detail, we first
informally present the features of separation logic
that are most important for this paper.

\subsubsection{Formulas as Access Tickets}

Separation logic~\cite{Reynolds02} combines the usual
logical operators with the points-to predicate $\PPPointsTo x f v$ and the
resource conjunction $F \LAnd G$. 

The predicate $\PPPointsTo x f v$ has a \emph{dual purpose}: 
firstly, it asserts that the object field $\GGet x f$ contains data value $v$
and, secondly, it represents a \emph{ticket} that grants permission to access
the field $\GGet x f$. This is formalized by separation logic's Hoare rules
for reading and writing fields (where $\PPPointsTo x f \wildcard$ is short
for $(\exists v)(\PPPointsTo x f v)$):
\begin{displaymath}
\set {\PPPointsTo x f \wildcard} 
\HdSet x f v
\set{\PPPointsTo x f v}
\qquad
\set {\PPPointsTo x f v}
\HdGet y x f
\set {\PPPointsTo x f v \ \LAnd\ v \jdeq y}
\end{displaymath}

The crucial difference to standard Hoare logic is that both these rules 
have a precondition of the form $\PPPointsTo x f
\wildcard$:
this formula functions as an \emph{access ticket} for $\GGet x f$. 
It is important that tickets are not duplicable: 
one ticket is not the same as
two tickets! 
Intuitively, the formula $F \LAnd G$ represents two access tickets $F$ and $G$ to \emph{separate} parts of the heap. 
In other words, the part of the heap that $F$ permits to access is \emph{disjoint} from the part of the heap that $G$ permits to access. 
As a consequence, separation logic's $\LAnd$ implicitly excludes interfering heap accesses through aliases: this is why
the Hoare rules shown above are sound. 
It is noteworthy that given two objects \texttt{a} and \texttt{b} with field \texttt{x}, the assertion $\PPPointsTo {\mathtt a} {\mathtt x} \whatever \LAnd \PPPointsTo {\mathtt b} {\mathtt x} \whatever$ does not mean the same as $\PPPointsTo {\mathtt a} {\mathtt x} \whatever \wedge \PPPointsTo {\mathtt b} {\mathtt x} \whatever$: the first assertion implies that \texttt{a} and \texttt{b} are distinct, while the second assertion can be satisfied even if \texttt{a} and \texttt{b} are aliases.

\subsubsection{Local Reasoning}

A crucial feature of separation logic is that it allows local
reasoning, as expresssed by the (Frame)
rule: 
\begin{displaymath}
  \AxiomC {$\hoaresimpl F \cmd {F'}$}
  \RightLabel{(Frame)}
  \UnaryInfC{$\hoaresimpl {F \LAnd G} \cmd {F' \LAnd G}$}
  \DisplayProof
\end{displaymath}

This rule expresses that given a command $\cmd$ that only accesses
the part of the heap described by $F$, one can reason locally about
command $\cmd$ ((Frame)'s premise) and deduce something globally, \ie
in the context of a bigger heap $F \LAnd G$ ((Frame)'s conclusion).
In this rule, $G$ is called the frame and represents the part of the
heap unaffected by executing $\cmd$. It is important that the (Frame)
rule can be added to our verification rules without harming soundness,
because it enables modular verification, and in particular it allows
one to verify method calls. When calling a method, from its
specification one can identify the (small) part of the heap accessed
by that method and use the frame rule to establish that the rest of
the heap is left unaffected.


\subsection{Contributions}

Using the aspects of separation logic described above, we have
developed a sound (but not complete) program logic for a concurrent
language with Java's main concurrency primitives.  Our logic combines
separation logic for Java~\cite{Parkinson05b} with fraction-based
permissions~\cite{Boyland03}. This results in an expressive and
flexible logic, which can be used to verify many realistic
applications. The logic ensures the absence of data races, but is not
overly restrictive, as it allows concurrent reads.  This subsection
summarizes our system and highlights our contributions; for a detailed
comparison with existing approaches, we refer to
Section~\ref{sec:relatedwork}.

Because of the use of fraction-based permissions, as
proposed by Boyland~\cite{Boyland03}, our program logic prevents data
races, but allows multiple threads to read a location
simultaneously. Permissions are fractions in the interval
$(0, 1]$. Each access to the heap is associated with a
permission. If a thread has full permission (\emph{i.e.}, with value
1) to access a location, it can write this location, because the
thread is guaranteed to have exclusive access to it. If a thread has a
partial permission (less than 1), it can read a location. However,
since other threads might also have permission to read the same
location, a partial permission does not allow to write a
location. Soundness of the approach is ensured by the guarantee that
the total permissions to access a location are never more than 1.

Permissions can be transferred from one thread to another upon
thread creation and thread termination. If a new thread is forked, the parent
thread transfers the necessary permissions to this new thread (and
thus the creating thread abandons these permissions, to avoid
permission duplication). Once a thread terminates, its permissions can
be transferred to the remaining threads. The mechanism for doing this
in Java is by joining a thread: if a thread $t$ joins another thread
$u$, it blocks until $u$ has terminated. After this, $t$ can
take hold of $u$'s permissions. In order to soundly account 
for permissions upon thread joining, a special
join-permission is used. Only threads that hold (a fraction of) this
join-permission can take hold of (the same fraction of) the
permissions that have been released by the terminating thread.
Note that, contrary to Posix threads, Java threads allow multiple joiners 
of the same thread, and our logic supports this.
For example, the logic can verify programs where multiple threads join the 
same thread $t$ in order to gain shared read-access to the part of the heap
that was previously owned by $t$.

Just as in O'Hearn's approach~\cite{OHearn07}, locks are associated
with so-called resource invariants. If a thread acquires a lock, it
may assume the lock's resource invariant and obtain access to the
resource invariant's footprint (i.e., to the part of the heap that the
resource invariant depends upon).

If a thread releases a lock, it has to establish the lock's resource invariant
and transfers access to the resource invariant's footprint back to the lock.
Previous variants of concurrent separation logic prohibit threads to
acquire locks that they already hold. In contrast, Java's locks are
reentrant, and our program logic supports this. To this end,
the logic distinguishes between initial lock entries and lock reentries.
Permissions are transferred upon initial lock entries only, but not upon reentries. 

Unfortunately, distinguishing between initial lock entries and reentries
is not well-supported by separation logic. The problem is 
that this distinction requires proving that, upon initial entry, a lock does not alias any
currently held locks. Separation logic, however, is designed to avoid depending
on such global aliasing constraints, and consequently does not provide good
support for reasoning about such. Fortunately, our logic includes a
rich type system that can be used towards proving global aliasing constraints
in many cases. The type system features
value-parametrized types, which naturally extend Java's type system
that already includes generic types.
Value parameters are used for static type
checking and static verification only, thus, do not change the dynamic semantics of Java. 
Value-parametrized types can be useful in many ways. For
instance, in~\cite{HaackH09} we use them to distinguish read-only
iterators from read-write iterators. Value-parametrized types can also 
express static object ownership relations, as done in parametric ownership
type systems (e.g., \cite{ClarkePN98,ClarkeD02}). Similar ownership type systems have 
been used in program verification systems to control aliasing (e.g, \cite{Muller02}). In
Section~\ref{subsec:examples:lock:lc}, we use type-based ownership 
towards proving the correctness of a fine-grained lock-coupling algorithms with 
our verification rules for reentrant locks. The type-based ownership relation 
serves to distinguish initial lock entries from lock reentries.

To allow the inheritance of resource invariants, we use abstract
predicates as introduced in Parkinson's object-oriented separation
logic~\cite{Parkinson05b}. 
Abstract predicates hide implementation details from clients but allow class implementers to use them. 
Abstract predicates are highly appropriate
to represent resource invariants: in class
\texttt{Object} a resource invariant with empty footprint is defined, 
and each subclass can extend this resource invariant to depend on
additional fields. 

\subsection{Earlier Papers and Overview}
This paper is based on several earlier papers, presenting parts of the proof system. The logic to reason about dynamic threads was presented at AMAST 2008~\cite{HaackH08b}, the logic to reason about reentrant locks was presented at APLAS 2008~\cite{HaackHH08}. However, compared to these earlier papers, the system has been unified and streamlined. In addition, novel specifications and implementations of sequential and parallel merge sort illustrate the approach.  The work as it is presented here is adapted from a part of Hurlin's PhD thesis~\cite{HurlinPhd}.

The remainder of this paper is organized as
follows. Section~\ref{sec:jll} presents the Java-like language, permission-based separation logic and basic proof rules for
single-threaded programs.
Section~\ref{sec:forkjoin} extends this to
multithreaded programs with dynamic thread creation and termination,
while Section~\ref{sec:locks} adds reentrant locks.  Finally,
Sections~\ref{sec:relatedwork} and~\ref{sec:conclusion} discuss
related work, future work and conclusions. 
The complete soundness proof for the system can be found in Hurlin's PhD thesis~\cite{HurlinPhd}.

\section{The Sequential Java-like language}
\label{sec:jll}

This section presents a \emph{sequential} Java-like (programming and
specification) language that models core features of Java: mutable
fields, inheritance and method overriding, and interfaces.  Notice
that we strongly base our work here on Parkinson's
thesis~\cite{Parkinson05b} and in particular reuse his notion of
abstract predicate.  Later sections will extend the language with
Java-like concurrency primitives.  Sequential programs written in the
Java-like language can be specified and verified with separation
logic.  However, to simplify the presentation of the program logic, we
assume that Java expressions are written in a form so that all
intermediate results are assigned to local read-only variables
(\cf~\eg~\cite{CraryWM99,SmithWM00,JiaW06,ParkinsonB08}).

\subsection{Syntax}
\label{subsec:syntax:jll}

 The language distinguishes between read-only variables~$\rvar \in \RdVarSet$,
read-write variables~$\lvar \in \RdWrVarSet$, and logical variables~$\logvar \in \LogVarSet$. 
Method parameters (including $\This$) are always read-only, and local 
variables can be both read-only  or read-write. Logical variables can 
only occur in specifications and types.
We treat read-only variables specially, because their use
often avoids the need for syntactical side conditions in the proof rules 
(see Section~\ref{subsec:sl:hoare}). 
The model language also includes class identifiers ($\ClIdSet$), interface
identifiers ($\IntIdSet$), field identifiers ($\FldIdSet$), method identifiers ($\MthIdSet$) and predicate identifiers ($\PredIdSet$).
Object identifiers ($\ObjIdSet$) are used in the operational semantics, but must not occur in 
source programs. Variable $\VarSet$ is the union of $\RdVarSet$, $\RdWrVarSet$ and $\LogVarSet$. In addition, type identifiers are defined as the union of $\ClIdSet$ and $\IntIdSet$.

Figure~\ref{fig:jll-syntax} defines syntax of our Java-like language.
(Open) values are integers, booleans, object
identifiers, $\Null$, and read-only variables. 
\emph{Open values} are values that are not variables.
Initially, \emph{specifications values} range over logical
variables and values; this will be extended in subsequent sections.
To simplify later developments, our grammar for writing programs imposes that (1)~every intermediate result is assigned to a local variable and (2)~the right hand sides of assignments contain no read-write variables.
Since interfaces and classes can be parametrized with specification values, object types are of the form $\TyApp t \specvals$.
We introduce two special operators: \texttt{instanceof} and
$\texttt{classof}$, where
$C\,\texttt{classof}\,v$ tests whether $C$ is $v$'s dynamic class.
Note that these last two operators depend on object types, as stored on the heap. 
Classes do not have constructors: fields are initialized to a default value when objects are created. 
Later, for clarity, methods that act as constructors are called {\tt init}.
\emph{Abstract predicates}~\cite{Parkinson05b,ParkinsonB05} and class \emph{axioms} are part of our specification language.
Interfaces may declare abstract predicates and classes may implement them by providing concrete definitions as separation logic formulas.
Appendix \ref{sec:add-defs} defines syntactic functions to lookup fields, axioms, method types and bodies, and predicate types and bodies. 

\begin{figure}[h]
\begin{renewcommand}{\ratio}{.3}
\begin{display}{}

\clause{
  \enn \in \IntSet \quad 
  u,v,w \in \ValSet \ \Is \ \Null \ \Or\ \enn \ \Or\ \bee \ \Or\ o \ \Or\ \rvar \qquad
  \bee \in \BoolSet \ = \ \set{\True,\False}
  }{}
  \\
\clause{
  \ClValSet \  = \ \ValSet\setminus\RdVarSet \quad
  \specval \in \SpecValSet \ \Is \ \logvar\ \Or\ v
  }{}
\\
\clause{  T,U,V,W \in \TySet \Is \Void \ \Or\  \IntTy \ \Or\  \BoolTy \ \Or\ \PermTy \ \Or\  \TyApp t \specvals}{}
\\
\clause{
\op  \supseteq  
  \set{\jjdeq,\texttt{!},\texttt{\&},\texttt{|}}   
  \ \cup\ 
  \setcomp {\,\isclassof C\,} {\,C \in \ClIdSet\,} 
  \ \cup\
  \setcomp {\,\iinstanceof T\,} {\,T\in\TySet\,}
}{}
\\
\clause{   e \in \ExpSet \Is \specval \ \Or\ \lvar\ \Or\ \op\lpa \es \rpa}{}
\label{idx:expressions}
\\
\clause{}{}
\\

\clause{\fd \Is \Fld T f;}{field\ declarations}
  \\
\clause{\pd \Is \NewPD P \Ts \alphas F;}{predicate\ definitions}
  \\
\clause{\ax \Is \AX F;}{class\ axioms}
  \\
\clause{\md \Is \MD \Ts \alphas \mspec U m \Vs \rvars c}{methods (scope of $\alphas,\rvars$ is $\Ts,\mspec,U,\Vs,c$)}
  \\
\clause{\mspec \Is \Mspec F F}{pre/postconditions}
  \\
\clause{F \in \PermFormSet}{specification\ formulas}
  \\
\clause{\cl \in \ClassSet \Is \CClass C \Ts \alphas U \Vs {\fds\;\pds\;\axs\;\mds}}{}
\\
\clause{}{classes(scope of $\alphas$ is $\Ts,U,\Vs,\fds,\pds,\axs,\mds$)}
 \\
\clause{}{}
  \\
\clause{\pdty \Is \PTNew P \Ts \logvars;}{predicate types}
\\
\clause{\mt \Is \MMT \Ts \logvars \mspec U m {\TypedVar \Vs {\rvars}}}{method types (scope of $\alphas,\rvars$ is $\Ts,\mspec,U,\Vs$)}
\\
\clause{
    \inter \in \InterfaceSet \Is 
    \Interface I \Ts \alphas \Us {\pdtys\;\axs\;\mts}
}{}
\\
\clause{}{interfaces (scope of $\alphas$ is $\Ts,\Us,\pdtys,\axs,\mts$)} 
\\
\clause{}{}
\\  
\clause{
	\begin{array}{rcl}
  \cmd \in \CmdSet & \Is &
  v
  \ \Or\ 
  \Dcl T \lvar \cmd
  \ \Or\ 
  \FinDcl T \rvar \lvar \cmd
  \ \Or\ 
  \Seq \hc \cmd
  \label{idx:command}
\\
  \hc \in \HdCmdSet & \Is & 
     \HdSetVar \lvar v 
     \ \Or\ 
     \HdOp \lvar \op \vs 
     \ \Or\ 
     \HdGet \lvar v f
     \ \Or \ 
     \HdFSetNoFin v f v
     \ \Or\     
     \HdNew \lvar C \specvals
     \ \Or\ 
\\ & & 
     \HdCall \lvar v m \vs 
     \ \Or\ 
     \HdCond v c c
     \ \Or\ 
     \HdWhile e c
     \label{idx:head:command}
\end{array}
  }{}
  \\
\clause{}{}
  \\
\clause{
  \begin{array}{c}
    \bop \in \set {\LLAnd,\LLImplies,\CCAnd,\CCOr}
    \ \qquad \
    \quant \in \set{\jkw{ex},\jkw{fa}}
    \ \qquad\
    \kappa \in \PredSet \Is P \Or \PAt P C
    \\
    F \in \PermFormSet \Is
    e
    \Or
    \PointsTo e f \specval {e}
    \Or
    \MathPred \specval \kappa {\specvals}
    \Or
    F\;\bop\;F
    \Or
    \Qt x T F
  \end{array}
}{}

\end{display}
\end{renewcommand}

\caption{Sequential Java-Like Language (JLL)}
\label{fig:jll-syntax}
\end{figure}

To write method contracts, we use \emph{intuitionistic} separation logic~\cite{IshtiaqO01,Reynolds02,Parkinson05b}. 
Contrary to classical separation logic, intuitionistic separation logic admits weakening \ie it is invariant under heap extension. 
Informally, this means that one can "forget" a part of the state, which makes it appropriate for garbage-collected languages.

The \emph{resource conjunction} $F \LAnd\, G$ (a.k.a\ \emph{separating conjunction}) expresses that resources $F$ and $G$ are independently available: using either of these resources leaves the other one intact. 
Resource conjunction is not idempotent: $F$ does \emph{not} imply $F \LAnd\, F$. Because Java is a garbage-collected language, we allow dropping assertions: $F \LAnd\, G$ implies $F$. 
\label{idx:star}

The \emph{resource implication} $F \LImplies G$ (a.k.a.\ \emph{linear implication} or \emph{magic wand}) means "consume $F$ yielding $G$''. 
Resource $F \LImplies G$ permits to trade resource $F$ to receive resource $G$ in return. 
Most related work omit the magic wand. 
Parkinson and Bierman~\cite{ParkinsonB05} entirely prohibit predicate occurrences in negative positions (\ie to the left of an odd number of implications).
We allow negative dependencies of predicate $P$ on predicate $Q$ as long as $Q$ does not depend on $P$ (cyclic predicate dependencies must be positive). 
We include it, because it can be added without any difficulties, and we found it useful to specify some typical programming patterns.
Blom and Huisman show how the magic wand is used to state loop invariants over pointer data structures ~\cite{BlomH13}, while Haack and Hurlin use the magic wand to capture the behaviour of the iterator~\cite{HaackH09}.
To avoid a proof theory with bunched contexts (see Section~\ref{subsec:sl:proof:theory}),
we omit the $\Rightarrow$-implication between
heap formulas (and did not need it in later examples). 

The \emph{points-to predicate} $\PointsTo e f \specval v$ is a textual representation for $\FPPointsTo e f \specval v$~\cite{BornatOCP05}. 
Superscript~$\specval$ must be a fractional permission~\cite{Boyland03} \ie a fraction $\frac{1}{2^n}$ ($n \geq 0$) in the interval $(0, 1]$.
The \emph{predicate application} $\MathPred \specval \kappa \specvals$ applies
abstract predicate $\kappa$ to its receiver parameter $\specval$ and the
additional parameters $\specvals$. As explained above, predicate definitions
in classes map abstract predicates to concrete definitions. 
Predicate definitions can be extended in subclasses to account for extended
object state. Semantically, $P$'s predicate extension 
in class~$C$ gets $\LLAnd$-conjoined with $P$'s predicate extensions in~$C$'s
superclasses. The \emph{qualified predicate} $\MathPred \specval {\PAt P C}
\specvals$ represents the $\LLAnd$-conjunction of $P$'s predicate extensions in
$C$'s superclasses, up to and including $C$.  The \emph{unqualified predicate}
$\MathPred \specval P \specvals$ is equivalent to $\MathPred \specval {\PAt P C}
\specvals$, where $C$ is $\specval$'s dynamic class. 
We allow predicates with missing parameters: semantically, missing parameters are existentially quantified.
Predicate definitions can be preceded by an optional {\tt public} modifier.
The role of the {\tt public} modifier is to export the definition of a predicate \emph{in a given class} to clients (see \eg the predicates in class \texttt{List} in the merge sort example in Section~\ref{subsec:sequential-mergesort}). 
For additional usage and formal definitions of {\tt public}, we refer to \cite[\S3.2.1]{HurlinPhd} and Sections~\ref{subsec:example-fj} and~\ref{sec:examples:lock}.

To be able to make mutable and immutable
instances of the same class, it is crucial to allow parametrization
of objects and predicates by permissions. For this, we include a
special type $\PermTy$ for fractional permissions.  Because class
parameters are instantiated by specification values, we extend
specification values with fractional permissions.  Fractional
permissions are represented symbolically: $1$ represents itself, and
if symbolic fraction $\specval$ represents concrete fraction $\fr$
then $\Split \specval$ represents~$\frac 1 2 \cdot \fr$.
\begin{displaymath}
  \begin{array}{rcl@{\qquad}rcl}
    & \specval \in \SpecValSet & \Is & \dots \Or\ \full\ \Or\ \Split \specval \ \Or \ \dots
  \end{array}
  \label{idx:specvals:slj}
\end{displaymath}

Quantified formulas have the shape  $\Qt \logvar T F$, where \textit{qt} is
a universal or existential quantifier, $\alpha$ is a variable whose scope is
formula $F$, and $T$ is $\alpha$'s type.
Because specification values $\specval$ and expressions $e$ may contain
logical variables $\alpha$,
quantified variables 
can appear in many positions: as type parameters; as
the first, third, and fourth parameter in {\tt PointsTo} predicates\footnote{Note that we forbid to quantify over
the second parameter of {\tt PointsTo} predicates, \ie the field name.
This is intentional, because this would complicate {\tt PointsTo}'s semantics.
We found this not to be a restriction, because we did not need this kind
of quantification in any of our examples.};
as predicate parameters \etc 

Class and interface declarations define \emph{class tables} (${\cls \ \subseteq\ \InterfaceSet \, \cup\,  \ClassSet}$) ordered by \emph{subtyping}.
We write $\dom\cls$ for the set of all type identifiers declared in~$\cls$. 
\emph{Subtyping} is defined in a standard way. 

We define several convenient \emph{derived forms} for specification formulas:
\begin{displaymath}
      \begin{array}[t]{c c}
	\multicolumn{2}{c}{
	  \PointsTo e f {\specval} T
	  \deq 
	  \Ex \logvar {T} {\;\Contains e f {\specval} \logvar}
	}
	\\
	\multicolumn{2}{l}{
	  \Perm e f {\specval}
	  \deq
	  \Ex \logvar {T} {\;\Contains e f {\specval} \logvar}
	  \quad \mbox{where $T$ is $\GGet e f$'s type}
	}
	\\
	\multicolumn{2}{c}{
	  F \LEquiv G
	  \deq 
	  \lpa F \MLI G \rpa \MCA \lpa G \MLI F \rpa
	}
         \\
	\multicolumn{2}{c}{
         \assures F G 
         \deq
         F \MLI \lpa F \LAnd G \rpa
        }
        \\
	\multicolumn{2}{c}{
	\ispartof F G
        \deq G \LImplies \lpa F \LAnd \lpa F \LImplies G \rpa \rpa 
        }
      \end{array}
\end{displaymath}

Intuitively, $\ispartof F G$ says that $F$ is a physical part of $G$:
one can take $G$ apart into $F$ and its complement $F \LImplies G$,
and can put the two parts together to obtain $G$ back.

\subsection{Operational Semantics}
\label{subsec:semantics:jll}

The operational semantics of our language is fairly standard, except
that the state does not contain a call stack, but only a single store
to keep track of the current receiver.  It operates on states,
consisting of a heap ($\HeapSet$), a command ($\CmdSet$), and a stack
($\StackSet$).  Section~\ref{sec:forkjoin} will extend the state, to
cope with multithreaded programs.  Given a heap $\hp$ and an object
identifier $o$, we write $\Fst {\hp(o)}$ to denote $o$'s dynamic type
and $\Snd {\hp(o)}$ to denote $o$'s store.  We use the following
abbreviation for field updates: $ \updtheap h o f v = \updtfun h o
{(\Fst{h(o)},\updtfun {\Snd{h(o)}} f v)} $. For initial states, we
define function ${\kw{init}} $ to denote a newly initialized object.
Initially, the heap and the stack are empty.

\begin{display}{Heap, Stack and State:} 
\clause{
  \begin{array}{cc}
  \ObjStoreSet = \ParFun \FldIdSet \ClValSet
&
  \hp \in \HeapSet = \ParFun \ObjIdSet {\TySet \times \ObjStoreSet}
\\
  \str \in \StoreSet = \ParFun \RdWrVarSet \ClValSet
&
  \stt \in \StateSet = \HeapSet \times \CmdSet \times \StoreSet
 \end{array}
}
\end{display}

The semantics of values, operators  and specification values are standard.
The formal semantics of the built-in operators is presented in \ref{sec:sem:op} and the formal semantics of specification values is defined in~\ref{sec:sem:specs}.
In addition, we allow one to use any further built-in operator that satisfies the following two axioms:

  \begin{enumerate}[label=\({\alph*}]
  \item\label{op-ax:hp-ext}
    If $\semo\op \hp ( \vs) =  w$ and $\hp \subseteq \hp'$, then
    $\semo\op {\hp'} ( \vs) =  w$.
  \item\label{op-ax:fld-updt}
    If $h' = \updtheap \hp o f u$,
    then $\semo\op {\hp} =  \semo\op{\hp'}$.
  \end{enumerate}

\noindent
The first of these axioms ensures that operators are invariant under heap extensions.
The second axiom ensures that operators do not depend on values stored on the heap.

\paragraph{Auxiliary syntax for method call and return.}
We introduce a derived form, $\Bind \lvar c {c'}$ that
assigns the result of a computation~$c$ to variable~$\lvar$. 
In its definition, we write $\fv c$ for the set of free
variables of $c$. Furthermore, we make use of 
some auxiliary syntax $\returnand \lvar v c$. This construct
is not meant to be used in source programs. Its purpose is to mark
method-return-points in intermediate program states. 
The extra $\kw{return}$ syntax allows us to associate a special proof rule with the post-state of method calls that characterizes this state. 
Technically, these definitions are chosen to support Lemma~\ref{lem:bind}, which is central for dealing with call/return in the preservation proof.
%
\begin{displaymath}
\begin{array}{rcl@{\quad}l}
  \Bind \lvar v c
  & \deq &
  \returnand \lvar v c
\\
  \Bind {\lvar} {(\Dcl T {\lvar'} c)} {c'}
  & \deq &
  \Dcl T {\lvar'} {\Bind {\lvar} c {c'}}
  & \mbox{if $\lvar'\not\in \fv{c'}$ and $\lvar' \neq \lvar$}
\\
  \Bind \lvar {(\FinDcl T {\rvar} {\lvar'} c)} {c'}
  & \deq &
  \FinDcl T {\rvar} {\lvar'} {\Bind \lvar c {c'}}
  & \mbox{if $\rvar \not\in\fv{c'}$}
\\
  \Bind \lvar {(\Seq \hc c)} c'
  & \deq &
  \Seq \hc {\Bind \lvar c {c'}}
\end{array}
\end{displaymath}
\begin{displaymath}
  \begin{array}{l}
    c \Is \dots\ \Or\ \returnand \lvar v c\ \Or\ \dots
    \\
    \mbox{\emph{Restriction:} This clause must not occur in source programs.}
  \end{array}
  \label{idx:return}
\end{displaymath}

\noindent
We can now also define sequential composition of commands as follows:
\begin{displaymath}
  \Seq c {c'}
  \Deq
  \Dcl \Void \lvar {\Bind \lvar \cmd {\cmd'}}
  \quad \mbox{where $\lvar \not\in \fv{c,c'}$}
\end{displaymath}

\paragraph{Small-step reduction.}
\label{sec:smallstep-reduction}
The state reduction relation $\rightarrow_\cls$ is given with respect
to a class table $\cls$. Where it is clear from the context, we omit
the subscript $\cls$.  
The complete set of the rules are defined in~\ref{subsec:apx:reductions}, here we only discuss the most important cases.

\renewcommand{\Thread}[3]{#2,#3}

\newcommand{\RuleRedDcl}[4]{#1
   { 
     \lvar \not\in\dom s
     \quad
     s' = \updtstore s \lvar {\df {#4}}
   }
   { 
     \State { 
              
              \Thread {#2} {\Dcl {#4} \lvar {#3}} s 
            } 
            \hp 
     \step
     \State { 
              
              \Thread {#2} {#3} {s'} 
            }
            \hp  
   }  
}

\newcommand{\RuleRedFinDcl}[4]{#1
   { 
       s(\lvar) = v
       \quad
       {#4} = \subst {v} \rvar {#3}
   }
   { 
     \State { 
              
              \Thread {#2} {\FinDcl T \rvar \lvar {#3}} s 
            } 
            \hp 
     \step
      \State { 
              
              \Thread {#2} {#4} {s} 
            }
            \hp  
   }  
}

\newcommand{\RuleRedVarSet}[4]{#1
   { 
     s' = \updtstore s \lvar {#4} 
   }
   { 
     \State { 
              
              \Thread {#2} {\SetVar \lvar {#4} {#3}} s 
            } 
            \hp 
     \step
     \State { 
              
              \Thread {#2} {#3} {s'} 
            }
            \hp  
   }
}

\newcommand{\RuleRedOp}[3]{#1
   { 
     \arity\op = |\vs|
     \quad
     \osem \op \hp \vs = w
     \quad
     s' = \updtstore s \lvar w 
   }
   { 
     \State { 
              
              \Thread {#2} {\Op \lvar \op \vs {#3}} s 
            } 
            \hp 
     \step
     \State { 
              
              \Thread {#2} {#3} {s'} 
            }
            \hp  
   }  
}

\newcommand{\RuleRedGet}[4]{#1
   { 
     s' = \updtstore s \lvar { \Snd {\hp (#4)}(f)} 
   }
   { 
     \State { 
              
              \Thread {#2} {\Get \lvar {#4} f {#3}} s 
            } 
            \hp 
     \step
     \State { 
              
              \Thread {#2} {#3} {s'}
            }
            \hp 
   }  
}

\newcommand{\RuleRedSet}[5]{#1
   { 
     h' = \updtheap h {#4} f {#5} 
   }
   { 
     \State { 
              
              \Thread {#2} {\FSetNoFin {#4} f {#5} {#3}} s
            } 
            \hp 
     \step
     \State { 
              
              \Thread {#2} {#3} s 
            }
            {\hp'} 
   }  
}

\newcommand{\RuleRedIfTrue}[5]{#1
  {
     \State {
              
	      \Thread {#2} {\Cond \True {#3} {#4} {#5}} s
            }
	    \hp 
     \step
     \State {
               \Thread {#2} {\Seq {#3} {#5}} s
            }
	    \hp 
  }
}

\newcommand{\RuleRedIfFalse}[5]{#1
  { 
     \State {
              
	      \Thread {#2} {\Cond \False {#3} {#4} {#5}} s
            }
	    \hp 
     \step
     \State {
               \Thread {#2} {\Seq {#4} {#5}} s
            }
	    \hp 
  }
}

\newcommand{\RuleRedWhileTrue}[5]{#1
  {
    \esem e h s = \True
  }
  {
     \State {
              
	      \Thread {#2} {\While {#3} {#4} {#5}} s
            }
	    \hp 
     \step
     \State {
               \Thread {#2} {\Seq {#4} {\While {#3} {#4} {#5}}} s
            }
	    \hp 
  }
}

\newcommand{\RuleRedWhileFalse}[5]{#1
  {
    \esem e h s = \False
  }
  {
     \State {
              
	      \Thread {#2} {\While {#3} {#4} {#5}} s
            }
	    \hp 
     \step
     \State {
               \Thread {#2} {#5} s
            }
	    \hp 
  }
}

\newcommand{\RuleRedCallSeq}[6]{#1
    {
      \DynTy h {#3} = \TyApp C {\specvals}
      \quad
      \mbody m {\TyApp C {\specvals}} 
      = \Mbody {\logvars,\logvars'} {\rvar_0;\rvars} {{#5}} 
      \quad
      {#6} = #5[#3/\rvar_0,\vs/\rvars]
    }
    { 
     \State { 
              
              \Thread {#2} {\Call \lvar {#3} m \vs {#4}} s 
            } 
            \hp 
     \step
     \State { 
              
              \Thread {#2} {\Bind \lvar {#6} {#4}} s
            }
            \hp  
   }  
}

\newcommand{\RuleRedAssert}[4]{#1
  {
    \State { \Thread {#2} {\Assert {#4} {#3}} s} \hp 
    \step
    \State { \Thread {#2} {#3} s} \hp 
  }
}

\newcommand{\RuleRedNoop}[3]{#1
  {
    \State { \Thread {#2} {\Seq \scmd {#3}} s} \hp 
    \step
    \State { \Thread {#2} {#3} s} \hp 
  }
}

\newcommand{\RuleRedReturn}[3]{#1
    { 
     \State {  \Thread {#2} {\returnand \lvar v {#3}} s } \hp  
     \step
     \State {  \Thread {#2} {\SetVar \lvar v {#3}} s } \hp  
   }
}

\newcommand{\RuleRedNew}[4]{#1
    {
      {#4} \notin \dom\hp 
      \quad
      \hp' = \updtfun \hp {#4} {(\TyApp C \perms,\init {\TyApp C \perms})}
      \quad
      s' = \updtstore s \lvar {#4}
    }
    {
     \State { 
              
              \Thread {#2} {\New \lvar C \perms {#3}} s 
            }
            \hp 
     \step
     \State { 
              
              \Thread {#2} {#3} {s'}
            } 
            {\hp'}
   }  
}

\begin{display}{State Reductions, $\stt \sstep \cls \stt'$:}
\RuleRedDcl {\stateaxiomcond{(Red Dcl)}} p c T
\label{rule:red-dcl}
\\[\jot]
\RuleRedFinDcl {\stateaxiomcond{(Red Fin Dcl)}} p c {c'}
\label{rule:red-fin-dcl}
\\[\jot]
\RuleRedNew {\stateaxiomcond{(Red New)}} p \cmd o
\label{rule:red-new}
\\[\jot]
\RuleRedCallSeq{\stateaxiomcond{(Red Call)}} p o c {c_m} {c'}
\label{rule:red-call}
\label{idx:op:sem}
\end{display}


In~\ref{rule:red-dcl}, read-write variables are initialized to a default value.
In~\ref{rule:red-fin-dcl}, declaration of read-only variables is handled by substituting
the right-hand side's value for the newly declared variable in the continuation.
In~\ref{rule:red-new}, the heap is extended to contain a new object. 
In~\ref{rule:red-call}, $\rvar_0$ is the formal method receiver and
$\rvars$ are the formal method parameters. Like for declaration of read-only variables,
both the formal method receiver and the formal method parameters are substituted by the actual
receiver and the actual method parameters.

\subsection{Validity of Resource Formulas}
\subsubsection{Augmented heaps}
\label{subsubsec:sl:augmented-heaps}

To define validity of our resource formulas, we augment the heap with a permission table.
\emph{Augmented heaps} $\ahp$ as models of our
formulas, 
\label{idx:augmented heaps} range over the set $\AugHeapSet$ with a
binary relation $\compatibleSym \subseteq \AugHeapSet \times \AugHeapSet$
(the \emph{compatibility relation}) and a partial binary operator
$\LLAnd : \Fun \compatibleSym \AugHeapSet$ (the \emph{augmented heap joining
  operator}) that is associative and commutative. 
Concretely, augmented heaps are pairs
$\ahp = \Cpl \hp \pt$ of 
a \emph{heap} $\hp$ and
a \emph{permission table} 
  $\pt \in \Fun {\ObjIdSet \times \FldIdSet} {[0,1]}$
\index{$\pt$ \hfill A permission table}.
To prove soundness of the verification rules for field updates and allocating new objects, augmented heaps must satisfy the following axioms:
\label{idx:star:join}
\begin{enumerate}
\item\label{rsc-pos}
  $\pt(o,f) > 0$ for all $o \in \dom h$ and $f \in \dom{\Snd{h(o)}}$.
\item\label{rsc-neg} 
  $\pt(o,f) = 0$ for all $o \not\in\dom h$ and all $f$.
\end{enumerate}

Axiom~\ref{rsc-pos} ensures that the (partial) heap $h$ only contains
cells that are associated with strictly positive
permissions. 
Axiom~\ref{rsc-neg} ensures that all unallocated objects have minimal permissions (with respect to the augmented heap order presented below).

Each of the two augmented heap components defines $\compatibleSym$ (compatibility) and $\LLAnd$ (joining) operators.
\emph{Heaps} are compatible if they agree on shared object types and memory content:
\begin{displaymath}
  \compatible h {h'}\;\text{iff}\;
  \left\{\begin{array}{l}
     (\forall o \in \dom h \cap \dom {h'})\ ( 
     \\
     \qquad\Fst{h(o)} = \Fst{h'(o)} \mbox{ and } \\
     \qquad(\forall f \in \dom {\Snd{h(o)}} \,\cap\, \dom {\Snd{h'(o)}})(\;\Snd {h(o)}(f) = \Snd {h'(o)}(f)\,)\ )
    \end{array}\right.
    \label{idx:comp:heap:slj}
\end{displaymath}

To define heap joining, we lift set union to deal with undefinedness:
$f \vee g = f \cup g$, $f \vee \undef = \undef \vee f = f$.
Similarly for types: $T \vee \undef = \undef \vee T = T \vee T = T$.
\begin{displaymath}
  \Fst {(h\;\LLAnd\;h')(o)} \deq
  \Fst{h(o)} \vee\, \Fst{h'(o)}
  \qquad
  \Snd {(h\;\LLAnd\;h')(o)} 
  \deq
  \Snd {\hp(o)} \vee\, \Snd {\hp'(o)} 
  \label{idx:star:heap:slj}
\end{displaymath}

\emph{Permission tables} join by point-wise addition: $(\pt \LAnd \pt')(o) \deq \pt(o) + \pt'(o)
$,
where compatibility ensures that the sums never exceed 1, \ie $\compatible \pt {\pt'}
\ \text{iff}\ 
(\forall o)(\pt(o) + \pt'(o) \leq 1)
$.

We define projection operators: ${\Cpl \hp \pt}_{\kw{hp}} \deq \hp$ and ${\Cpl \hp \pt}_{\kw{perm}} \deq \pt$.
Moreover, ordering on heaps, permission tables, and augmented heaps are defined as follows:
\begin{displaymath}
  \begin{array}{rcll}
      \hp \leq \hp' & \deq & (\exists \hp'')(\hp \LAnd \hp'' = \hp')
      & 
      : \hp \textit{ contains less memory cells than } \hp'
    \\
    \pt \leq \pt' & \deq & (\exists \pt'')(\pt \LAnd \pt'' = \pt')
	&
     : \pt\textit{'s permissions are less than }\pt'\textit{'s permissions}
    \\
    \ahp \leq \ahp' & \deq & (\exists \ahp'')(\ahp \LAnd \ahp'' = \ahp')
    &
     : \ahp\textit{'s components are all less than }\ahp'\textit{'s components}
  \end{array}
\end{displaymath}

\subsubsection{Meaning of Formulas}
\label{subsubsec:sl:semantics}

To define the meaning of
predicates, the notion of predicate environments is used. A predicate
environment $\pe$ maps predicate identifiers to concrete heap
predicates. Following Parkinson~\cite{Parkinson05b}, it is defined as a
least fixed point of an endofunction $\pef_{ct}$ on predicate environments. We do
not detail its definition further, but instead refer to Parkinson's
thesis.

An augmented heap $\ahp$ is well-formed under typing environment
$\Gamma$, \emph{i.e.}, $(\goodrsc \Gamma \ahp)$, whenever the heap
and the permission table are well-formed, \emph{i.e.} $\goodheap
\Gamma \ahph$ and $\pt(o,f) > 0$ implies $o \in \dom \Gamma$.
Furthermore, given formula $F$ and stack $s$, we say $(\OLDwtt
\Gamma \pe \ahp r s F)$ whenever the predicate environment is a least
fixed point, and the augmented heap, stack, and formula are
well-formed, \emph{i.e.}, $\goodrsc \Gamma \ahp$, $\goodstore \Gamma
s$, and $\goodform \Gamma F$, respectively\footnote{All typing
  judgments are defined in~\ref{subsubsec:types}.}.  Now we define a
forcing relation of the form $\holds \Gamma \pe \ahp s F$, which
intuitively expresses that if $\holds \Gamma \pe \ahp s F$ holds, then
the
augmented heap $\ahp$ is a state 
that is described by $F$.  The relation $(\OLDholds \Gamma \pe \ahp r
s F)$ is the unique subset of $(\OLDwtt \Gamma \pe \ahp r s F)$ that
satisfies the clauses in Figure~\ref{fig:forcing}.

\begin{figure}[t]
\begin{displaymath}
\begin{array}{rcl@{\  }c@{\ \ }l}
\Gamma \vdash \pe; \Cpl \hp \pt ;s 
& \models & e
& \hspace{-2cm}\iff & \hspace{-1cm} \esem e h s = \True
\medskip\\
\Gamma \vdash 
  \pe; \Cpl h \pt; s & \models & 
  \Contains e f \specval {e'} 
& \iff &
\left\{\begin{array}{l}
 \esem e h s = o, 
 \Snd{h(o)}(f) = \esem {e'} h s, \\
 \mbox{ and } \sem \specval \leq \pt(o,f)  
\end{array}\right.
\medskip\\
\Gamma \vdash \pe;\ahp;s & \models & \Pred \Null \kappa \specvals
& \hspace{-2cm}\iff & \hspace{-1cm}
\mbox{true}
\medskip\\
\Gamma \vdash \pe;\ahp;s & \models & \Pred o {\PAt P C} \specvals
& \hspace{-2cm}\iff & \hspace{-1cm}
\left\{\begin{array}{l}
  \Fst{\ahph(o)} \subty \TyApp C {\specvals'}
  \mbox{ and} \\ 
   \pe(\PAt P C)(\specvals',\ahp,o,\specvals) = 1 
\end{array}\right. 
\medskip\\
\Gamma \vdash \pe;\ahp;s & \models & \Pred o P {\specvals}
& \hspace{-2cm}\iff & \hspace{-1cm}
\left\{\begin{array}{l}
  (\exists \specvals'')(
  \Fst{\ahph(o)} = \TyApp C {\specvals'} 
  \mbox{ and } \\
  \pe(\PAt P C)(\specvals',\ahp,o,(\specvals,\specvals'')) = 1)
\end{array}\right. 
\medskip\\
\Gamma \vdash \pe;\ahp;s & \models & F \LAnd G
& \hspace{-2cm}\iff & \hspace{-1cm}
\left\{\begin{array}{l}
(\exists \ahp_1,\ahp_2)(\ahp = \ahp_1 \LAnd \ahp_2, \\
\OLDholds \Gamma \pe {\ahp_1} r s F \mbox{ and } \\
\OLDholds \Gamma \pe {\ahp_2} r s G)
\end{array}\right.
\medskip\\
 \Gamma \vdash \pe;\ahp;s & \models & F \LImplies G
& \hspace{-2cm}\iff & \hspace{-1cm}
\left\{\begin{array}{l}
(\forall \Gamma' \hpsup \Gamma, \ahp') (
\\\  \compatible{\ahp}{\ahp'} \mbox{ and } 
     \OLDholds {\Gamma'} \pe {\ahp'} r s F
\\\ 
 \Rightarrow\ \OLDholds {\Gamma'} \pe {\ahp \LAnd \ahp'}  r s G \ )
\end{array}\right.
\medskip\\
\Gamma \vdash \pe;\ahp;s & \models & F \CAnd G
& \hspace{-2cm}\iff & \hspace{-1cm}
  \OLDholds \Gamma \pe \ahp r s F
  \mbox{ and } \OLDholds \Gamma \pe \ahp r s G
\medskip\\
\Gamma \vdash \pe;\ahp;s & \models & F \COr G
& \hspace{-2cm}\iff & \hspace{-1cm}
  \OLDholds \Gamma \pe \ahp r s F 
  \mbox{ or } \OLDholds \Gamma \pe \ahp r s G
\medskip\\
\Gamma \vdash \pe;\ahp;s & \models & \Ex \logvar T F
& \hspace{-2cm}\iff & \hspace{-1cm}
\left\{\begin{array}{l}
  (\exists \specval)(\  
   \hpenv \Gamma \vdash \hastype \specval T
   \mbox{ and }
   \\
   \ \OLDholds \Gamma \pe \ahp r s {\subst \specval \logvar F}\ )
\end{array}\right.
\medskip\\
\Gamma \vdash \pe;\ahp;s & \models & \Fa \logvar T F
& \hspace{-2cm}\iff & \hspace{-1cm}
\left\{\begin{array}{l}
(\forall  \Gamma' \hpsup \Gamma, \ahp' \geq \ahp, \specval)( 
\\\ 
\goodheap {\hpenv {\Gamma'}} {\ahph'} \mbox{\ \ and\ \ } 
\hpenv{\Gamma'} \vdash \hastype \specval T 
\\\ 
\Rightarrow\ 
\OLDholds {\Gamma'} \pe {\ahp'} r s {\subst \specval \logvar F}\ )
\end{array}\right.
\end{array}
\end{displaymath}
\caption{Meaning of formulas}\label{fig:forcing}
\end{figure}

\subsection{Verification}
This section first presents the proof theory,
and next, Hoare triples to verify Java-like programs are introduced.

\subsubsection{Proof Theory}
\label{subsec:sl:proof:theory}

As usual, Hoare triples use a logical consequence judgment.
We define logical consequence proof-theoretically. The proof theory has two
judgments: 
\begin{displaymath}
\begin{array}{l@{\qquad}l}
\entails v \Gamma \Fs G &  
  \mbox{$G$ is a logical consequence of the $\LAnd$-conjunction of $\Fs$}
\\
\axiom v \Gamma F &    
  \mbox{$F$ is an axiom}
\end{array}
\end{displaymath}
where $\Fs$ is a \emph{multiset} of formulas, and parameter $v$
represents the \emph{current receiver}. 
  The receiver parameter is needed to determine the scope of predicate
definitions: a receiver $v$ knows the definitions of predicates of the form
$\GGet v P$, but not the definitions of other predicates. 
In source code verification, the receiver parameter is always $\This$ and can
thus be omitted. We explicitly include the receiver parameter in the general
judgment, because we want the proof theory to be closed under 
value substitutions.

\paragraph{Semantic Validity of Boolean Expressions.}
The proof theory depends on the relation  $\Gamma \models e$ (``$e$ is
valid in all well-typed heaps''), which we do not axiomatize (in an implementation, we would
use an external and dedicated theorem prover to decide this relation) but instead we define as validity over all \emph{closing substitutions}.
Let $\sigma$ range over \emph{closing substitutions},
i.e, elements of $\ParFun \VarSet \ClValSet$: 
\begin{displaymath}
\begin{array}{c}
\recallmath
  { 
    \dom \sigma = \dom \Gamma \cap \VarSet
    \quad
    (\forall x \in \dom\sigma)
    (\hpenv\Gamma \vdash \hastype {\sigma(x)} {\Gamma(x)[\sigma]})
  }
  {
    \goodsubst \Gamma \sigma
  }
\\[\GAP]
\ClSubstSet\Gamma  \deq 
  \setcomp {\ \sigma \ } {\ \goodsubst \Gamma \sigma\ } 
\end{array}
\end{displaymath}

We say that a heap $h$ is \emph{total} iff for all $o$ in $\dom h$ and all $f \in \dom{\fields{\Fst{h(o)}}}$,
 $f \in \dom{\Snd{h(o)}}$. Then we have: 
%
$
\HeapSet(\Gamma)  \deq 
  \setcomp {\ h\ } {\ 
             \goodheap {\hpenv \Gamma} h
             \mbox{ and $h$ is total}
            \ } 
$.
Now, we define $\Gamma \models e$ as follows: 
\begin{displaymath}
\begin{array}{c}
\Gamma \models e  \quad\mbox{iff}\quad 
\left\{\begin{array}{l}
  \Gamma \vdash \hastype e \BoolTy \mbox{ and }
  \\
  \ \forall \Gamma' \hpsup \Gamma, h \in \HeapSet(\Gamma'),
             \sigma \in \ClSubstSet{\Gamma'}:  \
  ~~(\ \esem {e[\sigma]} h \emptyset = \True\ ) 
\end{array}\right.
\end{array}
\end{displaymath}

\paragraph{Natural Deduction Rules.}
\label{subsec:natrualdeductionrules}

The logical consequence judgment of our Hoare logic is defined in a standard way based
on the natural deduction calculus of \emph{(affine) linear logic}
\cite{Wadler93}, which coincides with BI's natural deduction calculus
\cite{OHearnP99} on our restricted set of logical operators. The complete list is presented in~\ref{subsec:apx:verifications}.

\paragraph{Axioms.} 

In addition to the logical consequence judgment, sound
\emph{axioms} capture additional properties of our model. These
additions do not harm soundness, as shown by
Theorem~\ref{thm:sound-entails} below. Table~\ref{table:axioms}
presents the different axioms that we use:

\begin{table}[t]
\hspace{-0.5cm}
$
\begin{array}{c @{\;\;} l}
  \Gamma; v \vdash
  \PointsTo e f \specval {e'}
  \LEquiv
  \left(
    \begin{array}{c}
	     \PointsTo e f {\frac \specval 2} {e'} \\
	     \LAnd \\
	     \PointsTo e f {\frac \specval 2} {e'}
    \end{array}
  \right)
\labelandtext{(Split/Merge)}{axiom:split/merge}

\\\\[-1.25ex]

  \begin{array}{c}
    (\Gamma \vdash \hastype v {\TyApp C {\specvals''}}
    \wedge
    \pbody {\GGet v P} {\specvals,\specvals'} C {\specvals''} = 
    \Pbody F {\TyApp D {\specvals'''}})
    \\
    \Rightarrow\ 
    \axiom v \Gamma 
    {      
      \Pred v {\PAt P C} {\specvals,\specvals'}
      \LEquiv 
      \lpa F \LAnd \Pred v {\PAt P D} {\specvals} \rpa
    }
  \end{array}
\labelandtext{(Open/Close)}{axiom:open/close}

\\\\[-0.5ex]

\Gamma;v \vdash
\Pred \specval P \specvals
\LEquiv
\Ex \logvars \Ts {\Pred \specval P {\specvals,\logvars}}
\labelandtext{(Missing Parameters)}{axiom:missing-parameters}

\\

\Gamma;v \vdash
  {\ispartof {\Pred \specval {\PAt P C} \specvals}
             {\Pred \specval P {\specvals}}}
\labelandtext{(Dynamic Type)}{axiom:dynamic-type}

\\

    C \subcl D
    \ \Rightarrow\ 
    \axiom v \Gamma {\ispartof {\Pred \specval {\PAt P D} \specvals} {\Pred \specval {\PAt P C} {\specvals,\specvals'}} } 
\labelandtext{({\tt ispartof} Monotonic)}{axiom:ispartofmono}

\\

\axiom v \Gamma 
       {\lpar \Pred \specval {\PAt P C} \specvals \,\LAnd\; 
        \isclassof C \; \specval \rpar 
        \,\LImplies\, \Pred \specval P \specvals}
\labelandtext{(Known Type)}{axiom:known-type}

\\

  \axiom v \Gamma {\Pred \Null \kappa \specvals}
\labelandtext{(Null Receiver)}{axiom:null-receiver}

\\

\axiom v \Gamma \True
\labelandtext{(True)}{axiom:true}

\\
\axiom v \Gamma {\False \LImplies F}
\labelandtext{(False)}{axiom:false}


\\\\[-0.5ex]

\begin{array}{c}
(\Gamma \vdash \hastype {e,e'} T \ \wedge\ 
 \goodform {\Gamma,\hastype x T} F) 
\\ 
\Rightarrow
\axiom v \Gamma {\lpa \subst e x F \LAnd e \jdeq e'\rpa
               \LImplies
               \subst {e'} x F}
\end{array}
\labelandtext{(Substitutivity)}{axiom:substitutivity}

\\\\[-1ex]

(\Gamma \models \jnot {e_1} \jor \jnot {e_2} \jor e')
\ \Rightarrow\ 
\axiom v \Gamma {\lpa e_1 \LAnd e_2 \rpa \LImplies e'}
\labelandtext{(Semantic Validity)}{axiom:semantic_validity}

\\\\[-0.5ex]

\axiom v \Gamma
  {\begin{array}{c}
	{\lpa \PointsTo e f \specval {e'}
                  \,\CAnd\, 
                  \PointsTo e f {\specval'} {e''} \rpa }
	\\
   \tjkw{assures}\;\;
            { e' \jdeq e'' } 
  \end{array}}
\labelandtext{(Unique Value)}{axiom:unique}

\\\\[-0.5ex]

(\Gamma \vdash \hastype e T)
\ \Rightarrow\ 
\axiom v \Gamma \Ex \logvar T {e \jdeq \logvar}
\labelandtext{(Well-typed)}{axiom:welltyped}

\\

\axiom v \Gamma \lpa F \CAnd e \rpa  \LImplies \lpa F \LAnd e \rpa
\labelandtext{(Copyable)}{axiom:copyable}

\\
( \Gamma \vdash   \hastype \pi {\TyApp t {\specvals'}} \ \wedge \ \kw{axiom}\lpar {\TyApp t {\specvals'} } \rpar = F ) \Rightarrow\ \Gamma ; v \vdash F[\pi / \java{this} ]
\labelandtext{(Class)}{axiom:class}

\end{array}
$
\caption{Overview of Axioms}
\label{table:axioms}
\end{table}

\begin{itemize}
\item ~\ref{axiom:split/merge} regulates permission accounting
(where $v$ denotes the current receiver and
$\frac \specval 2$ abbreviates $\Split \specval$). 
\item ~\ref{axiom:open/close} allows predicate receivers to
toggle between predicate names and predicate definitions (where~--~as
defined in~\ref{sec:lookup}~--~ $\pbody {\GGet o P} {\specvals'} C
\specvals$ looks up $\Pred o P {\specvals'}$'s definition in the type
$\TyApp C \specvals$ and returns its body $F$ together with $\TyApp C
\specvals$'s direct superclass $\TyApp D {\specvals''}$): Note that
the current receiver, as represented on the left of the $\vdash$, has
to match the predicate receiver on the right. This rule is the only
reason why our logical consequence judgment tracks the current
receiver. Note that $\PAt P C$ may have more parameters than
$\PAt P D$: following Parkinson~\cite{Parkinson05b} we allow
subclasses to extend predicate arities. 

\item \ref{axiom:missing-parameters} expresses that missing predicate
  parameters are existentially quantified. 
\item ~\ref{axiom:dynamic-type} states that a predicate at a
receiver's dynamic type (\ie without \tjkw{@}-selector)
is stronger than the predicate at its static type. In combination with
the axiom \ref{axiom:open/close}, this allows us to open
and close predicates at the receiver's static type. The axiom
\ref{axiom:ispartofmono} is similar. 
\item ~\ref{axiom:known-type} allows one to drop the class
modifier $C$ from $\GGet \specval {\PAt P C}$ if we know that~$C$ is
$\specval$'s dynamic class.
\item Axioms \ref{axiom:null-receiver}, \ref{axiom:true} and \ref{axiom:false} define the semantics of predicates with
$\Null$-receiver, and of $\True$ and $\False$, respectively.

\item The \ref{axiom:substitutivity} axiom allows to replace expressions by
equal expressions, while \ref{axiom:semantic_validity} lifts semantic
validity of boolean expressions to the proof theory.
\item \ref{axiom:unique} captures the fact that fields point to
a unique value.  Recall that we write "$\assures F G$'' to abbreviate
"$F \LImplies \lpa F \LAnd G\rpa$'' (see Section~\ref{subsec:syntax:jll}).
\item \ref{axiom:welltyped} captures that all well-typed
closed expressions represent a value (because built-in operations are
total). 
\item \ref{axiom:copyable} expresses copyability of
boolean expressions.

\item \ref{axiom:class} allows the application of class axioms where $\kw{axiom}\lpar {\TyApp t {\specvals'} } \rpar$ is the $\LAnd$-conjunction of all class axioms in $\TyApp t {\specvals'}$ and its supertypes.
\end{itemize}

\paragraph{Soundness of the proof theory.}
We define {semantic entailment} $\sentails \Gamma \pe \Fs G$:
\begin{displaymath}
\begin{array}{rcl}
\holds \Gamma \pe \ahp s {F_1,\dots,F_n}
& \hspace{-0.5ex}\mbox{iff} & \hspace{-0.5ex}
\holds \Gamma \pe \ahp s {F_1 \LAnd \cdots \LAnd F_n}
\\
\sentails \Gamma \pe \Fs G
& \hspace{-0.5ex}\mbox{iff} & \hspace{-0.5ex}
(\forall \Gamma,\ahp,s)
(\holds \Gamma \pe \ahp s \Fs 
 \ \Rightarrow\  
 \holds \Gamma \pe \ahp s G)
\end{array}
\end{displaymath}

Now, we can express the proof theory's soundness:

\begin{thm}[Soundness of Logical Consequence]
\label{thm:sound-entails}
If $\pef_\cls(\pe) = \pe$ and $(\entails o \Gamma \Fs G)$,
then $(\sentails \Gamma \pe \Fs G)$.
\end{thm}

\proof
The proof of the theorem is by induction on $(\entails v \Gamma \Fs G)$'s proof tree.
The pen and paper proof can be found in~\cite[\S R]{HaackH08}.
\qed

\paragraph{Remark.} Note that the receiver parameter $o$ is only used
in the assumption, and does not play a role in the semantics of
logical consequence.  The reason why we included the receiver
parameter in the logical consequence judgment is the
\ref{axiom:open/close} axiom.  This axiom permits the opening/closing
of only those abstract predicates that are defined in the
receiver-parameter's class.  While limiting the visibility of
predicate definitions is not needed for soundness of logical
consequence, it is important from a software engineering standpoint,
because it provides a well-defined abstraction boundary.

\subsubsection{Hoare Triples}
\label{subsec:sl:hoare}

Next we present Hoare rules to verify programs written in Section~\ref{subsec:syntax:jll}'s language. 
Appendix B of Hurlin's PhD thesis~\cite{HurlinPhd} lists the complete collection of Hoare rules, presented here and in the next sections.
 Hoare triples for \emph{head commands} have the following form: 
$
  \hhoare v \Gamma F \hc G 
$.
Our judgment for \emph{commands} combines typing and Hoare triples:
$
\hoare v \Gamma F c T {(U \ \logvar)(G)}
$
where $G$ is the postcondition, $\logvar$ refers to the return value, and
$T$ and $U$ are types of the return value (possibly supertypes of the return value's dynamic type).
In derivable judgments, it is always the case that $U \subty T$.

Here we explain some important rules listed in Figure~\ref{fig:hoare:triples}. The rest of the rules are standard and provided in~\ref{subsec:apx:verifications}.
The field writing~\ref{rule:hoa-fld-set} requires the full permission ($1$) on the object's field 
and it ensures that the heap has been updated with the value assigned. 
The rule for field reading~\ref{rule:hoa-get} requires a {\tt
  PointsTo} predicate with \emph{any} permission $\perm$.  The rule
for creating new objects~\ref{rule:hoa-new} has precondition {\tt
  true}, because we do not check for out of memory errors.  After
creating an object, all its fields are writeable: the $\GGet \lvar
\initpred$ predicate (formally defined
in~\ref{sec:lookup}) 
$\LLAnd$-conjoins the predicates $\PointsTo \lvar f 1 {\df T}$ for all
fields $\Fld T f$ in $\lvar$'s class, \ie expressing that all fields
have their default values.  The rule for method
calls~\ref{rule:hoa-call} is verbose, but standard.  Importantly, our
system includes the~\ref{rule:hoa-frame} rule, which allows to reason
locally about methods. To understand this rule, note that $\fv F$ is
the set of free variables of $F$ and that we write $x \not\in F$ to
abbreviate $x \not \in \fv F$.  Furthermore, we write $\writes \hc$
for the set of read-write variables $\lvar$ that occur freely on the
left-hand-side of an assignment in $\hc$.  \ref{rule:hoa-frame}'s side
condition on variables is standard~\cite{OHearn07,Parkinson05b}.
Bornat showed how to get rid of this side condition by treating
variables as resources~\cite{BornatCY05}.  We should stress here that
the rule \ref{rule:hoa-con} applies \emph{only for} head commands.
Therefore, the correctness proof for a method body can never end in an
application of a rule of \ref{rule:hoa-con}.  However, it is possible
to apply this rule at the caller site and in the proof of the method
body at any point before applying the \ref{rule:hoa-val} rule that
introduces the outer existential.  Notice that since we do not have
the \emph{conjunction rule} in our rule set, we do not need the preciseness
condition of the resource invariant~\cite{Gotsman:2011:PCR}.

\begin{figure}[hbtp] 

\begin{center}
\staterulelabel{(Fld Set)}
  {
    \Gamma \vdash \hastype {u,w} {U,W}
    \quad
    \Fld W f \in \fields U
  }
  {
    \hhoare v \Gamma 
       {\Contains u f \one W} 
       {\HdFSetNoFin u f w}
       {\Contains u f \one w}
  }
\label{rule:hoa-fld-set}
\end{center}

\begin{center}
\staterulelabel{(Get)}
   {  
     \Gamma \vdash \hastype {u,\specval,w} {U,\PermTy,W}
     \quad
     \Fld W f \in \fields U
     \quad
     W \subty \Gamma(\lvar)
    }
   {
     \hhoare v \Gamma {\Contains u f \perm w} 
                      {\HdGet \lvar u f} 
		      {
                        \Contains u f \perm w 
                        \LAnd 
			\lvar \jdeq w
                      }
   } 
\label{rule:hoa-get}
\end{center}
%

\begin{center}
\staterulelabel{(New)}
  {
    \TyApp C {\TypedVar \Ts \logvars} \in \cls
    \quad
    \Gamma \vdash \hastype \specvals {\subst \specvals\logvar \Ts}
    \quad
    \TyApp C \perms \subty \Gamma(\lvar)
  }
  {
      \hhoare v \Gamma 
         \True
         {\HdNew \lvar C \perms}
        {
        \GGet \lvar \initpred
        \,\LAnd\, \iisclassof C \lvar
        }
  }
\label{rule:hoa-new}
\end{center}
%

\begin{center}
\staterulelabelbis{(Call)}
  {
  \begin{array}{c}
    \mtype m {\TyApp t \perms} 
    = \TyAbs {\TypedVar \Ts \alphas}\; {\Mspec {G} 
					{ \lpa {\logvar'} \rpa \lpa {G'} \rpa} }
   \\ \hspace{1cm}U\;m\;\lpa{\TyApp t \perms}\;{\rvar_0} , \Ws \ \rvars \rpa
   \\
    \quad	
    \sigma = (u/\rvar_0,\perms'/\logvars,\ws/\rvars)
    \quad
    \Gamma \vdash \hastype {u,\specvals',\ws} 
        {\TyApp t \perms,\Ts[\sigma],\Ws[\sigma]}
    \quad
    U[\sigma] \subty \Gamma(\lvar) 
  \end{array}}
  {
     \hhoare v \Gamma {u \jneq \Null \LAnd\, G[\sigma]}
        {\HdCall {\lvar} u m \ws} 
     {
       \Ex {\logvar'} {U[\sigma]\;} {
         \logvar' \jdeq \lvar 
         \,\LAnd\;
         G'[\sigma] }
     }
  }
 \label{rule:hoa-call}
\end{center}

\begin{center}
\staterulelabel{(Frame)}
  { 
    \hhoare v \Gamma F \hc G
    \quad
    \goodform \Gamma H
    \quad
    \fv H \cap \writes \hc = \emptyset
  } 
  {
    \hhoare v \Gamma {F \LAnd H} \hc {G \LAnd H}
  }
\label{rule:hoa-frame}
\end{center}

\begin{center}
\staterulelabelbis{(Consequence)}
  { 
    \begin{array}{l}
    \hhoare v \Gamma {F'} \hc {G'}
    \\
    \entails v \Gamma F {F'}
    \quad
    \entails v \Gamma {G'} G
    \end{array}
  }
  {
    \hhoare v \Gamma F \hc G
  }
\label{rule:hoa-con}
\staterulelabel{(Exists)}
  {
    \hhoare v {\Gamma,\hastype \logvar T} F \hc G
  }
  {
    \hhoare v \Gamma {\Ex \logvar T F} \hc {\Ex \logvar T G}
  }
\label{rule:hoa-exists}
\staterulelabel{(Val)}
  {
   \entails v \Gamma F {\subst w \logvar G}
   \quad
   \Gamma \vdash \hastype w {U \subty T} 
   \quad
   \goodform{\Gamma,\hastype \logvar U} G
  }
  {
    \hoare v \Gamma F w T {\lpa U \ \logvar \rpa \lpa G \rpa } 
  }
\label{rule:hoa-val}

\end{center}


\caption{Hoare triples}
\label{fig:hoare:triples}
\end{figure}

The following lemma states that Hoare proofs can be normalized for head commands, which is needed in the preservation proof in order to deal with structural rules.

\begin{lem}[Proof Normalization]
\label{lem:normalization}
If  $ \hhoare v \Gamma F \hc G $ is derivable, then it has a proof where every path to the proof goal ends in zero or more applications of \ref{rule:hoa-con} and \ref{rule:hoa-exists} preceded by exactly one application of \ref{rule:hoa-frame}, preceded by a rule that is not a structural rule (\ie a rule different from \ref{rule:hoa-frame}, \ref{rule:hoa-con} and \ref{rule:hoa-exists}.)
\end{lem}
\proof
See~\cite[Chap. 6]{HurlinPhd}.
\qed

\subsection{Verified Programs}
\label{subsec:verified:ic}

To prove soundness of the logic, we need to define the notion of a verified
program. We first define judgments for
verified interface and classes, which in turn depend on the notions of
method and predicate subtyping, and soundness of axioms.

\paragraph{Subtyping.}

We define method and predicate subtyping. 
We present the method subtyping rule in its full generality, accounting for logical parameters:

\begin{displaymath}
\recallmath
{\begin{array}{l}
   \entails {\rvar_0} {\Gamma,\hastype {\rvar_0} {V_0}} \True 
     \Fa {\Ts'}\;\logvars  
     \Fa {\Vs'}\;{\rvars} \lpa 
     F' \LImplies \\
      \hspace{4.45cm} \Ex {\logvars'} \Ws {F \LAnd 
       \Fa {\result} U {G\LImplies G'} } \rpa
\end{array}}
  {
  \begin{array}{l}
   \Gamma \vdash
   \SMT {\TypedVar \Ts \logvars, \TypedVar \Ws {\logvars'}} F 
        G
	U m
        {\TypedVar {V_0} {\rvar_0}, \TypedVar \Vs {\rvars}} 
   \\
   \qquad\quad
   \subty
   \SMT {\TypedVar {\Ts'} \logvars} {F'} 
        {G'}
	{U'} m
        {\TypedVar {V_0'} {\rvar_0}, \TypedVar {\Vs'} {\rvars}} 
  \end{array}
  }
\end{displaymath}

Predicate type $\pdty$ is a subtype of $\pdty'$,
if $\pdty$ and $\pdty'$ have the same name and $\pdty$'s parameter signature
``extends'' $\pdty'$'s parameter signature:
\begin{displaymath}
\recallmath
  {}
  { 
    \pred\; 
    \TyApp P {\TypedVar \Ts \logvars, \TypedVar {\Ts'} {\logvars'}}
    \subty
    \pred\;
    \TyApp P {\TypedVar \Ts \logvars}
  }
\end{displaymath}

\paragraph{Soundness of Class Axioms.} 
So far {\tt axiom}s are used to export useful relations between predicates to clients. 
A class is $\kw{sound}$ if all its axioms are sound (the lookup function for axioms ($\kw{axiom}$) is defined in~\ref{sec:lookup}).
To prove soundness of axioms, we define a restricted logical consequence judgment that disallows the application of class axioms for proving their soundness, in order to avoid circularities:

\begin{smallmath}
\vdash'\ \  \Deq\ \ \vdash \mbox{ without class axioms }
\end{smallmath}

\paragraph{Verified Interfaces and Classes.}
\label{subsec:verifiedinterfacesclasses}

Next, we define same sanity conditions on classes and interfaces, which are later used to ensure that we only verify sane programs.
Judgment $\Extends C \Ts \logvars U$ expresses that: 
(1) class $C$ does not redeclare inherited fields, and (2) methods and predicates
overridden in class $C$ are subtypes of the corresponding methods and predicates implemented in class $U$.
Judgment $\TypeExtends I \Ts \logvars U$ expresses that: 
 methods and predicates overridden in interface $I$ are subtypes of the corresponding methods and predicates
declared in $U$.
Judgment $\Implements C \Ts \logvars U$ expresses that:
(1) methods and predicates declared in interface $U$ are implemented in $C$, and
(2) methods and predicates implemented in $C$ are subtypes of the corresponding
methods and predicates declared in $U$. These judgments are defined formally in~\ref{subsec:apx:verifications}.

Finally, verified methods, verified interfaces and verified classes are defined formally in~\ref{subsec:apx:verifications}. 
Later, when we verify a user-provided program, we will assume that the class table is verified.

\paragraph{Soundness of the Program Logic}
\label{subsec:verified:programs}

We now have all the machinery to define what is a verified program.
To do so, we extend our verification rules to runtime states. 
Of course, the extended rules are never used in verification, but instead
define a global state invariant, $\goodstate \stt$, which is preserved by
the small-step rules of our operational semantics.
Our forcing relation $\models$ from Section~\ref{subsubsec:sl:semantics} assumes formulas without 
logical variables: we deal with those by substitution, ranged over by
$\sigma \in \ParFun \LogVarSet \SpecValSet$. We let 
$(\Gamma \vdash \hastype \sigma {\Gamma'})$ whenever $\dom\sigma =
\dom{\Gamma'}$ and $(\Gamma[\sigma] \vdash \hastype {\sigma(\logvar)}
{\Gamma'(\logvar)[\sigma]})$ for all $\logvar$ in $\dom\sigma$. 

Now, we extend the Hoare triple judgment to states: 

\begin{center}
\staterulelabelbis{(State)}
  {
    \begin{array}{c}
      \Gamma \vdash \hastype \sigma {\Gamma'}
      \qquad
      \dom{\Gamma'} \cap \cfv c = \emptyset
      \qquad
      \goodstore {\Gamma,\Gamma'} s
      \\
      \holds {\Gamma[\sigma]} \pe \ahp s {F[\sigma]}
      \qquad
      \hoare r {\Gamma,\Gamma'} F c \Void {G}
    \end{array}
  }
  {\goodstate {\State s {\hp,c}}}
\label{rule:state}
\label{rule:thread}
\end{center}
where $\cfv c$ denotes the set of variables that occur freely in an object creation command in $c$.

The rule for states ensures that there exists an
augmented heap $\ahp$ to satisfy the state's command. 
The object identifier $r$ in the Hoare triple (last premise) is the 
current receiver, needed to determine the scope of abstract predicates.
Rule~\ref{rule:state} enforces the current command to be verified with precondition $F$
and postcondition $G$. No condition is required on $F$ and $G$, but note that, by the semantics
of Hoare triples, $F$ represents the state's allocated memory before executing $\cmd$:
if $\cmd$ is not a top level program (\ie some memory should be allocated for $\cmd$ to execute correctly),
choosing a trivial $F$ such as $\True$ is incorrect. Similarly, $G$ represents the state's memory after
executing $\cmd$.

The judgment $(\hastype \cls \ok)$ is the top-level judgment of our source
code verification system, to be read as ``class table $\cls$ is verified''.
Before presenting the preservation theorem, we first give the following lemma, which illustrates how we handle method calls in the preservation proof.

\begin{lem}
\label{lem:bind}
If $\ (\ \Gamma;o \vdash \{\ F \ \} \ c:T \ \{\ \lpar \java{ex} \ T \ \alpha \rpar \lpar G\rpar \ \} \ \  )$ , 
$T<:\Gamma( \lvar )\ $ 
and $(\ \Gamma; p \vdash \{\ \lpar \java{ex} \ T \ \alpha \rpar \lpar \alpha == \lvar \LAnd G \rpar\ \} \ c':U\ \{\ H\ \}\ )$ 
then $(\ \Gamma ; o \vdash \{\ F\ \} \ \Bind \lvar c {c'}:U \ \{\ H\ \}\ )$.
\end{lem}
\proof
By induction on the structure of $c$.\medskip
\qed

The following theorem shows that the Hoare rules from Section~\ref{subsec:sl:hoare} are sound.

\begin{thm}[Preservation]
\label{thm:preservation}
If $(\hastype \cls \ok)$, $(\goodstate \stt)$ and $\stt \sstep {\cls} \stt'$, then $(\goodstate {\stt'})$.
\end{thm}

\proof

In order to deal with structural rules we need Lemma \ref{lem:normalization} in the preservation proof. 
Based on the assumptions and Lemma \ref{lem:normalization} there is a proof tree for $\goodstate \stt$ ending in \ref{rule:hoa-con}, \ref{rule:hoa-exists} or \ref{rule:hoa-frame}.
Using case analysis on the shape of the head command we prove that there exists a proof tree for $\goodstate {\stt'}$ in all the cases \ref{rule:hoa-con}, \ref{rule:hoa-exists} and \ref{rule:hoa-frame}. 
Details can be found in~\cite[Chap. 6]{HurlinPhd}.
\qed

From the preservation theorem, we can draw two corollaries:
verified programs never dereference $\Null$ and
verified programs satisfy partial correctness.
To formally state these theorems, we say that a class table $\cls$ together with a ``main'' program $\cmd$ is sound (written
$\hastype {\Cpl \cls c} \diamond$) iff ($\hastype \cls \diamond$ and
$\hoare \emptyset \Null \True \cmd \Void \True$). In the latter judgment,
$\emptyset$ represents that the type environment is initially empty,
$\Null$ represents that the receiver is initially vacuous, and
$\True$ represents that the top level program has $\True$
both as a precondition and as a postcondition.
Notice that $\True$ is a correct precondition for top level programs (Java's {\tt main}), because
when a top level program starts to execute, the heap is initially empty.

\begin{lem}
\label{lem:good-init}
If $\hastype {(\cls,c)} \ok$, then $\hastype {\initcmd c} \ok$.
\end{lem}
\proof
See~\cite[Chap. 6]{HurlinPhd}.
\qed

We can now state the first corollary (no $\Null$ dereference) of the preservation theorem.
A head command $\hc$ is called a \emph{null error} iff it tries to
dereference a null pointer, \emph{i.e.}, 
$\hc = (\HdGet \lvar \Null f)$ 
or $\hc = (\HdFSetNoFin \Null f v)$ 
or $\hc = (\EHdCall \lvar \Null m \specvals \vs)$
for some $\lvar, f, v, m, \specvals, \vs$.

\begin{thm}[Verified Programs are Null Error Free] 
\label{theorem:null-error}
If $\hastype {(\cls,c)} \ok$ 
and $\initcmd c \ssteps \cls {\stt = \State s {\hp,\hc;c'}}$,
then $\hc$ is not a null error.
\end{thm}

\proof
By $\hastype {\initcmd c} \ok$ (lemma \ref{lem:good-init}) and
preservation theorem (Theorem \ref{thm:preservation}), we have
$\goodstate \stt$. Suppose by contradiction that $\hc$ is a null
error. An inspection of the last rules of ($\goodstate \stt$)'s
derivation reveals that there must be an environment, predicate
environment, augmented heap, stack and value such that the result is a
null error. But by definition of $\models$ this is not possible (details in~\cite[Chap. 6]{HurlinPhd}).
\qed

To state the second corollary of the preservation theorem,
we extend head commands with \emph{specification commands}. Specification
commands $\scmd$ are used by the proof system, but are ignored at runtime.
The specification command $\HdAssert F$ makes the proof system check
that $F$ holds at this program point:

\begin{displaymath}
\begin{array}{rcl}
  \hc \in \HdCmdSet & \Is & 
     \dots
     \ \Or\ 
     \scmd
     \ \Or\ 
     \dots
     \\
     \scmd \in \SpecCmdSet & \Is & \HdAssert F
\end{array}
\label{idx:spec:command}
\end{displaymath}

We update Section~\ref{subsec:semantics:jll}'s operational semantics to deal with
specification commands. Operationally, specification commands are no-ops:

\begin{display}{State Reductions, $\stt \sstep \cls \stt'$:}
  \dots
  \quad
  \RuleRedNoop {\stateaxiom{(Red No Op)}} p \cmd 
  \label{rule:red-noop}
  \quad
  \dots
  \label{idx:op:sem:slj}
\end{display}

Now, we can state the partial correctness theorem. It expresses
that if a verified program contains a specification command $\HdAssert F$, then
$F$ holds whenever the assertion is reached at runtime:

\begin{thm}[Partial Correctness]
\label{theorem:partial-correctness} \mbox{ }\\
If $\hastype {(\cls,c)} \ok$ 
and 
$\initcmd c \ssteps \cls {\stt = \State s {\hp,{\Assert F c}}}$,
then $(\holds \Gamma \pe {(h,\pt)} s F[\sigma])$
for some $\Gamma,\pe = \pef_\cls(\pe),\pt$ 
and $\sigma \in \ParFun \LogVarSet \SpecValSet$.
\end{thm}

\proof
By $\hastype {\initcmd c} \ok$ (lemma \ref{lem:good-init}) and preservation theorem (Theorem \ref{thm:preservation}), we know that $\goodstate \stt$.
An inspection of the last rule of ($\goodstate \stt$)'s derivation reveals that there must be $\Gamma, \pe=  \pef_\cls(\pe), \ahp, \sigma \in \ParFun \LogVarSet \SpecValSet$ such that $(\holds \Gamma \pe {(h,\pt)} s F[\sigma])$ .
\qed

\subsection{Example: Sequential Mergesort}
\label{subsec:sequential-mergesort}

To show how the verification system works, we specify a
(sequential) implementation of mergesort.  In the next section, when
we add multithreading, we extend this example to
parallel mergesort and we verify the parallel implementation w.r.t its specification.

Since our model language has no arrays, we use linked lists. For
simplicity, we use integers as values. Alternatively, as in the Java
library, values could be objects that implement the {\tt Comparable}
interface.
Our example contains two classes: {\tt List} and {\tt MergeSort}, defined\footnote{For clarity
of presentation, these classes are written using a more flexible language
than our formal language. 
E.g. we allow reading of fields in conditionals and write chains of fields dereferencing.}
 and specified in Figures~\ref{fig:list}, \ref{fig:ListPredicates}, \ref{fig:ListSpecification}, and~\ref{fig:mergesort-seq}.
{\small
\begin{figure}[t]
\begin{tabbing}
\tt class List extends Object\lbr\\
~~\=\tt int val;  List next;\\
\> \\
\> \tt void init(val v)\lbr~val = v;~\rbr\\
\> \\
\> \tt void append(int v)\lbr\\
\>~~\=\tt List rec; rec = this;\\
\>\>\tt while(rec.next!=null)\lbr~rec = rec.next;~\rbr \\
\>\> \tt List novel = new List; novel.init(v); rec.next = novel; \\
\>\rbr\\
\> \\
\> \tt void concatenate(List l,int i)\lbr\\
\>~~\=\tt List rec; rec = this;\\
\>\>\tt while(rec.next!=null)\lbr~rec = rec.next;~\rbr \\
\>\>\tt while(i>0)\lbr~List node = new List; node.init(l.val); \\
\>\>~~\=\tt~~~~~~~~~~~rec.next = node; l = l.next; rec = rec.next; i = i-1; \rbr\\
\>\rbr\\
\> \\
\> \tt List get(int i)\lbr\\
\>\>\tt List res;\\
\>\>\tt if(i==0) res = this; \\
\>\>\tt if(i > 0) res = next.get(i-1); \\
\>\>\tt res;\\
\>\rbr\\
\rbr
\end{tabbing}
\caption{Implementation of class \texttt{List}}\label{fig:list}
\end{figure}
}

\paragraph{Class {\tt List}}
Figure~\ref{fig:list} contains the implementation of class {\tt
  List}. This class has three methods): method {\tt append} adds a
value to the tail of the list; method {\tt concatenate(l,i)}
concatenates the {\tt i}-th first elements of list {\tt l} to the
receiver list; and method {\tt get} returns the sub-tail of the
receiver starting at the {\tt i}-th element. Note that
these methods use iteration in different ways. In method {\tt append}'s
loop, iteration is used to reach the tail of the receiver list, while in
method {\tt concatenate}'s second loop, iteration is used to reach
elements \emph{up to a certain length} of list {\tt l}.  
This means that, in the first case, the executing method should have permission to
access the whole list, while in the second case, it suffices to
have access to the list up to a certain length. 
To capture this, class \texttt{List} defines t two state predicates (see Figure~\ref{fig:ListPredicates}): 
(1)~{\tt state<n,p,q>} gives access to the first {\tt n} elements of the
receiver list with permissions {\tt p} on the field {\tt next} and
{\tt q} on the field {\tt val}; and (2)~{\tt state<n,l,p,q>}
additionally requires the successor of the {\tt n}-th element to point to list {\tt l}. Both predicates
ensure that the receiver list is at least of length {\tt n}, because of the test for non-nullness
on the next element ({\tt lb!=null}). As a consequence, predicate {\tt state<n,null,p,q>}
represents a list of \emph{exact length} {\tt n}.

\begin{figure}
{\small
\begin{tabbing}
\tt class List extends Object\lbr\\[1ex]
~~\tt\xy public pred state<nat n,perm p, perm q> = (n==0 -* True) * \\
~~\tt\xy~~~~~~~(n==1 -* [ex List l.~PointsTo(next,p,l) * Perm(val,q)]) * \\
~~\tt\xy~~~~~~~(n>1~~-* [ex List lb.~PointsTo(next,p,lb) * Perm(val,q) * \\
~~\tt\xy~~~~~~~~~~~~~~~~~lb!=null * lb.state<n-1,p,q>]);\\[1ex]

~~\tt\xy public pred state<nat n,List l, perm p, perm q> = (n==0 -* True) * \\
~~\tt\xy~~~~~~~(n==1 -* [PointsTo(next,p,l) * Perm(val,q)]) * \\
~~\tt\xy~~~~~~~(n>1~~-* [ex List lb.~PointsTo(next,p,lb) * Perm(val,q) * \\
~~\tt\xy~~~~~~~~~~~~~~~~~lb!=null * lb.state<n-1,l,p,q>]);\\[1ex]
\rbr
\end{tabbing}
}
\caption{{\tt List} state predicates} \label{fig:ListPredicates}
\end{figure}

\begin{figure}
{\small
\begin{tabbing}
\tt class List extends Object\lbr\\[1ex]

~~\=\xy requires init; ensures state@List<1,null,1,1>;\\
\>\tt List init(val v)\\[1ex]

\>\xy requires state<i,null,1,q> * i>0; ensures state<i+1,null,1,q>;\\
\>\tt void append(int v)\\[1ex]

\>\xy requires state<j,null,1,q> * j>0 * l.state<k,1,q> * k>=i;\\
\>\xy ensures~ state<j+i,null,1,q> * l.state<k,1,q>;\\
\>\tt void concatenate(List l,int i)\\[1ex]

\>\xy requires state<j,p,q> * j>=i * i>=0;\\
\>\xy ensures~ state<i,result,p,q> * result.state<j-i,p,q>;\\
\>\tt List get(int i)\\[1ex]

\rbr
\end{tabbing}
}
\caption{Method contracts of class {\tt List}} \label{fig:ListSpecification}
\end{figure}

Finally,
Figure~\ref{fig:ListSpecification}) provides the method specifications
for the methods in class {\tt List}. 
We should note here that in the specifications provided for the methods, the binders for logical variables are considered implicit.
Method \texttt{init}'s postcondition refers explicitly to the {\tt List}
class. This might look like breaking the abstraction provided by subtyping.
However, because method \texttt{init} is meant to be called right after object creation
({\tt new List}),
{\tt init}'s postcondition can be converted into a form that does not
mention the {\tt List} class. E.g.
after calling {\tt $l$ = new List} and {\tt $l$.init()}, the caller knows
that \texttt{List} is $l$'s dynamic class (recall that~\ref{rule:hoa-new}'s postcondition
includes an {\tt classof} predicate) and can therefore
convert the access ticket \texttt{$l$.state@List<1,null,1,1>} to
\texttt{$l$.state<1,null,1,1>} (using axiom~\ref{axiom:known-type}). 
Because they are standard, we do not detail the proofs of the methods in class {\tt List}.

{
  \paragraph{Class {\tt MergeSort}} Figure~\ref{fig:mergesort-seq}
  present the mergesort algorithm. Class {\tt MergeSort} has two
  fields: a pointer to the list to be inspected, and an integer
  indicating how many nodes to inspect.  The algorithm itself is
  implemented by methods {\tt sort} and {\tt merge}. For space
  reasons, we omit the full implementation, as it is standard: method
  {\tt sort} distinguishes three cases: (i)~if there is only one node
  to inspect, nothing is done; (ii)~if there are only two nodes to
  inspect, the value of the two nodes are compared and swapped if
  necessary; and (iii)~if the list's length is greater than $2$, two
  recursive calls are made to sort the left and the right parts of the
  list.  The next section will present both the implementation and the
  proof outline of the parallel mergesort algorithm in more detail.
}

We have proved that mergesort is memory safe (references point to valid memory locations) and that the length of
the sorted list is the same as the input list's length. We do not
prove, however, that sorting is actually performed.
This would require heavier machinery, because we would have to include mathematical lists
in our specification language.

Instances of class {\tt MergeSort}
are parameterized by the number of nodes they have to inspect. This is
required to show that the input list's length is preserved by the
algorithm after the two recursive calls in
method {\tt sort()}.

\begin{figure}
{\small
\begin{tabbing}
\tt class MergeSort<int length> extends Object\lbr\\[1ex]

~~\=\tt List list; int num;\\
\> \\
~~\=\tt\xy  pred state = PointsTo(list,1,l) * PointsTo(num,1,n) *\\
\xy~~~~~~~~~~~~~~~~~~~~~~~\tt l!=null * n >= 1 * n==length * l.state<length,1,1>;\\
\>\\
\>\xy  requires init * l.state<length,1,1> * i>=1 * i==length * l!=null;\\
\>\xy  ensures~ state@MergeSort;\\
\>\tt  init(List l, int i)\lbr~list = l; num = i; \rbr\\
\> \\
\>\xy  requires state; ensures~ result.state<length,1,1>;\\
\>\tt  List sort()\lbr ~/* {\it uses {\tt merge}, sorts the elements } */ \rbr\\
\> \\
\>\xy  requires ll.state<lenleft,1,1> * rl.state<lenright,1,1> * \\
\> ~~\= \xy lenleft+lenright==length;\\
\>\xy  ensures~ result.state<length,1,1>;\\
\>\tt  List merge(List ll,int lenleft,List rl,int lenright)\lbr ~... \rbr\\
\rbr
\end{tabbing}
\caption{Specification of sequential mergesort algorithm}\label{fig:mergesort-seq}
}
\end{figure}

In the proof (and also in the proof of the parallel version presented
in the next section) we use two special-purpose axioms. Axiom \ref{axi:split}
states that a list of length {\tt n} can be split into a list of
length {\tt m1} and a list of length {\tt m2} if \un~ {\tt m1+m2==n}
and \deux~{\tt m1}'s tail points to {\tt m2}'s head.  It can be proved
by induction over {\tt n}.  Axiom \ref{axi:forget-tail} relates the two
versions of predicate {\tt state}.  This allows - for example - to
obtain the access ticket {\tt state<1,1,1>} after a call to {\tt init}
(in combination with axiom~\ref{axiom:known-type}).

\begin{table}[h]
\hspace{-0.5cm}
$
\begin{array}{l}
\begin{array}{l @{\;\;} l}
\text{\tt (m1+m2==n  * state<n,p,q>)  *-*} 
\\
\qquad \text{\tt (ex List l. state<m1,l,p,q> * l.state<m2,p,q>) } \qquad
\labelandtext{(Split)}{axi:split}
\end{array}
\\
\begin{array}{l @{\;\;} l}
\text{\tt state<n,l,p,q>  -*   state<n,p,q>} \qquad
\labelandtext{(Forget-tail)}{axi:forget-tail}
 \end{array}
 \end{array}
$

\end{table}

\section{Separation Logic for dynamic threads}
\label{sec:forkjoin}

\renewcommand{\Thread}[3]{\IdThread {#1} {(\BareThread {#2} {#3})}}

This section extends Section~\ref{sec:jll}'s language with
threads with fork and join primitives, \`a la Java. The assertion language and verification rules are extended
to deal with these primitives. The rules support permissions transfer
between threads upon thread creation and termination. The
resulting program logic is sound, and its use is illustrated on two examples:
a parallel implementation of the mergesort algorithm and an
implementation of a typical signal-processing pattern.

\paragraph{Convention:} In formal material, we use grey background to highlight
what are the changes compared to previous sections.

\subsection{A Java-like Language with Fork/Join}
\label{subsec:language:fj}

\paragraph{Syntax.} First, we extend the syntax of Section~\ref{subsec:syntax:jll}'s language with fork and join
primitives. We assume that class tables always contain the 
declaration of class {\tt Thread}, where class {\tt Thread} contains methods
{\tt fork}, {\tt join}, and {\tt run}:

\begin{tabbing}\tt
cl\tt ass Thread extends Object\lbr 
\\
~~\tt \hlspec{final} void fork();
\\
~~\tt \hlspec{final} void join(); 
\\
~~\tt void run() \lbr\ null \rbr 
\\
\rbr 
\end{tabbing}

As in Java, the methods $\fork$ and $\join$
are assumed to be implemented natively and their behavior is specified
by the operational semantics as follows:
$o.\fork \lpa \rpa$ 
creates a new thread, whose thread identifier is~$o$, and executes 
$o.\run \lpa \rpa$ in this thread. 
Method {\tt fork} should not be called more than
once on $o$. Any subsequent call results in blocking of the calling thread.  
A call $o.\join \lpa \rpa$ blocks until thread $o$ has terminated. 
The \texttt{run}-method is meant to be overridden, while methods {\tt
join} and {\tt fork} cannot be overridden (as indicated by the {\tt
final} modifiers).  
In Java, {\tt fork} and {\tt join} are not final,
because in combination with super calls, this is useful for custom
{\tt Thread} classes. 
However, we leave the study of overrideable {\tt fork} and {\tt join} methods together with super calls as future work.

\paragraph{Runtime Structures.}

In Section~\ref{subsec:semantics:jll}, our operational semantics
$\rightarrow_\cls$ is defined to operate on states consisting of a
heap, a command, and a stack.  To account for multiple threads, states
are modified to contain a heap and a \emph{thread pool}.  A thread
pool maps object identifiers (re\-presenting \texttt{Thread} objects)
to threads. Threads consist of a thread-local stack $s$ and a
continuation $\cmd$.  For better readability, we write ``$\BareThread
c s$'' for threads $t = (s,c)$, and ``$\IdThread {o_1} {t_1} \parpop
\cdots \parpop \IdThread {o_n} {t_n}$'' for thread pools $\tpool =
\set{o_1 \mapsto t_1,\dots, o_n \mapsto t_n}$:
\begin{displaymath}
  \begin{array}{rclcl}
    \thr \in \ThreadSet & = & \changed{\StoreSet \times \CmdSet} & ::= & \changed{\BareThread c s}
    \\
    \tpool \in \ThreadPoolSet & = & \ParFun \ObjIdSet \ThreadSet & ::= &
    \IdThread {o_1} {t_1} \parpop \cdots \parpop \IdThread {o_n} {t_n}
    \\
    \stt \in \StateSet & = & \multicolumn{3}{l}{\changed{\HeapSet \times \ThreadPoolSet}}
  \end{array}
  \label{idx:thread:threadpool:state}
\end{displaymath}

\paragraph{Initialization.}

The definition of the initial state of a program is extended to account for multiple threads.
Below, $\main$ is some distinguished object identifier for the main thread.
The main thread has an empty set of fields (hence the first $\emptyset$),
and its stack is initially empty (hence the second $\emptyset$):

\begin{displaymath}
  \initcmd c = \changed{\State {\IdThread {\main} {(\BareThread c {\emptyset})}} {\set{\main \mapsto (\ThreadTy,\emptyset)} }}
  \label{idx:initcmd:fj}
\end{displaymath}

\paragraph{Semantics.} The \emph{operational semantics} defined in Section~\ref{subsec:semantics:jll}
is straightforwardly modified to deal with multiple threads. In each case,
\emph{one} thread proceeds, while the other threads remain untouched.
In addition, to model {\tt fork} and {\tt join}, we change the
reduction step~\ref{rule:red-call} to model that it does
not apply to {\tt fork} and {\tt join}. Instead, {\tt fork} and {\tt join}
are modeled by two new reductions steps (\ref{rule:red-fork} and~\ref{rule:red-join}): 

\newcommand{\RuleRedCall}[6]{#1
    {
      \changed{m \not\in \set{\fork,\join} }
    }
    {
      \DynTy h {#3} = \TyApp C {\specvals}
      \quad
      \mbody m {\TyApp C {\specvals}} 
      = \Mbody {\logvars,\logvars'} {\rvar_0;\rvars} {{#5}} 
      \quad
      {#6} = #5[#3/\rvar_0,\vs/\rvars]
    }
    { 
     \State { 
              \tpool \parpop
              \Thread {#2} {\Call \lvar {#3} m \vs {#4}} s 
            } 
            \hp 
     \step
     \State { 
              \tpool \parpop 
              \Thread {#2} {\Bind \lvar {#6} {#4}} s
            }
            \hp  
   }  
}

\newcommand{\RuleRedFork}[7]{#1
    {
    \begin{array}{l}
      \hspace{-2ex}
      \DynTy h {#3} = \TyApp C {{#7}} 
      \quad
      {#3} \notin (\dom{\tpool}\cup\set #2)
      \\
      \hspace{-2ex}
      \mbody \run {\TyApp C {{#7}}} = \Mbody {} {#6} {c_r}
      \quad
      {#5} = \subst {#3} {#6} {c_r}
    \end{array}
    }
    {
     \State { 
              \tpool \parpop
              \Thread {#2} {\Call {\lvar} {#3} \fork {} {#4}} s 
            }
            \hp 
     \step
     \State { 
              \tpool \parpop
              \Thread {#2} {\SetVar \lvar \Null {#4}} {s} \parpop
              \Thread {#3} {#5} {\emptyset}
            } 
            \hp
   }  
}

\newcommand{\RuleRedJoin}[5]{#1
  { 
  }
  { 
  \begin{array}{l}
    \State { 
      {#5} \parpop
      \Thread {#2} {\Call {\lvar} {#3} \join {} {#4}} s  \parpop
      \Thread {#3} v {s'}
    }
    \hp 
    \step \\
    \qquad\State { 
      {#5} \parpop
      \Thread {#2} {\SetVar \lvar \Null {#4}} {s} \parpop
      \Thread {#3} v {s'}
    } 
    \hp
  \end{array}
  }
} 

\begin{small}
  \begin{display}{State Reductions, $\stt \sstep \cls \stt'$:}
    \dots
    \\[\jot]
    \RuleRedCall{\stateaxiomshortlongcond{(Red Call)}} p o c {c_m} {c'}
    \label{rule:red-call:fj}
    \\[\jot]
    \RuleRedFork {\stateaxiomcond{(Red Fork)}} p o c {c_o} \rvar
    {\specvals} 
    \label{rule:red-fork}
    \\[\jot]
    \RuleRedJoin {\stateaxiomcond{(Red Join)}} p o c \tpool
    \label{rule:red-join}
    \\[\jot]
    \dots
    \label{idx:op:sem:fj}
  \end{display}
\end{small}

In~\ref{rule:red-fork}, a new thread $o$ is forked. Thread $o$'s state
consists of an empty stack $\emptyset$ and command $\cmd_o$. Command $\cmd_o$
is the body of method {\tt run} in $o$'s dynamic type where the formal receiver $\This$
and the formal class parameters have been substituted by the actual receiver and the actual class parameters.
In~\ref{rule:red-join}, thread $p$ joins the terminated thread $o$. Our rule
models that {\tt join} finishes when $o$ is terminated, \ie its current command is reduced to a single return value.
However, notice that the semantics blocks on an attempt to join $o$, if $o$ has not yet been started. This is different from real Java programs.

\subsection{Assertion Language for Fork/Join}
\label{subsec:sl:fj}
\label{sec:sl:fj}

This section extends the assertion language to deal with {\tt fork} and {\tt join}
primitives. To this end, we introduce a {\tt Join} predicate that controls
how threads access postconditions of terminated threads. We also introduce {\tt group}s, which
are a restricted class of predicates. 

\subsubsection{The {\tt Join} predicate}

To model {\tt join}'s behavior,
we add a new formula $\Join e \specval$ to the assertion language.
The intuitive meaning of $\Join e \specval$ is as follows: it allows one
to pick up fraction $\specval$ of thread $e$'s postcondition after $e$
terminated. As a specific case, if $\specval$ is $1$, the thread in
which the {\tt Join} predicate appears can pick up thread $e$'s entire
postcondition when $e$ terminates.
Thus this formula is used (see Section~\ref{subsec:hoare:fj}) to govern exchange of permissions
from terminated threads to alive threads:
\begin{displaymath}
  F \Is \dots \parpop \Join e \specval \parpop \dots
  \label{idx:formulas:fj}
\end{displaymath}

Notice that the same approach can be used to model other
synchronisation mechanisms where multiple threads can obtain part of
the shared resources.

When a new thread is created, a {\tt Join} predicate is emitted for it.
To model this, we redefine the $\initpred$ predicate (recall
that $\initpred$ appears in~\ref{rule:hoa-new}'s postcondition) for subclasses of {\tt Thread}
and for other classes. We do that by
(1) adding the following clause to the definition of predicate lookup:
\begin{displaymath}
  \plkup {\initpred,\ThreadTy} = \changed{\NewPDNoArg {} \initpred {\Join \This 1}\;\kw{ext}\;\Object}
\end{displaymath}
and (2) adding $C \neq \ThreadTy$ as a premise to the original
definition~\ref{rule:plkup-init}.  Intuitively, when an object $o$
inheriting from {\tt Thread} is created, a {\tt Join($o,1$)} ticket is
issued.

\label{idx:augmented heaps:fj} 
\paragraph{Augmented Heaps.} To express the semantics of the {\tt Join} predicate,
we need to change our definition of augmented heaps. Recall that,
in Section~\ref{subsubsec:sl:augmented-heaps}, augmented heaps were pairs of a
heap and a permission table of type $\Fun {\ObjIdSet \times \FldIdSet}
{[0,1]}$. Now, we modify permission tables so that they have type
$\Fun {\ObjIdSet \times (\FldIdSet \cup \set \join)} {[0,1]}$.  The
additional element in the domain of permission tables keeps track of
how much a thread can pick up of another thread's postcondition.  
Obviously, we forbid {\tt join} to be a program's field identifier.

In addition, we add an additional element to augmented heaps; so that they become
triples of a heap, a permission table, and a \emph{join table} $\calj \in \Fun \ObjIdSet {[0,1]}$. 
Intuitively, for a thread $o$,
$\calj(o)$ keeps track of how much of $o$'s postcondition has been
picked up by other threads: when a thread gets joined, its entry in
$\calj$ drops.  The compatibility and joining operations on join
tables are defined as follows:
\label{idx:jointable}
\begin{displaymath}
  \compatible \calj {\calj'} \text{ iff } \calj = \calj'
  \qquad
  \calj \LAnd \calj' \deq \calj
  \label{idx:join:fj}
\end{displaymath}

Because $\compatibleSym$ is equality, join tables are ``global'': in 
the preservation proof, all augmented heaps will have the same join table\footnote{This suggests
that join tables could be avoided all together in augmented heaps. It is unclear, however, if an alternative approach
would be cleaner because rules~\ref{rule:state:fj},~\ref{rule:cons-pool}, and~\ref{rule:thread:fj}
would need extra machinery.}.
As usual, we define a projection operator:
%
$
  {\Tpl \hp \pt \calj}_{\kw{join}} \deq \calj
$.

Further, we require augmented heaps to satisfy these additional axioms:

\begin{enumerate}[label=\({\alph*}]
\setcounter{enumi}{2}
\item\label{rsc-neg:fj} 
  For all $o \not\in\dom h$ and all $f$ (including {\tt join}), 
  $\pt(o,f) = 0$ and $\calj(o) = 1$.
\item\label{rsc-leq}
  $\forall\;o\;.\;\pt(o,{\tt join}) \leq \calj(o)$.
\end{enumerate}

Axiom~\ref{rsc-neg:fj} ensures that all unallocated objects 
have minimal permissions, which is needed to prove soundness of the verification rule for
allocating new objects.
Axiom~\ref{rsc-leq} ensures that a thread will never try to pick up more than is available of a thread's postcondition. 


%


\paragraph{Semantics.} We update the predicate environments with an axiom to ensure that when a thread is joined, its corresponding entry drops in all join tables.
The semantics of the {\tt Join} predicate is as follows:
\begin{displaymath}
\begin{array}{rcl@{\  }c@{\ \ }l}
\Gamma \vdash \Tpl \hp \pt \calj;s 
& \models & \Join e \perm
& \iff & \esem e h s = o \text{ and } \sem \perm \leq \pt(o,\join) 
\end{array}
\end{displaymath}

\paragraph{Axiom.} In analogy with the {\tt PointsTo} predicate,
we have a split/merge axiom for the {\tt Join} predicate:
\begin{displaymath}
  \refstepcounter{rule}%
  \addToLabel{(Split/Merge Join)}%
  \begin{array}{c c}
  \Gamma; v \vdash
  \Join e \specval
  \LEquiv
  \lpa\,
    \Join e {\frac \specval 2}
    \;\LAnd\;
    \Join e {\frac \specval 2}
  \rpa
  &
  \qquad
  \text{(Split/Merge Join)}
\end{array}
\label{axiom:split/merge:join}
\end{displaymath}

\subsubsection{Groups}

In order to ensure that multiple threads can join a terminated
thread, we introduce the notion of \emph{groups}. Groups are special
predicates, denoted by keyword {\tt group} that satisfy an
additional split/merge axiom.  Formally, {\tt group} desugars to
a {\tt pred}icate and an {\tt axiom}:
\begin{displaymath}
  \mathtt{group}~\TyApp P {\TypedVar \Ts \xs} \jeq F
  \Deq
  \begin{array}{l}
    \mathtt{pred}~ \TyApp P {\TypedVar \Ts \xs} \jeq F\tjkw{;}
    \\
    \jkw{axiom}\
    \TyApp P \xs \LEquiv \lpa\TyApp P {\Split {\Ts,\xs}} \LAnd \TyApp P {\Split {\Ts,\xs}}\rpa
  \end{array}
  \label{idx:group}
\end{displaymath}

\noindent where {\tt split} is extended to pairs of type and parameter, so that
it splits parameters of type {\tt perm} and leaves other parameters unchanged:
\begin{displaymath}
  \Split {T,x} \deq \left\{
  \begin{array}{l l}
    \Split x & \text{iff $T = \PermTy$} \\
    x & \text{otherwise}
  \end{array}\right.
\end{displaymath}

The meaning of the axiom for groups is as follows: (1) splitting (reading $\LLEquiv$ from left to right) $P$'s parameters
splits predicate $P$ and (2) merging (reading $\LLEquiv$ from right to left) $P$'s parameters
merges predicate $P$.

\subsection{Contracts for Fork and Join}
\label{subsec:hoare:fj}

Next, we discuss how the verification logic for the sequential
language, presented in Section~\ref{subsec:sl:hoare} is adapted to
cater for the multithreaded setting with {\tt fork} and {\tt join}
primitives. Since we can specify contracts in the program logic for
{\tt fork} and {\tt join} in class {\tt Thread}, we do not need to
give new Hoare rules for them.
Instead, rules for {\tt fork} and {\tt join} are simply
instances of the rule for method call~\ref{rule:mth}.  The
contracts for {\tt fork} and {\tt join} model how permissions to
access the heap are exchanged between threads. 
Intuitively, newly created threads obtain a part of the heap from
their parent thread. Dually, when a terminated thread is joined, (a part
of) its heap is transferred to the joining thread(s).


\paragraph{Class {\tt Thread}.}

In Section~\ref{subsec:language:fj}, we introduced class {\tt Thread} but did not
give any specifications. Class {\tt Thread} is specified as follows:

\begin{tabbing}\tt
cl\tt ass Thread extends Object\lbr \\[\jot]
~~\tt\xy \hlspec{pred ~preFork = true;} \\
~~\tt\xy \hlspec{group postJoin<perm p> = true;}
\\[\jot]
~~\tt\xy \hlspec{requires preFork; ensures true;} \\
~~\tt \hlspec{final} void fork();
\\[\jot]
~~\tt\xy \hlspec{requires Join(this,p); ensures postJoin<p>;} \\
~~\tt \hlspec{final} void join(); 
\\[\jot]
~~\tt\xy \hlspec{final requires  preFork; ensures postJoin<1>;} \\
~~\tt void run() \lbr\ null \rbr 
\\[\jot]
\rbr 
\end{tabbing}


Predicates {\tt preFork} and {\tt postJoin} describe the pre- and
postcondition of {\tt run}, respectively.  Notice that the contracts
of {\tt fork}, {\tt join}, and {\tt run} are tightly related: (1) {\tt
fork}'s precondition is the same as {\tt run}'s precondition and (2)
{\tt run}'s postcondition is the predicate {\tt postJoin<1>} while
{\tt join}'s postcondition is {\tt postJoin<p>}. 
Point (1) models that when a thread is forked, part of the parent thread's state is transferred to the forked thread.  
Point (2) expresses that 
threads joining a thread pick up a part of the joined thread's state.
The fact that permission {\tt p} appears both as an argument to {\tt
  Join} and to {\tt postJoin} (in {\tt join}'s contract) models that
joining threads pick up that part of the terminated thread's state
which is \emph{proportional} to {\tt Join}'s argument. Because one
{\tt Join($o,1$)} predicate is issued per thread $o$, and this cannot
be duplicated, our system enforces that all threads joining $o$
together do not pick up more than thread $o$'s postcondition. 
The signal-processing example below illustrates reasoning about
multiple joins.

Notice that defining {\tt
postJoin} as a group is needed because {\tt join}'s
postcondition (\ie~{\tt postJoin}) is split among several threads, and
by declaring it as a group, we make sure that this splitting is sound.



Although method {\tt run} is meant to be overridden, we require that
method {\tt run}'s contract cannot be modified in subclasses of {\tt
Thread} (as indicated by the {\tt final} modifier).
Subclasses are able to redefine {\tt preFork} and {\tt postJoin} and add parameters to customize the specification.
In our examples, this proved to be convenient, however we have not
investigated consequences of this choice on more intricate
examples. 
Enforcing {\tt run}'s contract to be fixed allows to express that {\tt join}'s
postcondition is proportional to the second parameter of {\tt Join}'s
predicate in an easy way (because we can assume that {\tt run}'s
postcondition is always {\tt postJoin<1>}).  

Since {\tt run}'s contract is fixed, {\tt run}'s contract cannot be
parameterized by logical parameters. 
But this is unproblematic; in fact it would be
unsound to allow logical parameters for method {\tt run}. As {\tt
run}'s pre and postconditions are interpreted in different threads,
one cannot guarantee that logical parameters are instantiated in a
similar way between callers to {\tt fork} and callers to {\tt
join}. Hence, logical parameters have to be forbidden for {\tt run}.

We highlight that method {\tt run} can also be called directly, without forking a new thread. Our system
allows such behavior which is used in practice to flexibly control concurrency (cf Java's {\tt Executor}s~\cite{Java}).


\paragraph{Alternative Solutions.}

Alternatively, we could allow arbitrary contracts for {\tt run}, as we
did in our earlier AMAST paper~\cite{HaackH08b}.
Yet another solution would be to combine
(1) our approach of specifying {\tt fork}, {\tt join}, and {\tt run} with predicates in class {\tt Thread} and
(2) to use scalar multiplication as a new \emph{constructor} for formulas (\ie not a derived form) to express
that {\tt run}'s postcondition can be split among joiner threads. This solution, however, requires
a thorough study, because having scalar multiplication as a new constructor for formulas may raise
semantical issues (as studied by Boyland~\cite{Boyland07}).
Finally, as we stated before, our~\ref{rule:red-join} rule is slightly different from the Java behaviour when a thread tries to join a thread which is not in the thread pool.
The contracts proposed here for {\tt fork}, {\tt join} and {\tt run} are adapted in~\cite{AmighiBDHMZ14} for real Java programs.

\subsection{Verified Programs}
\label{subsec:verified:programs:fj}

To extend the definition of a verified program to the multithreaded
setting, we have to update Section~\ref{subsec:verified:programs}'s
rules for verified programs to account for multiple threads. First, we
craft rules for thread pools:
\begin{center}
\staterulelabel{(Empty Pool)}
  {
  }
  {
    \goodpool \ahp \emptyset
  }
\label{rule:empty-pool}
\qquad
\staterulelabel{(Cons Pool)}
  {
    \goodthread {\ahp} t
    \qquad
    \goodpool {\ahp'} \tpool
  }
  {
    \goodpool {\ahp \LAnd \ahp'} {t \parpop \tpool}
  }
\label{rule:cons-pool}
\end{center}

For sequential programs, the core rule extended Hoare triple judgments
to states. In the multithreaded setting, this is done in two steps:
(1)~the rule for states ensures that there exists an augmented heap $\ahp$ to
satisfy the thread pool $\tpool$, and (2)~a rule for individual thread
states corresponds to the original state rule for sequential programs
(as defined in Section~\ref{subsec:verified:programs}). The new state
rule looks as follows.
\begin{center}
  \staterulelabel{(State)}
  {
    \begin{array}{c}
    h = \ahph
      \qquad
      \goodpool {\ahp} \tpool
      \end{array} 
  }
  {
    \goodstate {\State \tpool \hp}
  }
\label{rule:state:fj}
\end{center}

The rule
for individual threads additionally has to model that threads have a fraction of
$\TyApp \postJoin \one$ as postcondition.  Therefore, we introduce
\emph{symbolic binary fractions} that represent numbers of the forms
$1$ or $\sum_{i=1}^n \bit_i \cdot
\frac{1} {2^i}$: 
\begin{displaymath}
  \bit \in \set{0,1}
  \quad
  \bits \in \BitsSet \Is 1 \ \Or\ \bit,\bits 
  \quad
  \binfr \in \BinFrSet \Is \All \ \Or \ \BinFr {} \ \Or\ \BinFr \bits
\end{displaymath}

Intuitively, we use symbolic binary fractions to speak about finite formulas of the
form $\Pred r P 1 \LAnd \Pred r P {\frac 1 2} \LAnd \Pred r P {\frac 1 8} \LAnd \dots$.
Formally, we define the scalar multiplication $\scalar \binfr {\Pred r P \perm}$ as follows:
%
\begin{displaymath}
  \begin{array}{r @{\;\;=\;\;} l}
  \scalar \All {\Pred r P \perm} & \Pred r P \perm
  \\[1ex]
  \scalar {\BinFr{}} {\Pred r P \perm} & \True
  \\[1ex]
  \scalar {\BinFr 1} {\Pred r P \perm} & \Pred r P {\Split \perm}
  \\[1ex]
  \scalar {\BinFr{0,\bits}} {\Pred r P \perm} & \scalar {\BinFr\bits} {\Pred r P {\Split \perm}}
  \\[1ex]
  \scalar {\BinFr{1,\bits}} {\Pred r P \perm} & \Pred r P {\Split \perm} \LAnd \scalar {\BinFr\bits} {\Pred r P {\Split \perm}}
  \end{array}
\end{displaymath}

For instance, $\scalar {\BinFr {1,0,1}} {\Pred r P 1}
\,\LEquiv\, \lpa \Pred r P {\frac 1 2} \LAnd \Pred r P {\frac 1 8} \rpa$.
The map ${\sem \cdot} : \BinFrSet \\\rightarrow \Rationals$
interprets symbolic binary fractions as concrete rationals:
\begin{displaymath}
\begin{array}{c}
  \sem \All \deq 1
  \quad
  \sem {\BinFr {}} \deq 0
  \quad
  \sem {\BinFr 1} \deq \frac 1 2
\\~\\
  \sem {\BinFr {0,\bits}} \deq \frac 1 2 \sem {\BinFr\bits}
  \quad
  \sem {\BinFr {1,\bits}} \deq \frac 1 2 + \frac 1 2 \sem{\BinFr\bits}
\end{array}
\end{displaymath}

Now, the rule for individual threads is as follows:

\begin{center}
\staterulelabelbisecart{(Thread)}
  {
   \begin{array}{c} 
   \changed{\ahpg(o) \leq \sem \binfr}
   \quad
   \Gamma \vdash \hastype \sigma {\Gamma'}
   \quad
   \goodstore {\Gamma,\Gamma'} s
   \quad
   \cfv c \cap \dom{\Gamma'} = \emptyset
   \\
   \holds {\Gamma[\sigma]} \pe \ahp s {F[\sigma]}    
   \quad
   \hoare r {\Gamma,\Gamma'} F c \Void {\changed{\scalar \binfr {\Pred o \postJoin 1}}}
   \end{array}
  }
  {\goodthread \ahp {\Thread o c s}}
\label{rule:thread:fj}
\end{center}

In rule~\ref{rule:thread:fj}, $\binfr$ should be
bigger than the thread considered's entry in the global join table
(condition $\ahpg(o) \leq \sem \binfr$). This forces joining threads
to take back a part of a terminated thread's postcondition which is not larger
than the terminated thread's ``remaining'' postcondition. This
follows from the semantics of the {\tt Join} predicate and the semantics
of join tables: $\Gamma \vdash \Tpl \hp \pt \calj;s
\models \Join e \perm$ holds iff $\esem e h s = o \text{ and } \sem \perm \leq \pt(o,\join)$.
Moreover, 
we have that $\pt(o,\join) \leq \calj(o)$ (see axiom~\ref{rsc-leq} on page~\pageref{rsc-leq}).


As in Section~\ref{subsec:verified:programs},
we have shown that the preservation theorem
(Theorem~\ref{thm:preservation}) holds, 
and we have shown that  verified programs satisfy the following properties:
null error freeness and partial correctness. 
To adapt the proof of Theorem~\ref{thm:preservation} to our new settings the only change for existing cases is that there is an extra level of indirection between the top level augmented heap and the augmented heap for each thread regarding this fact that states in multithreading settings include a thread pool.
Then using this fact that the proof tree for  ($\goodstate \stt$) ends in an application of \ref{rule:state:fj}, preceded by an application of \ref{rule:cons-pool}, preceded by an application of \ref{rule:thread:fj} we do case analysis for two additional reduction rule \ref{rule:red-fork} and \ref{rule:red-join}. The other cases remains the same (details in~\cite[Chap.~6]{HurlinPhd}).

In addition, verified programs
are \emph{data race} free.
A pair $(\hc,\hc')$ of head commands is called a \emph{data race} iff 
$\hc = (\HdFSetNoFin o f v)$ 
and either $\hc' = (\HdFSetNoFin o f {v'})$ 
or $\hc' = (\HdGet \lvar o f)$ for some $o,f,v,v',\lvar$.

\begin{thm}[Verified Programs are Data Race Free]
\label{theorem:data-race} 
If $\hastype {(\cls,c)} \ok$ 
and $\initcmd c \ssteps \cls
	{\stt = 
    \State 
     { \tpool 
       \parpop 
       \IdThread {o_1} {(\BareThread {\hc_1;c_1} {s_1})} 
       \parpop 
       \IdThread {o_2} {(\BareThread {\hc_2;c_2} {s_2})} }
     { \hp }}$,
then $(\hc_1,\hc_2)$ is not a data race.
\end{thm}

\proof
 By $\hastype {init(c)} \ok $ (Lemma \ref{lem:good-init}) and the adapted preservation theorem (Theorem \ref{thm:preservation}), we know that ($\goodstate \stt$). 
Suppose, towards a contradiction, that $(\hc_1,\hc_2)$ is a data race. 
An inspection of the last rules of $(\goodstate \stt)$'s derivation reveals that there must be augmented heaps $\ahp , \ahp'$ and a heap cell $o.f$ such that: 
${\goodthread \ahp {\Thread {o_1} {\hc_1; \cmd_1} {s_1}}}$, 
${\goodthread {\ahp'} {\Thread {o_2} {\hc_2; \cmd_2} {s_2}}}$, 
$\compatible \ahp {\ahp'}$, 
$\ahp_{perm}(o,f)=1$ and $\ahp'_{perm}(o,f)>0$. 
But then $\ahp_{perm}(o,f)+\ahp'_{perm}(o,f)>1$, in contradiction to $\compatible \ahp {\ahp'}$.
  
\qed

\begin{figure}[!h]
\begin{small}
\begin{tabbing}
\tt class MergeSort<int length> extends Thread\lbr\\[1ex]

~~\=\tt List list; int num;\\[1ex]

\>\tt void init(List l, int i)\lbr~list = l; num = i; \rbr\\[1ex]

\>\tt void run()\lbr\\
\>~~\tt if(num == 1)\lbr\rbr\\
\>~~\tt else\lbr~if(num == 2)\lbr\\
\>~~~\=\tt        if(list.val > list.next.val)\lbr \\
\>\>~~\=\tt	  int lval = list.val;
\tt	  list.val = list.next.val;
\tt	  list.next.val = lval;\\
\>\>\tt        \rbr~
\tt      else\lbr\\
\>\>\>\tt        if(num > 2)\lbr\\
\>\>\>~~\=\tt	int lenleft; int lenright;\\[1ex]
\>\>\>\>\tt	if(num \% 2 == 0)\lbr~lenleft = num / 2; lenright = lenleft; \rbr\\
\>\>\>\>\tt	else \lbr~lenleft = (num - 1) / 2; lenright = lenleft + 1; \rbr\\[1ex]
\>\>\>\>\tt	List tail = list.get(lenleft);\\
\>\>\>\>\tt	MergeSort<lenleft> left = new MergeSort;\\
\>\>\>\>\tt     left.init(list,lenleft); left.fork();\\[0.75ex]
\>\>\>\>\tt	MergeSort<lenright> right = new MergeSort;\\
\>\>\>\>\tt     right.init(tail,lenright); right.fork();\\[0.75ex]
\>\>\>\>\tt	left.join();	right.join();	merge(left,right);\\
\>\>\>\tt	\rbr \rbr \rbr \\[0.1ex]
\>\tt void merge(MergeSort left, MergeSort right)\lbr~\dots {\it /* standard */} ~\rbr\\
\rbr\\
\end{tabbing}
\end{small}
\caption{Implementation of parallel mergesort algorithm
}\label{fig:mergesort}
\end{figure}

\begin{figure}[!h]
{\small
\begin{tabbing}
\tt class MergeSort<int length> extends Thread\lbr\\[1ex]

~~\=\tt\xy  pred preFork = PointsTo(list,1,l) * PointsTo(num,1,n) *\\
\xy~~~~~~~~~~~~~~~~~~~~~~~~~\tt l!=null * n >= 1 * n==length * l.state<length,1,1>;\\
\>\tt\xy  group postJoin<perm p> = PointsTo(list,p,l) * PointsTo(num,p,n) *\\
\xy~~~~~~~~~~~~~~~~~~~~~~~~~\tt l!=null * n >= 1 * n==length * l.state<length,p,1>;\\
\> \\
\>\xy  requires init * l.state<length,1,1> * length>=1 * i==length * l!=null;\\
\>\xy  ensures~ Join(this,1) * preFork@MergeSort;\\
\>\tt  void init(List l, int i)  \lbr \dots \rbr\\
\> \\
\>\xy  requires preFork; ensures~ postJoin<1>;\\
\>\tt  void run()  \lbr \dots \rbr\\
\> \\
\>\xy  requires Perm(list,1) * left.postJoin<1> * right.postJoin<1> * nl+nr==length;\\
\>\xy  ensures~ PointsTo(list,1,l) * l.state<length,1,1>;\\
\>\tt  void merge(MergeSort<nl> left, MergeSort<nr> right)  \lbr \dots \rbr\\
\rbr
\end{tabbing}
}
\caption{Specification of class \texttt{MergeSort} (parallel version)}\label{fig:mergesort-spec}
\end{figure}

\subsection{Examples of Reasoning}
\label{subsec:example-fj}
To illustrate the use of verification system in a multithreaded setting, we discuss two examples here. First we discuss the verification of a parallel implementation of mergesort, concentrating in particular on on the changes in specification and verification because of the use of threads. 
Second we discuss the verification of a typical digital signal-processing algorithm using multiple joins.

\begin{figure}[p]
\begin{small}
\begin{tabbing}
~~\ \tt(\textit{Let {\tt F} be the abbreviation of {\tt PointsTo(list,1,l) * PointsTo(num,1,n)}})\\
~~\=\xy \lbr~ F * l!=null * n >= 1 * l.state<length,1,1> * n==length \rbr\\
\tt if(num > 2)\lbr \\
\>\tt int lenleft; int lenright;\\
\>\tt if(num \% 2 == 0)\lbr\\
\>~~\=\tt lenleft = num / 2; lenright = lenleft; \\
\>\>~~\tt (\textit{(Split) axiom with} m1 == m2 == n/2 == lenleft == lenright)\\
\>\>~~\=\xy \lbr~F * n > 2 * n==length * \\
\>\>\>~~\=~\xy l.state<lenleft,r,1,1> * r.state<lenright,1,1> * lenleft+lenright==length~\rbr\\
\>\rbr\tt~else \lbr~lenleft = (num - 1) / 2; lenright = lenleft + 1; \\
\>\>\>\tt (\textit{(Split) axiom with} m1 == (n-1)/2 and m2==[(n-1)/2]+1) \\
\>\>\>\xy\lbr~F * n > 2 * n==length * \\
\>\>\>~\xy~l.state<lenleft,r,1,1> * r.state<lenright,1,1> * lenleft+lenright==length~\rbr\\
\>\rbr\\
\>\> \tt(\textit{In both cases, we have:})\\
\>\> \xy \lbr~F * n > 2 * n==length * \\
\>\> \xy~~l.state<lenleft,r,1,1> * r.state<lenright,1,1> * lenleft+lenright==length~\rbr\\
\>\>\tt (\textit{(Split) axiom from right to left}) \\
\>\>\xy\lbr~F * n > 2 * n==length * l.state<n,1,1> * lenleft+lenright==length \rbr\\
\>\>\tt(\textit{This matches {\tt get}'s precondition, because 1/ {\tt n>=lenleft} follows from}\\
\>\>\tt~\textit{{\tt lenleft+lenright==length} and 2/ {\tt lenleft>=0} follows from}\\ 
\>\>\tt~\textit{{\tt num==length} and {\tt length>=0} (not shown in this proof outline).})\\
\>\tt List tail = list.get(lenleft);\\
\>\>\tt(\textit{Let {\tt G} be the abbreviation of {\tt n>2 * lenleft+lenright==length * n==length}})\\
\>\>  \xy \lbr~F * G * l.state<lenleft,tail,1,1> * tail.state<n-lenleft,1,1> \rbr\\
\>\>\tt (\textit{axiom {\tt (Forget-tail)} and arithmetic ({\tt n-lenleft==lenright})})\\
\>\>  \xy \lbr~F * G * l.state<lenleft,1,1> * tail.state<lenright,1,1> \rbr\\
\>\tt MergeSort<lenleft> left = new MergeSort; left.init(list,lenleft);\\
\>\>  \xy \lbr~F * G * tail.state<lenright,1,1> * left.preFork * Join(left,1) \rbr\\
\>\tt left.fork();\\
\>\>  \xy \lbr~F * G * tail.state<lenright,1,1> * Join(left,1) \rbr\\
\>\tt MergeSort<lenright>  right = new MergeSort; right.init(tail,lenright);\\
\>\tt right.fork();\\
\>\>  \xy \lbr~F * G * Join(left,1) * Join(right,1) \rbr\\
\>\tt left.join();\\
\>\>  \xy \lbr~F * G * left.postJoin<1> * Join(right,1) \rbr\\
\>\tt right.join();\\
\>\>  \xy \lbr~F * G * left.postJoin<1> * right.postJoin<1> \rbr\\
\>\>\tt (\textit{This matches {\tt merge}'s precondition because (1) the type system}\\
\>\>\tt~\textit{tells us:}~left~:~MergeSort<lenleft> \textit{and} right~:~MergeSort<lenright>\\
\>\>\tt~\textit{(2) {\tt F} entails {\tt Perm(list,1)}, and}\\
\>\>\tt~\textit{(3) {\tt G} entails {\tt lenleft+lenright==length}})\\
\>\tt merge(left,right);\\
\>\>  \xy \lbr~F * G * l.state<length,1,1> \rbr\\
\>\>\tt (\textit{Close})\\
\>\>\xy \lbr~postJoin<1> \rbr
\end{tabbing}
\end{small}
\caption{Correctness proof of method \texttt{run} in class \texttt{MergeSort}}\label{fig:mergesort-proof}
\end{figure}

\paragraph{ Parallel Mergesort}
The parallel mergesort algorithm is a perfect example of disjoint
parallelism, because the different threads all modify the \emph{same}
list simultaneously but in different
places. Figure~\ref{fig:mergesort} shows the parallel mergesort
implementation, spread over the methods {\tt run} and {\tt merge}. It
reuses class {\tt List} from
Section~\ref{subsec:sequential-mergesort}.  Similar to the sequential
implementation, the class has two fields: a pointer to the list to be
inspected, and an integer indicating how many nodes to inspect.
Again, method {\tt run} distinguishes three cases, however, in the
third case, two new threads are created to sort the left and the right
parts of the list, and the parent thread waits for the two new threads
and {\tt merge}s their results.  Figure~\ref{fig:mergesort-spec} shows
the adapted specifications for the parallel version.

Finally, Figure~\ref{fig:mergesort-proof} outlines the correctness
proof of method \texttt{run}.  It illustrates how in the
recursive case, the two child threads both receive access to
\emph{part of} the parent thread's list.  We use the (Split) axiom
(defined in Section~\ref{subsec:sequential-mergesort}) to specify this
behaviour in the proof.  This requires some arithmetic reasoning,
because threads all have access to the same global list, but then we
can conclude that each thread's access is confined to a limited number
of nodes in the list.

\begin{figure}[b]
\begin{small}
\begin{tabbing}
\= \tt class MVList extends List \{ \\
\>~~\= \tt	int outa; int outb; \\
\>\> \tt \\
\>\> \tt\xy public pred adata<nat n, perm p, perm q> = (n==0 -* true) * \\
\>\>~~\= \tt\xy 	( n==1 -* [ex List l. PointsTo(next, p, l) * Perm(outa,q) ] ) * \\
\>\>\> \tt\xy 		( n>=1 -* [ex List l. PointsTo(next, p, l) * Perm(outa,q) ] *  \\
\>\>\> \tt\xy 		l!=null * l.adata<n-1,p,q> ); \\
\>\> \tt \\
\>\> \tt\xy public pred bdata<nat n, perm p, perm q> = (n==0 -* true) * \\
\>\>~~\= \tt\xy 	( n==1 -* [ex List l. PointsTo(next, p, l) * Perm(outb,q) ] ) * \\
\>\>\> \tt\xy 		( n>=1 -* [ex List l. PointsTo(next, p, l) * Perm(outb,q) ] *  \\
\>\>\> \tt\xy 		l!=null * l.bdata<n-1,p,q> ); \\
\>\> \tt \\
\>\> \tt\xy public pred plt<nat n, perm p> = (n==0 -* true) * \\
\>\>~~\= \tt\xy 	( n==1 -* [ex List l. PointsTo(next, p, l) ] ) * \\
\>\>\> \tt\xy 		( n>=1 -* [ex List l. PointsTo(next, p, l) ] *  \\
\>\>\> \tt\xy 		l!=null * l.plt<n-1,p> ); \\
\>\> \tt \\
\>\> \tt\xy public pred mvstate<nat n,perm p> = super.state<n,p/4,p> * \\
\>\> \tt\xy~~~~~~~adata<n,p/4,p> * bdata<n,p/4,p> * plt<n,p/4>;\\
\>\> \tt \\
\> \tt \} 
\end{tabbing}
\end{small}
\caption{ Multi-valued list extending class {\tt List}} \label{fig:mult-val-list} 
\end{figure}

\begin{figure}
\begin{small}
\begin{tabbing}
\= \tt class  Sampler<int len> extends Thread \{ \\
\>~~\= \tt	MVList<len> lst; \\
\>\> \tt \\
\>\> \tt\xy  pred preFork = PointsTo(lst,1,l) * l!=null * l.state<len,1/4,1>; \\
\>\> \tt\xy  group postJoin<perm p> = PointsTo(lst,p,l) * l!=null * l.state<len,p/4,p>; \\
\>\> \tt \\
\>\> \tt \xy requires init * Perm(lst,1) * l!=null; \\
\>\> \tt \xy ensures Join(this,1) * PointsTo(lst,1,l) * l!=null * l.state<len,1/4,1>; \\
\>\> \tt	void init(MVList l) \lbr~ lst = l;  \rbr \\
\>\> \tt \\	
\>\> \tt \xy requires preFork; ensures postJoin<1>; \\
\>\> \tt	void run() \lbr~	 sample(); \rbr \\
\>\> \tt \\
\>\> \tt \xy requires lst.state<len, 1/4, 1>; ensures lst.state<len,1/4, 1>;\\
\>\> \tt	void sample() \lbr~ { ... /*\textit{(fills raw data fields ( {\tt{val}}s )  with samples)}*/ } \rbr \\
\> \tt \} \\
\>\> \\
\end{tabbing}
\end{small}
\caption{The Sampler thread} \label{fig:sampler}
\end{figure}

\begin{figure}
\begin{small}
\begin{tabbing}
\= \tt class  AFilter<int len> extends Thread \{ \\
\>~~\= \tt	MVList<len> lst; Sampler<len> sampler; \\
\>\> \tt \\	
\>\> \tt\xy pred preFork = PointsTo(lst,1,l) * l!=null * l.adata<len,1/4,1> * \\
\>\>\xy~~~~~~~~~~~~~~\tt Perm(sampler,1) * Join(sampler,1/2) ; \\
\>\> \tt\xy group postJoin<perm p> = Perm(sampler,p) * PointsTo(lst,p,l) * l!=null * \\
\>\>\xy~~~~~~~~~~~~~~\tt l.state<len,p/8,p/2> * l.adata<len,p/4,p>; \\
\>\> \tt \\
\>\> \tt \xy requires init * Perm(sampler,1) * Perm(lst,1) * l.adata<len,1/4,1> * l!=null; \\
\>\> \tt \xy ensures Join(this,1) * lst.adata<len,1/4,1> * len >= 1 * lst!=null; \\
\>\> \tt	void init(MVList l, Sampler s) \lbr~ lst = l; sampler = s; \rbr \\	
\>\> \tt \\
\>\> \tt \xy requires preFork; ensures postJoin<1>;\\	
\>\> \tt	void run() \lbr \\
\>\>~~\=\tt \xy \lbr~ Join(sampler,1/2) * lst.adata<len,1/4,1> \rbr  \\
\>\>\> \tt		sampler.join(); \\
\>\>~~\=\tt \xy \lbr~ lst.state<len,1/8,1/2> * lst.adata<len,1/4,1> \rbr \\
\>\>\> \tt		processA(); \\	
\>\>~~\=\tt \xy \lbr~ lst.state<len,1/8,1/2> * lst.adata<len,1/4,1> \rbr \\
\>\> \tt	\rbr \\
\>\> \tt \\
\>\> \tt \xy requires lst.state<len,p,q> * lst.adata<len,r,1>; \\
\>\> \tt \xy ensures lst.state<len,p,q> * lst.adata<len,r,1>; \\	
\>\> \tt	void processA() \lbr~ ... {\textit{/* using raw data computes {\tt{outa}} fields. */}} 	\rbr \\
\> \tt \} \\
\end{tabbing}
\end{small}
\caption{Processing thread A} \label{fig:filters} 
\end{figure}

\begin{figure}
\begin{small}
\begin{tabbing}
\= \tt class Process<int len> extends Object\{ \\
\>~~\= \tt	MVList<len> data;  \\
\>\> \tt \\
\>\> \tt \xy requires Perm(data,1) * l!=null * l.mvstate<len,1>; \\
\>\> \tt \xy ensures PointsTo(data,1,l) * l!=null * l.mvstate<len,1>;\\
\>\> \tt void init(MVList l) \lbr~ data = l; \rbr \\
\>\> \tt \\
\>\>\tt \xy requires PointsTo(data,1,l) * l !=null * l.mvstate<len,1>; \\
\>\>\tt \xy ensures PointsTo(data,1,l) * l !=null * l.mvstate<len,1>; \\
\>\> \tt	void process(MVList lst) \lbr  \\
\>\>~~\= \tt {\it (Let abbreviate: {\tt data.state<len,1/4,1>} with {\tt S}, {\tt data.adata<len,1/4,1>} with {\tt A} } \\ 
\>\>\>~~\= \tt {\it  {\tt data.bdata<len,1/4,1>} with {\tt B}, {\tt data.plt<len,1/4>} with {\tt P} } \\ 
\>\>\> \tt\xy \lbr~ S * A * B * P  \rbr \\
\>\>\> \tt		Sampler<len> smp = new Sampler; smp.init(data);  \\
\>\>\> \tt \xy \lbr~  Join(smp,1) * smp.preFork * A * B * P \rbr \\
\>\>\> \tt		AFilter<len> af = new AFilter; af.init(data, smp);  \\
\>\>\> \tt \xy \lbr~   Join(smp,1/2) *  Join(af,1) * smp.preFork * af.preFork * B * P \rbr \\
\>\>\> \tt		BFilter<len> bf = new BFilter; bf.init(data, smp);  \\
\>\>\> \tt \xy \lbr~  Join(af,1) * Join(bf,1) * smp.preFork * af.preFork * bf.preFork * P \rbr \\
\>\>\> \tt		Plotter<len> plt = new Plotter; plt.init(data,af,bf);  \\
\>\>\> \tt \xy \lbr~  Join(plt,1) * smp.preFork * af.preFork * bf.preFork * plt.preFork \rbr \\
\>\>\> \tt		smp.fork();  af.fork(); bf.fork(); plt.fork(); \\
\>\>\> \tt \xy \lbr~ Join(plt,1) \rbr \\
\>\>\> \tt		plt.join();  \\
\>\>\> \tt \xy \lbr~ plt.postJoin<1> \rbr \\
\>\> \tt	\}  \\
\}  \\
\end{tabbing}
\end{small}
\caption{The main process}\label{fig:signal-processor} 
\end{figure}

\paragraph{Data Plotter}

Our next example uses a typical pattern of signal-processing
applications  to demonstrate how we reason about multiple joins to
the same thread.  The application has four threads: a sampler, filter
processes A and B, and a plotter.  The sampler collects the raw data
and delivers it to the two processors, which process the raw data
in parallel, and stores their results in an appropriate
field. Finally, the plotter prints all data (raw and processed). What
is important for our example is that both processes A and B join the
sampler process, to obtain (read) access to the raw data. To store the
data, Figure~\ref{fig:mult-val-list} extends class {\tt List} to
contain multiple values.  The list structure is captured by predicate
{\tt mvstate<n,p>}, which defines a list with length {\tt n} and
permission {\tt p} on all the fields of each node of the list. This
predicate is defined in terms of predicates {\tt state}, {\tt
  adata}, {\tt bdata} and {\tt plt}. Predicate {\tt state<n,p/4,p>}
is inherited from class {\tt List}; it provides permission {\tt p/4} to
the links and permission {\tt p} to the field {\tt val} of each node in the
list. Similarly, {\tt adata} and {\tt bdata} define permissions to
probe the list and access the fields {\tt outa} and {\tt outb},
respectively. Finally the predicate {\tt plt} provides
permissions to visit all nodes in the list.

Figures~\ref{fig:sampler} and~\ref{fig:filters} show the Sampler
thread and process A, respectively, with a proof outline for method
{\tt run} of process A. This shows how process A exchanges the half
join ticket on Sampler for half of the {\tt postJoin} predicate.
Process B is not given, as it similar process A. For space reason, we
also do not give the code for the plotter: essentially it joins
processes A and B to obtain full access to all data.

Figure~\ref{fig:signal-processor} shows the main application, with a
proof outline of the {\tt process} method.  Each thread issues a {\tt
  Join(this,1)} ticket upon initialization. Method {\tt process}
splits the ticket emitted by the sampler and passes each half to the
processing threads. Additionally, the join-tickets for the processing
threads are transferred to the plotter. The method then waits for the
plotter to finish, after which it obtains all permissions on the list
back again.  Notice that this enables the main thread to iterate on
the whole processing chain (not shown here).


\section{Separation Logic for Reentrant Locks}
\label{sec:locks}

This section presents verification rules for Java's reentrant locks.
Together with {\tt fork} and {\tt join}, reentrant locks are a crucial
feature of Java for multithreaded programs. 
In particular, locks serve to synchronize threads
and to control access to resources.

Reentrant locks can be acquired more than once by the same thread.
They are a convenient tool for programmers, but they also 
require extra machinery in the verification system, because
initial acquirements have to be distinguished from reentrant
acquirements.

After a short background discussion on modeling single-entrant locks
in separation logic, we discuss how syntax and semantics are extended
to model reentrant locks. We develop appropriate verification rules,
and discuss how their soundness can be proven. We finish the section by 
some examples that illustrate reasoning about re-entrant locks and the
wait-notify mechanisms.

\subsection{Separation Logic and Single-Entrant Locks}
\label{sec:background:locks}

Separation logic for a programming language with locks as a concurrency primitive
has been first explored by O'Hearn~\cite{OHearn07} in which he
elegantly adapted an old idea from concurrent programs with shared variables \cite{Andrews91}.
Each lock is associated with a \emph{resource invariant} that describes the part
of the heap that the lock guards. When a lock is acquired, it lends its resource
invariant to the acquiring thread. Dually, when a lock is released, it takes back
its resource invariant from the releasing thread. This is formally
expressed by the following Hoare rules:

\begin{center}
  \AxiomC{$I$ is $x$'s resource invariant}
  \UnaryInfC{$\hoaresimpl \True {\HdLock x} {I}$}
  \DisplayProof
  \qquad
  \AxiomC{$I$ is $x$'s resource invariant}
  \UnaryInfC{$\hoaresimpl I {\HdUnlock x} \True$}
  \DisplayProof
\end{center}

While these rules are sound for single-entrant locks, they are unsound
for reentrant locks, because they allow ``duplicating'' a lock's resource invariant:
\begin{tabbing}
~~\tt \lbr~$\True$~\rbr\\
\tt $\HdLock x$; // $I$ is $x$'s resource invariant\\
~~\tt \lbr~$I$~\rbr \\
\tt $\HdLock x$; \\
~~\tt \lbr~$I \LAnd I$~\rbr~$\leftarrow$ wrong! 
\end{tabbing}

\subsection{A Java-like Language with Reentrant Locks}
\label{sec:language:locks}

To recover soundness in the presence of reentrant locks, we design
proof rules that distinguish between initial acquirement and reentrant
acquirement of locks. This allows transferring a lock's resource
invariant to an acquiring thread only at initial acquirement.  In
contrast to existing work that studies simple "while" languages and
"C-like" languages~\cite{OHearn07,HoborAZ08,GotsmanBCRS07}, we also
handle reentrancy.

\paragraph{Syntax.} First we extend the syntax and the semantics of our Java-like language to model reentrant locks.
We extend the list of head commands defined in Section~\ref{subsec:syntax:jll} as follows:
\begin{displaymath}
  \begin{array}{rcl}
    \hc \in \HdCmdSet & \Is & 
    \dots
    \ \Or\ 
    \HdLock v
    \ \Or\ 
    \HdUnlock v
    \ \Or\ 
    \dots
  \end{array}
  \label{idx:head:commands:locks}
\end{displaymath}

Just as class invariants must be initialized before method calls, resource
invariants must be initialized before the associated locks can be
acquired. In O'Hearn's simple concurrent language~\cite{OHearn07}, the set of
locks is static and initialization of resource invariants is achieved in a
global initialization phase. This is not possible when locks are created
dynamically. 
Conceivably, we could tie the initialization of resource
invariants to the end of object constructors. However, this is problematic
because Java's object constructors are free to leak references to partially
constructed objects (e.g., by passing $\This$ to other methods). Thus, in
practice we have to distinguish between initialized and uninitialized objects
semantically. Furthermore, a semantic distinction enables late initialization
of resource invariants, which can be  useful for objects that remain
thread-local for some time before getting shared among threads. 

Therefore, we distinguish between \emph{fresh} locks and \emph{initialized}
locks. A fresh lock does not yet guard its resource invariant and thus it is not ready
to be acquired yet. An initialized lock, however, is ready to be acquired.
Initially, locks are fresh and they might become initialized later (and then will remain initialized). We require programmers
to explicitly change the state of locks (from fresh to initialized) with a {\tt commit} command:
\begin{displaymath}
  \begin{array}{rcl}
    \scmd \in \SpecCmdSet & \Is &
    \dots \Or \HdCommit \specval \Or \dots
  \end{array}
  \label{idx:spec:command:locks}
\end{displaymath}

Operationally, $\HdCommit \specval$ is a no-op; semantically it checks
that $\specval$ is fresh and changes $\specval$'s state to initialized.
For expressiveness {\tt commit}'s receiver ranges over specification variables, 
which include both program variables and logical variables (such as class parameters).
In real-world Java programs, a possible default would be to add a {\tt commit} command
at the end of constructors.

We assume that class tables always contain the following class declaration:
\begin{tabbing}
\tt class Object \lbr
\\[\jot] 
~~\tt \hlspec{pred inv = true;} 
\\[\jot] 
~~\tt \hlspec{final} void wait(); 
\\[\jot] 
~~\tt \hlspec{final} void notify(); 
\\[\jot] 
~~\tt \hlspec{final} void notifyAll(); \\[\jot]
\rbr 
\end{tabbing}

The distinguished {\tt inv} predicate assigns to each lock a resource
invariant. The definition {\tt true} is a default and objects meant to be used as locks
should extend {\tt inv}'s definition in subclasses of {\tt Object} (just like any other abstract predicate). 
As usual~\cite{OHearn07}, the resource invariant $o.\monInvpred$ can be assumed when $o$'s
lock is acquired non-reentrantly and must be established when $o$'s lock is
released with its reentrancy level dropping to~$0$. 

The methods \texttt{wait}, \texttt{notify} and \texttt{notifyAll} do not have Java
implementations, but are implemented natively. To model this, our
operational semantics (page~\pageref{idx:op:sem:locks}) specifies their behavior explicitly.
If $o.\java{wait} \lpa \rpa$ is called when object $o$ is
locked at reentrancy level $n$, then $o$'s lock is released and the
executing thread temporarily stops executing. 
If $o.\java{notify} \lpa \rpa$ is called, one thread that is stopped (because this
thread called $o.\java{wait}\lpa\rpa$ before) resumes and starts
competing for $o$'s lock. When a resumed thread reacquires $o$'s lock,
its previous reentrancy level is restored.
The method \texttt{notifyAll} performs similar to \texttt{notify} except that it resumes all the waiting threads.
Since we can specify method contracts for {\tt wait} and {\tt notify}, we do
not put them in our set of commands 
(see Section~\ref{sec:hoare:locks}).
In contrast, {\tt lock}, {\tt unlock}, and {\tt commit} are put in our
set of commands, because the Hoare rules for these methods cannot be expressed
using the syntax of contracts available to programmers: we need extra expressivity (see Section~\ref{sec:hoare:locks}).

\paragraph{Runtime Structures and Initialization.} To represent locks in the operational semantics, we
use a \emph{lock table}. \emph{Lock tables} map objects $o$ to either the symbol $\free$,
or to the thread object $t$ that currently holds $o$'s lock and a number that
counts how often $t$ currently holds $o$.
Accordingly states and the initial state of a program are extended with a lock table (see Figure~\ref{fig:settings:locks}).
Initially, the lock table of a program is empty (hence the second $\emptyset$).

\begin{figure}
$\begin{array}{ll}
  \lm \in \LockTableSet = \ParFun \ObjIdSet {\set\free \uplus {(\ObjIdSet \times \N)}} 
  \labelandtext{(Lock Table)}{idx:locktable}
  \\
  \stt \in \StateSet = \HeapSet \times \changed{\LockTableSet} \times \ThreadPoolSet
    \labelandtext{(States)}{idx:state:locks}
  \\
   \initcmd c = \Statelm {\IdThread {\main} {(\BareThread c {\emptyset})}} {\set{\main \mapsto (\ThreadTy,\emptyset)} } {\changed{\emptyset}}
    \labelandtext{(Initialization)}{idx:initcmd:locks}
\end{array}
$
\caption{Run-time structure, states and initialization in presence of locks}  \label{fig:settings:locks}

\end{figure}

\paragraph{Operational Semantics.} We modify the \emph{operational semantics} defined in Section~\ref{subsec:language:fj}
to deal with locks. 
To represent states in which threads are waiting to be notified, we could
associate each object with a set of waiting threads (the ``wait set''). 
However, we prefer to avoid introducing yet another runtime structure,
and therefore represent waiting states syntactically as special head commands:
\begin{displaymath}
  \begin{array}{l}
    \hc \Is \dots \ \Or\ \HdWaiting o n \ \Or\ \HdResume o n\ \Or\ \dots
    \\
    \mbox{\emph{Restriction:} These clauses must not occur in source programs.}
  \end{array}
  \label{idx:head:commands:locks:b}
\end{displaymath}

  If thread $p$'s head command is $\HdWaiting o n$,
  then $p$ is waiting to be notified. If thread $p$'s head command is $\HdResume o n$,
  then $p$ has been notified to resume competition for $o$'s lock at
  reentrancy level $n$, and is now competing for this lock.

Below we list the existing cases of the operational semantics that are
slightly modified: \ref{rule:red-new} and \ref{rule:red-call:fj}, and
the  cases  that  are  added: \ref{rule:red-lock},   \ref{rule:red-unlock},   \ref{rule:red-wait},  \ref{rule:red-notify-one},
\ref{rule:red-skip-notify}, \ref{rule:red-notify-all},
\ref{rule:red-skip-notify-all} and \ref{rule:red-resume}.
Except that a lock table is added, most of the existing cases of the operational
semantics are left untouched.

\renewcommand{\RuleRedNew}[4]{#1
    {
    \begin{array}{l}
      \hspace{-2ex}
      {#4} \notin \dom\hp 
      \ \;
      \hp' = \updtfun \hp {#4} {(\TyApp C \perms,\init {\TyApp C \perms})}
      \\
      \hspace{-2ex}
      s' = \updtstore s \lvar {#4}
      \ \;
      \changed{\lm' = \updtfun \lm o \free}
    \end{array}
    }
    {
     \Statelm { 
              \tpool \parpop
              \Thread {#2} {\New \lvar C \perms {#3}} s 
            }
            \hp 
            {\changed\lm}
     \step
     \Statelm { 
              \tpool \parpop
              \Thread {#2} {#3} {s'}
            } 
            {\hp'}
            {\changed{\lm'}}
   }  
}

\renewcommand{\RuleRedCall}[6]{#1
    {
    m \not\in \set{\java{fork},\java{join},\changed{\java{wait},\java{notify},\java{notifyAll}}} 
    }
    {
      \DynTy h {#3} = \TyApp C {\specvals}
      \quad
      \mbody m {\TyApp C {\specvals}} 
      = \Mbody {\logvars,\logvars'} {\rvar_0,\rvars} {{#5}} 
      \quad
      {#6} = #5[#3/\rvar_0,\vs/\rvars]
    }
    { 
     \Statelm { 
              \tpool \parpop
              \Thread {#2} {\Call \lvar {#3} m \vs {#4}} s 
            } 
            \hp
            {\changed\lm}
     \step
     \Statelm { 
              \tpool \parpop
              \Thread {#2} {\Bind \lvar {#6} {#4}} s
            }
            \hp 
            {\changed\lm}
   }  
}

\newcommand{\RuleRedLock}[4]{#1
    {
      (\lm(#4) = \free,\ \lm' = {\updtfun \lm {#4} {\Cpl  {#2} 1}})
      \mbox{ or }
      (\lm(#4) = \Cpl {#2} n,\ \lm' = {\updtfun \lm {#4} {\Cpl {#2} {n+1}}})
    }
    { 
     \Statelm { 
              \tpool \parpop
              \Thread {#2} {\Seq {\HdLock {#4}} {#3}} s 
            }
            \hp 
            \lm
     \step
     \Statelm { 
              \tpool \parpop
              \Thread {#2} {#3} {s}
            } 
            \hp
            {\lm'}
   }  
}

\newcommand{\RuleRedUnlock}[5]{#1
    {
    \begin{array}{l}
      \lm(#4) =  \Cpl  {#2} n
      \quad
      n = 1 \Rightarrow \lm' = \updtfun \lm {#4} \free
      \\
      n > 1 \Rightarrow \lm' = \updtfun \lm {#4} {\Cpl {#2} {n-1}}
    \end{array}
   }
    { 
     \Statelm { 
              \tpool \parpop
              \Thread {#2} {\Seq {\HdUnlock {#4}} {#3}} s 
            }
            \hp 
            \lm
     \step
     \Statelm { 
              \tpool \parpop
              \Thread {#2} {#3} s 
            } 
            \hp
            {\lm'}            
   }  
}

\newcommand{\RuleRedWait}[4]{#1
    {
      \lm(#4) = \Cpl {#2} n 
      \quad
      \lm' = \updtfun \lm {#4} \free
    }
    { 
     \begin{array}{l}
     \Statelm { 
              \tpool \parpop
              \Thread {#2} {\Wait \lvar {#4} {#3}} s 
            }
            \hp 
            \lm
     \\
     \qquad
     \step
     \Statelm { 
              \tpool \parpop
              \Thread {#2} {\Waiting {#4} n {\Resume {#4} n {#3}}} s
            } 
            \hp
            {\lm'}
    \end{array}
   }  
}

\newcommand{\RuleRedNotifyOne}[7]{#1
    {
      \lm(#4) = \Cpl {#2} n
   }
    { 
     \begin{array}{l}
     \Statelm { 
              \tpool 
              \parpop
              \Thread {#2} {\Notify \lvar {#4} {#3}} s 
              \parpop
              \Thread {#5} {\Waiting {#4} {n'} {#6}} {#7}
            }
            \hp 
            \lm
     \\
     \qquad
     \step
     \Statelm { 
              \tpool \parpop
              \Thread {#2} {#3} s 
              \parpop
              \Thread {#5} {#6} {#7}
            } 
            \hp
            \lm
     \end{array}
   }  
}

\newcommand{\RuleRedNotifyAll}[7]{#1
    {
      \lm(#4) = \Cpl {#2} n
   }
    { 
     \begin{array}{l}
     \Statelm { 
              \tpool 
              \parpop
              \Thread {#2} {\NotifyAll \lvar {#4} {#3}} s 
              \parpop
              \Thread {#5} {\Waiting {#4} {n'} {#6}} {#7}
            }
            \hp 
            \lm
     \\
     \qquad
     \step
     \Statelm { 
              \tpool \parpop
              \Thread {#2} {\NotifyAll \lvar {#4} {#3}} s 
              \parpop
              \Thread {#5} {#6} {#7}
            } 
            \hp
            \lm
     \end{array}
   }  
}

\newcommand{\RuleRedSkipNotify}[4]{#1
    {
      \lm(#4) = \Cpl {#2} n
    }
    { 
     \Statelm { 
              \tpool 
              \parpop
              \Thread {#2} {\Notify \lvar {#4} {#3}} s 
            }
            \hp 
            \lm
     \step
     \Statelm { 
              \tpool \parpop
              \Thread {#2} {#3} s 
            } 
            \hp
            \lm
   }  
}
\newcommand{\RuleRedSkipNotifyAll}[4]{#1
    {
      \lm(#4) = \Cpl {#2} n
    }
    { 
     \Statelm { 
              \tpool 
              \parpop
              \Thread {#2} {\NotifyAll \lvar {#4} {#3}} s 
            }
            \hp 
            \lm
     \step
     \Statelm { 
              \tpool \parpop
              \Thread {#2} {#3} s 
            } 
            \hp
            \lm
   }  
}

\newcommand{\RuleRedResume}[4]{#1
    {
      \lm(#4) = \free 
      \quad
      \lm' = {\updtfun \lm {#4} {\Cpl {#2} n}}
    }
    { 
     \Statelm { 
              \tpool \parpop
              \Thread {#2} {\Resume {#4} n {#3}} s 
            }
            \hp 
            \lm
     \step
     \Statelm { 
              \tpool \parpop
              \Thread {#2} {#3} {s}
            } 
            \hp
            {\lm'}
   }  
}

\newcommand{\RuleRedCommit}[4]{#1
   { 
     \Statelm { 
              \tpool \parpop
              \Thread {#2} {\Seq {\HdCommit {#4}} {#3}} s
            } 
            \hp
            \lm
     \step
     \Statelm { 
              \tpool \parpop
              \Thread {#2} {#3} s 
            }
            \hp
            \lm
   }  
}

\begin{small}
  \begin{display}{State Reductions, $\stt \sstep \cls \stt'$:}
    \dots
    \\[\jot]
    \RuleRedNew {\stateaxiomcond{(Red New)}} p \cmd o
    \label{rule:red-new:locks}
    \\[\jot]
    \RuleRedCall{\stateaxiomshortlongcond{(Red Call)}} p o c {c_m} {c'}
    \label{rule:red-call:locks}
    \\[\jot]
    \RuleRedLock {\stateaxiomcond{(Red Lock)}} p \cmd o
    \label{rule:red-lock}
    \\[\jot]
    \RuleRedUnlock {\stateaxiomcond{(Red Unlock)}} p \cmd o q
    \label{rule:red-unlock}
    \\[\jot]
    \RuleRedWait {\stateaxiomcond{(Red Wait)}} p \cmd o
    \label{rule:red-wait}
    \\[\jot]
    \RuleRedNotifyOne {\stateaxiomcond{(Red Notify)}} p \cmd o q {\cmd'} {s'}
    \label{rule:red-notify-one}
    \\[\jot]
    \RuleRedNotifyAll {\stateaxiomcond{(Red Notify All)}} p \cmd o q {\cmd'} {s'}
    \label{rule:red-notify-all}
    \\[\jot]
    \RuleRedSkipNotify {\stateaxiomcond{(Red Skip Notify)}} p \cmd o
    \label{rule:red-skip-notify}
    \\[\jot]
    \RuleRedSkipNotifyAll {\stateaxiomcond{(Red Skip Notify All)}} p \cmd o
    \label{rule:red-skip-notify-all}
    \\[\jot]
    \RuleRedResume {\stateaxiomcond{(Red Resume)}} p \cmd o
    \label{rule:red-resume}
    \\[\jot]
    \dots
  \end{display}
  \label{idx:op:sem:locks}
\end{small}


  Rule~\ref{rule:red-lock} distinguishes two cases: 
  (1) lock $o$ is acquired for the first time ($\lm(o) = \free$) and
  (2) lock $o$ is acquired reentrantly ($\lm(o) = \Cpl p n$).
  Similarly, rule~\ref{rule:red-unlock} distinguishes two cases: 
  (1) lock $o$'s reentrancy level decreases but $o$ remains acquired 
  ($\lm(o) = \Cpl p n$ and $n > 1$) and
  (2) lock $o$ is released ($\lm(o) = \Cpl p 1$).
  Rule~\ref{rule:red-wait} fires only if the thread considered previously acquired
  {\tt wait}'s receiver. In this case, {\tt wait}'s receiver is released
  and the thread enters the {\tt waiting} state. The thread's reentrancy level is
  stored in {\tt waiting}'s argument.

  Like rule~\ref{rule:red-wait}, the rules~\ref{rule:red-notify-one}, \ref{rule:red-notify-all}, \ref{rule:red-skip-notify} and \ref{rule:red-skip-notify-all}
  fire only if the thread considered previously acquired {\tt notifyAll()}'s receiver.
  The rules \ref{rule:red-notify-one} and \ref{rule:red-notify-all} fires if there \emph{exists} at least one thread waiting on {\tt notify}'s receiver.
  In case of \ref{rule:red-notify-one}, one of the waiting threads (arbitrarily) is resumed, 
  while \ref{rule:red-notify-all} awakes all the waiting threads.
  If there is no thread waiting on {\tt notify}'s receiver, based on the calling command, either \ref{rule:red-skip-notify} or \ref{rule:red-skip-notify-all} fires.
  In this case, the call to {\tt notify} and {\tt notifyAll} has no effect on other threads.
  In Java, if {\tt $o$.wait()}, {\tt $o$.notify()} and {\tt $o$.notifyAll()} are called
  by a thread that does not hold $o$, an {\tt IllegalMonitorState} exception is raised.
  In our semantics, this is modeled by being stuck. 
  In Section~\ref{sec:hoare:locks}, we will give preconditions for {\tt wait} and {\tt notify}
  that ensure that verified programs 
  would never throw an {\tt IllegalMonitorState }  exception (got stuck in our model).
  Rule~\ref{rule:red-resume} resumes a thread that previously waited on some lock and restores the reentrancy level.

\subsection{Separation Logic for Reentrant Locks}
\label{sec:sl:locks}

In this section, we describe the new formulas that we add to the specification language of
Section~\ref{sec:sl:fj}.

As explained earlier, a proof system for reentrant locks must keep track
of the locks that the current thread holds.
To this end, we enrich our specification language:
\begin{displaymath}
\begin{array}{rcl}
  \specval \in \SpecValSet & \Is & 
  \dots \ \Or\  \nil \ \Or\  \MsCup \specval \specval \ \Or\ \dots
  \\
  F \in \PermFormSet & \Is & \dots
  \ \Or\  \Lockset \specval
  \ \Or\  \contains \specval e
  \ \Or\ \dots
\end{array}
  \label{idx:formulas:locks}
\end{displaymath}

It is convenient to allow using objects as singleton locksets (rather than introducing explicit syntax for converting from objects to singleton locksets).
We classify the new formulas into
\emph{copyable} and \emph{non-copyable} ones. Copyable formulas represent
\emph{persistent state properties} (\ie properties that hold forever, once
established), whereas non-copyable formulas represent \emph{transient state 
properties} (\ie properties that hold temporarily). For copyable $F$, we
postulate the axiom $\lpa G \,\CAnd F \rpa \LImplies \lpa G \,\LAnd F \rpa$,
whereas for non-copyable formulas we postulate no such axiom. 
Note that this axiom implies $F \LImplies \lpa F \LAnd F \rpa$, hence the term 
``copyable''.
 The new specification values and formulas have the following intuitive meaning:
\begin{itemize}
\item 
  $\nil$: the empty multiset.
\item 
  $\MsCup \specval {\specval'}$: 
  the multiset union of multisets $\specval$ and $\specval'$.
\item 
  $\Lockset \specval$: 
  $\specval$ is the multiset of locks held by the current thread. 
  Multiplicities record the current reentrancy level.
  \emph{(non-copyable)}
\item
  $\contains \specval {e}$: multiset $\specval$ contains object $e$.
  \emph{(copyable)}
\end{itemize}


\paragraph{Initialization
}~~When~~verifying~~the~~body~~of~~\texttt{Thread.run()},~~we~~assume
$\Lockset\nil$ as a precondition.
As explained before, 
resource invariants must be initialized before the associated locks can be
acquired. 
To keep track of the state of locks
in our verification system, we introduce two more formulas:  
%

\begin{displaymath}
\begin{array}{rcl}
  F \in \PermFormSet & \Is & \dots
  \ \Or\  \Fresh e
  \ \Or\  \Initialized e
  \ \Or\ \dots
\\
\multicolumn{3}{l}{%
\mbox{\emph{Restriction:} $\Initialized e$ must not occur in
  negative positions.}}
\end{array}
  \label{idx:formulas:locks:b}
\end{displaymath}

\begin{itemize}
\item
  $\Fresh e$: $e$'s resource invariant is not yet initialized. 
  \emph{(non-copyable)}
\item
  $\Initialized e$: $e$'s resource invariant has been initialized.
  \emph{(copyable)}
\end{itemize}

Because $\Initialized e$ is copyable, {\tt initialized} formulas can ``spread'' to all
threads, allowing all threads to try to acquire locks ({\tt initialized} will be a precondition for (initial) lock acquirement; see Section~\ref{sec:hoare:locks}).  

\paragraph{Types.} We add a type to represent locksets and we postulate $\Object \subty \LocksetTy$:
\begin{displaymath}
  T \Is \dots \Or \LocksetTy \Or \dots
  \label{idx:types:locks}
\end{displaymath}

\label{idx:subtyping:locks}

Because we allow arbitrary specification values (including locksets) 
as type parameters, we consider that
types with semantically equal type parameters are type-equivalent.
Technically, we let $\specvaleq$ be the least equivalence relation on
specification values that satisfies the standard multiset axioms:
%
\begin{display}{Equivalence of Specification Values: $\specval \specvaleq \specval$}
\clause{
\begin{array}{c}
  \MsCup \nil \specval \specvaleq \specval
  \qquad
  \MsCup \specval {\specval'} 
  \specvaleq 
  \MsCup {\specval'} \specval
  \qquad
  \MsCup {\lpa \MsCup {\specval} {\specval'} \rpa} {\specval''}
  \specvaleq
  \MsCup {\specval} {\lpa \MsCup {\specval'} {\specval''} \rpa}
\end{array}
}
  \label{idx:specvals:locks}
\end{display}
Then we postulate that $\TyApp t {\specvals} \subty \TyApp t {\specvals'}$
when $\specvals \specvaleq \specvals'$.  

\paragraph{Augmented heaps.} 
\label{idx:augmented heaps:locks} 
 To express the semantics of the new formulas,
we need to extend augmented heaps with three new components.
From now on, augmented heaps are $6$-tuples of a heap, a permission table, a join table,
an \emph{abstract lock table} $\lt \in \ParFun \ObjIdSet \Bag\ObjIdSet$,
a \emph{fresh set} $\calf \subseteq \ObjIdSet$,
and an \emph{initialized set} $\cali \subseteq \ObjIdSet$. 
\label{idx:abstractlocktables}

\emph{Abstract lock tables} map thread identifiers to locksets.
Just as permission tables are an abstraction of heaps, abstract lock tables are
an abstraction of lock tables.
The compatibility relation captures that distinct threads cannot hold
the same lock (we use $\sqcap$ to 
denote bag intersection, $\sqcup$ for bag union, and $\emptybag$ for the 
empty bag):
\begin{displaymath}
  \compatible \lt {\lt'} \text{ iff }
  \left\{\begin{array}{l}
      \dom\lt \cap \dom{\lt'} = \emptyset
  \\      
      (\forall o \in\dom\lt, p \in \dom{\lt'})
      (\lt(o) \sqcap \lt'(p) = \emptybag)
  \end{array}\right.
  \quad \text{; and} \quad
  \lt \LAnd \lt' \deq \lt \cup \lt'
  \label{idx:join:locks}
\end{displaymath}

\emph{Fresh set} $\calf$ keeps track of allocated but not yet
initialized objects, while \emph{initialized set}~$\cali$ keeps track
of initialized objects. We define $\compatibleSym$ for fresh sets as
disjointness to mirror that $\Fresh o$ is non-copyable,
and for initialized sets as equality to mirror that 
$\Initialized o$ is copyable:
\begin{displaymath}
  \begin{array}{r c l @{\quad} r c l}
    \compatible \calf {\calf'} & \text{ iff } & \calf \cap \calf' =
    \emptyset 
    &
    \quad \text{; and} \quad \calf \LAnd \calf' & \deq & \calf \cup \calf'
    \\
    \compatible \cali {\cali'} & \text{ iff } & \cali = \cali'
    &
    \text{; and} \quad
    \cali \LAnd \cali' & \deq & \cali \ (= \cali') 
  \end{array}
\end{displaymath}

We require augmented heaps to satisfy the following axioms
(in addition to Section~\ref{sec:sl:fj}'s axioms):
\begin{enumerate}[label=\({\alph*}]
 \setcounter{enumi}{4}
  \item\label{rsc-fresh-init}  $\calf \cap \cali = \emptyset$.
  \item\label{rsc-locks} If $o \in \lt(p)$ then $o \in \cali$.
\end{enumerate}

Axiom~\ref{rsc-fresh-init} ensures that 
an object can never be
both fresh \emph{and} initialized.
Axiom~\ref{rsc-locks} ensures that locked objects are initialized.

As usual, we define projection operators:
\begin{displaymath}
  {\Sex \hp \pt \calj \lt \calf \cali}_{\kw{lock}} \deq \lt
  \quad
  {\Sex \hp \pt \calj \lt \calf \cali}_{\kw{fresh}} \deq \calf
  \quad
  {\Sex \hp \pt \calj \lt \calf \cali}_{\kw{init}} \deq \cali
\end{displaymath}

%

\paragraph{Semantics of Values.} Before defining the semantics of formulas,
we extend the semantics of values to locksets.
Recall that $\SemValSet$ is the set of semantic values (defined in Section~\ref{subsubsec:sl:semantics}).
Formerly,
$\SemValSet = \set \Null \,\cup\, \ObjIdSet \,\cup\, \IntSet \,\cup\,
\BoolSet \, \cup (0,1]$. We extend this set to include semantic domains for
locksets: 
%
\begin{displaymath}
  \semval \in \SemValSet \Deq \changed{(\set \Null \,\cup\, \ObjIdSet \,\cup\, \IntSet \,\cup\,\BoolSet \, \cup (0,1] \,\cup\ \Bag \ObjIdSet) / \equiv}
  \label{idx:semval:locks}
\end{displaymath}

\noindent
where $\equiv$ is the least equivalence relation on 
$\SemValSet$ such that 
$o \equiv [o]$ for all object ids $o$. That is, 
$\equiv$ is the least equivalence relation that identifies object identifiers
  with singleton bags.

Let $\WtClSpecValSet$ be the set of well-typed, specification
values:
\begin{displaymath}
  \WtClSpecValSet \deq
  \setcomp {\ \specval\ } 
    {\ (\exists \Gamma,T)(\dom\Gamma \subseteq \ObjIdSet
       \mbox{ and } 
       \Gamma \vdash \hastype \specval T)\ }
\end{displaymath}

To define the semantics of well-typed, open specification values, we simply
define the semantics of the two new specification values: 
%
\begin{displaymath}
  \hastype {\sem .} {\Fun \WtClSpecValSet \SemValSet}
  \qquad
  \sem \nil \deq \emptybag
  \qquad
  \sem {\MsCup \specval {\specval'}}
  \deq 
  \sem \specval \sqcup \sem{\specval'}
\end{displaymath}

\paragraph{Semantics of Formulas.} We now state the semantics
of formulas introduced to deal with reentrant locks:
\begin{displaymath}
\begin{array}{rcl@{\quad }c@{\quad}l}
\Gamma \vdash \pe;\Sex \hp \pt \calj \lt \calf \cali;s & \models & \Lockset \specval
& \iff &
\lt(o) = \sem \specval 
\mbox{ for some $o$}
\\[\jot]
\Gamma \vdash \pe;\Sex \hp \pt \calj \lt \calf \cali;s & \models & 
\contains \specval {e}
& \iff &
\esem {e} h s \in \sem\specval 
\\[\jot]
\Gamma \vdash \pe;\Sex \hp \pt \calj \lt \calf \cali;s & \models & \Fresh e
& \iff &
\esem e h s \in \calf
\\[\jot]
\Gamma \vdash \pe;\Sex \hp \pt \calj \lt \calf \cali;s & \models & \Initialized e
& \iff &
\esem e h s \in \cali
\end{array}
\end{displaymath}

These clauses are self-explanatory, except perhaps the existential
quantification in the clause for $\Lockset \specval$. Intuitively, this clause
says that there exists a thread identifier $o$ in $\lt$'s domain such
that $\specval$ denotes the current lockset associated with $o$. 
When we interpret an assertion for a single thread, we restrict the models to the ones where $\lt$ only contains a single entry for the current thread.  
Hence, the existential can only choose the current thread id. This restriction is in the (Thread) rule on page~\pageref{rule:thread:locks}. 

\begin{table}
$\begin{array}{cl}
\axiom v \Gamma {\jnot \lpa \contains \nil e \rpa}
\labelandtext{(Member Nil)}{axiom:member:nil}
\\
\axiom v \Gamma 
{ \contains {\lpa \MsCup \specval {\specval'} \rpa} {e}
	\LEquiv
		\lpa \contains {\specval}{e} \;\jor\; \contains {\specval'} {e} \rpa
}
\labelandtext{(Member Rec)}{axiom:member:rec}
\\
\specval \specvaleq \specval'
\ \Rightarrow\ 
\axiom v \Gamma {\specval \jdeq \specval'}
\labelandtext{(Eq Bag)}{axiom:eq:bag}
\\
\axiom v \Gamma {\specval \jdeq \specval}
\labelandtext{(Eq Refl)}{axiom:eq:refl}
\\
\axiom v \Gamma {\specval \jdeq \specval'}
\ \Rightarrow\ 
\axiom v \Gamma {\specval' \jdeq \specval}
\labelandtext{(Eq Sym)}{axiom:eq:sym}
\\
\axiom v \Gamma {\specval \jdeq \specval' \CAnd \specval' \jdeq \specval''}
\ \Rightarrow\ 
\axiom v \Gamma {\specval \jdeq \specval''}
\labelandtext{(Eq Trans)}{axiom:eq:trans}
\\
\begin{array}{c}
G \in \set {e,\changed{\ \contains \specval e,\ \Initialized e}}
\\
\Downarrow
\\
\axiom v \Gamma \lpa F \CAnd G \rpa  \LImplies \lpa F \LAnd G \rpa
\quad
\end{array}
\labelandtext{(Copyable)}{axiom:copyable:locks}
\end{array}$
\caption{Axioms to reasons about bags and copyable formulas}
\label{table:locks:axioms}
\end{table}

\paragraph{Axioms.} Table~\ref{table:locks:axioms} lists the new
axioms that can be used as an extension to the logical consequence
judgement (similar to the axioms in Table~\ref{table:axioms}. These
axiomatize bag membership (\ref{axiom:member:nil}
and~\ref{axiom:member:rec}); bag equality (\ref{axiom:eq:bag});
copyability and equality between specification values
(\ref{axiom:eq:refl},~\ref{axiom:eq:sym},~\ref{axiom:eq:trans}).
Axiom~\ref{axiom:copyable:locks} updates
Section~\ref{subsec:sl:proof:theory}'s \ref{axiom:copyable} axiom
about copyability of formulas. It is straightforward to extend
Theorem~\ref{thm:sound-entails} to these new axioms.

\subsection{Hoare Triples}
\label{sec:hoare:locks}

In this section, we modify the Hoare triple~\ref{rule:hoa-new} for allocating new objects
and we present Hoare triples for the new commands of our language.

We modify rule~\ref{rule:hoa-new} so that it emits the {\tt fresh} predicate
in its postcondition:

\begin{center}
  \staterulelabel{(New)}
  {
      \TyApp C {\TypedVar \Ts \logvars} \in \cls
      \quad
      \Gamma \vdash \hastype {\specvals} {\subst \specvals\logvar \Ts}
      \quad
      \TyApp C \perms \subty \Gamma(\lvar)
  }
  {
    \Gamma;v \vdash
    \begin{array}{c}
      \set \True
      \\
      {\HdNew \lvar C \perms}
      \\
      \set
      {
      \GGet \lvar \initpred
      \LAnd\, \iisclassof C \lvar
      \LAnd\, \changed{\varoast_{\Gamma(u) \subty \Object} {\lvar \jneq u}\,}
      \LAnd\, \changed{\Fresh \lvar}
      }
    \end{array}
  }
  \label{rule:new}
\end{center}

In addition to the usual {\tt init} and {\tt classof} predicates,
\ref{rule:new}'s postcondition records that the newly created object is
distinct from all other objects that are in scope. This postcondition is
usually omitted in separation logic, because separation logic avoids
explicit reasoning about the absence of aliasing. Unfortunately, we need this kind of reasoning when establishing the precondition for
the rule \ref{rule:lock} below, which requires
that the lock is \emph{not} already held by the current thread. 

The specification command $\GGet {\specval}{\java{commit}}$
triggers $\specval$'s transition from the \texttt{fresh} to 
the \texttt{initialized} state, provided $\specval$'s resource invariant is
established: 

\begin{center}
  \staterulelabel{(Commit)}
  {
    \Gamma \vdash \hastype {\specval,\specval'} {\Object,\LocksetTy}
  }
  {
  \Gamma;v \vdash
    \begin{array}{c}
      \set
      {
        \Lockset {\specval'}
        \LAnd \MonInv \specval
        \LAnd \Fresh \specval
      }
      \\
      {\HdCommit \specval}
      \\
      \set
      {
        \Lockset {\specval'}
        \LAnd \jnot \lpa \contains {\specval'} \specval \rpa
        \LAnd \Initialized \specval
      }
    \end{array}
  }
  \label{rule:commit}
\end{center}

Intuitively, the fact that {\tt $\specval$.inv} appears in~\ref{rule:commit}'s precondition
but does not appear in~\ref{rule:commit}'s postcondition indicates that after being
committed, lock $\specval$ \emph{guards} its resource invariant: the resource
invariant {\tt $\specval$.inv} has been given to lock $\specval$ and 
{\tt $\specval$.inv} is not available anymore to the executing thread. Furthermore,
because $\Fresh \specval$ only holds if $\specval \jneq \Null$, this rule
ensures that only non-null locks can become initialized.

The precondition of rule~\ref{rule:commit} ensures that monitor invariants cannot mention {\tt Lockset} predicates as it mentions both
a {\tt Lockset} predicate and the lock's monitor invariant {\tt inv}. 
This follows from the semantics of the {\tt Lockset} predicate
and the semantics of the $\LLAnd$ operator: two {\tt Lockset} predicates cannot be $\LLAnd$-conjoined. 
This is important because {\tt Lockset} predicates are interpreted \wrt the current thread.

There are two rules each for locking and unlocking, depending on whether or
not the \texttt{lock}/\texttt{unlock} is associated with an initial entry or a reentry.

First, we present the two rules for locking:

\begin{center}
    \staterulelabel{(Lock)}
    {
    \Gamma \vdash \hastype {u,\specval} {\Object,\LocksetTy}
    }
    {
      \begin{array}{c}
        \Gamma;v \vdash
        \set
        {
          \Lockset \specval
          \LAnd \jnot\lpa \contains \specval u \rpa
          \LAnd \Initialized u
        }
        \\
        {
          \HdLock u
        }
        \\
        \set
        {
          \Lockset {\MsCup u \specval}
          \LAnd
          \MonInv u
        }
    \end{array}
    }
    \label{rule:lock}
\end{center}

\begin{center}
  \staterulelabel{(Re-Lock)}
    {
      \Gamma \vdash \hastype {u.\specval} {\Object,\LocksetTy}
    }
    {
      \hhoare v \Gamma 
      {
        \Lockset {\MsCup u \specval}
      }
      {
        \HdLock u
      }
      {
        \Lockset {\MsCup u {\MsCup u \specval}}
      }
   }
\label{rule:re-lock}
\end{center}


The rule~\ref{rule:lock} applies when lock $u$ is acquired non-reentrantly, as
expressed by the precondition $\Lockset \specval \LAnd \jnot \lpa 
\contains \specval u \rpa$. The precondition $\Initialized u$ makes sure that 
\un threads only acquire locks whose resource invariant is initialized, and
\deux no null-error can happen (because initialized values are non-null).
The postcondition adds $u$ to the current thread's lockset, and assumes $u$'s
resource invariant. The resource invariant obtained is $\MonInv u$ (without $\tjkw{@}$
selector). 

Proving~\ref{rule:lock}'s precondition requires reasoning about
aliases because one has to prove $\jnot \lpa \contains \specval u
\rpa$. In practice, this assertion is proven by showing that $u$ is
different from all elements of lockset $\specval$. Such a reasoning is
a form of alias analysis. On one hand this is unfortunate, because
separation logic's power comes from the fact that it does not need to
reason about aliases. On the other hand, this seems
unavoidable. Whether this is problematic in practice needs to be
investigated on large case studies. In Section
\ref{sec:examples:lock}, the lock coupling example illustrates how ownership can be used as a
possible solution to the problem.

The rule~\ref{rule:re-lock} applies when a lock is acquired reentrantly.
The precondition of~\ref{rule:re-lock}, contrary to~\ref{rule:lock},
does not require $\Initialized u$,
because this follows from $\Lockset {\MsCup u \specval}$
(locksets contain only initialized values).


Based on rule~\ref{rule:re-lock} one expects the lock's resource invariant to hold. To have a more accurate feedback, in practice, one may define a derived rule like~\ref{rule:re-lock:accurate} to enforce the existence of the resource invariant: 


\begin{center}
    \staterulelabel{(Re-Lock-Accurate)}
    {
      \Gamma \vdash \hastype {u,\specval} {\Object,\LocksetTy}
    }
    {
        \Gamma; v \vdash 
	\begin{array}{c}
	  \set {
	    \Lockset {\MsCup u \specval} \LAnd \MonInv u
	  }
	  \\
	  {
	    \HdLock u
	  }
	  \\
	  \set
	  {
	    \Lockset {\MsCup u {\MsCup u \specval}} \LAnd \MonInv u
	  }
	\end{array}
    }
    \label{rule:re-lock:accurate}
\end{center}

Next, we present the two rules for unlocking:

\begin{center}
    \staterulelabel{(Re-Unlock)}
    {
      \Gamma \vdash \hastype {u,\specval} {\Object,\LocksetTy}
    }
    {
      \hhoare v \Gamma 
      {
        \Lockset {\MsCup u {\MsCup u \specval}}
      }
      {\HdUnlock u}
      {
        \Lockset {\MsCup u \specval}
      }
    }
    \label{rule:re-unlock}
\end{center}

\begin{center}
  \staterulelabel{(Unlock)}
    {
      \Gamma \vdash \hastype {u,\specval} {\Object,\LocksetTy}
    }
  {
    \hhoare v \Gamma 
    {
      \Lockset {\MsCup u \specval}
      \LAnd \MonInv u
    }
    {\HdUnlock u}
    {
      \Lockset \specval
    }
  }
  \label{rule:unlock}
\end{center}

The  rule \ref{rule:re-unlock}  applies  when  $u$'s  current  reentrancy  level  is  at  least  $2$  and
\ref{rule:unlock} applies when $u$'s resource invariant 
holds in the precondition.

\paragraph{Other Hoare Rules that Do Not Work.}
One might wish to avoid the 
inequalities in \ref{rule:new}'s postcondition. Several approaches
for this come to mind.  
First, one could drop the inequalities in \ref{rule:new}'s
postcondition, and rely on \ref{rule:commit}'s postcondition $\jnot
\lpa \contains {\specval'} \specval \rpa$ to establish
\ref{rule:lock}'s precondition.  While this would be sound, in general
it is too weak, as we are unable to lock $\specval$ if we first
lock some other object $x$ (because from $\jnot\lpa \contains
{\specval'} \specval \rpa$ we cannot derive $\jnot\lpa \contains {x
\cdot \specval'} \specval \rpa$ unless we know $\specval \jneq
x$). Second, the \texttt{Lockset} predicate could be abandoned
altogether, using a predicate $\Held \specval n$ instead, that specifies
that the current thread holds lock $\specval$ with reentrancy level
$n$. In particular, $\Held \specval 0$ means that the current thread
does not hold $\specval$'s lock at all. We could reformulate the rules
for locking and unlocking using the $\texttt{Held}$-predicate, and
introduce $\Held \lvar 0$ as the postcondition of \ref{rule:new},
replacing the inequalities. However, this approach does not
work, because it grants only the object creator permission to lock the
created object! While it is possible that a clever program logic
could somehow introduce $\Held \specval 0$-predicates in other ways
(besides introducing it in the postcondition of \ref{rule:new}), we
have not been able to come up with a workable solution along these lines.
Besides, it does not solve the aliasing problem.

\paragraph{Wait and notify.} Methods {\tt wait}, {\tt notify} and {\tt notifyAll} in class {\tt Object} (introduced in Section~\ref{sec:language:locks}) are specified as follows: 
  \begin{tabbing}
  \tt class Object\lbr
  \\[\jot] 
  ~~\tt \hlspec{pred inv = true;} 
  \\[\jot] 
  ~~\xy \hlspec{requires Lockset(S) * S contains this * inv;} \\
  ~~\xy \hlspec{ensures ~Lockset(S) * inv;} \\
  ~~\tt \hlspec{final} void wait(); 
  \\[\jot] 
  ~~\xy \hlspec{requires Lockset(S) * S contains this;} \\
  ~~\xy \hlspec{ensures ~Lockset(S);} \\
  ~~\tt \hlspec{final} void notify(); 
  \\[\jot] 
  ~~\xy \hlspec{requires Lockset(S) * S contains this;} \\
  ~~\xy \hlspec{ensures ~Lockset(S);} \\
  ~~\tt \hlspec{final} void notifyAll(); \\[\jot]
  \rbr 
  \end{tabbing}

The preconditions for \texttt{wait}, \texttt{notify} and \texttt{notifyAll} require that
the receiver is locked, thus ensuring that if a program can be
verified, it will never throw an
\texttt{IllegalMonitorState} exception (or be stuck, according to our
formal semantics). 
The postcondition of $\GGet o {\texttt{wait()}}$ ensures 
$\GGet o {\texttt{inv}}$, because $o$ is locked again just before
$\GGet o {\texttt{wait()}}$ terminates. 

\paragraph{Auxiliary Syntax.} Recall that in Section~\ref{sec:language:locks},
we added two new head commands {\tt waiting} and {\tt resume}
to represent waiting states. The Hoare rules for these commands are as follows:

\begin{center}
  \staterulelabel{(Waiting)}
    {
      \Gamma \vdash \hastype {\specval,o} {\LocksetTy,\Object}
    }
    {
    \Gamma; r \vdash
    \begin{array}{c}
      \set {
        \Lockset \specval
        \LAnd \Initialized o
      }
      \\
      {\HdWaiting o n}
      \\
      \set
      {
        \Lockset \specval
        \LAnd \Initialized o
      }
    \end{array}
    }
  \label{rule:hoa-waiting}
  \\[\GAP]
  \staterulelabel{(Resume)}
    {
      \Gamma \vdash \hastype {o,\specval} {\Object,\LocksetTy}
    }
    {
    \Gamma; r \vdash
    \begin{array}{c}
      \set {
        \Lockset \specval
	\LAnd \Initialized o
      }
      \\
      {\HdResume o n}
      \\
      \set
        {
          \Lockset {\MsCup {\MsCupN o n} \specval}
	  \LAnd \MonInv o \rpa
	}
      \end{array}
    }
    \label{rule:hoa-resume}
\end{center}

In \ref{rule:hoa-resume}, $\MsCupN o n$ denotes the multiset with $n$
occurrences of $o$. 
Of course,
the rules \ref{rule:hoa-waiting} and \ref{rule:hoa-resume} are never used in
source code verification, because source programs do not contain the auxiliary
syntax. Instead, the rules~\ref{rule:hoa-waiting} and~\ref{rule:hoa-resume}
are used to state and prove the preservation theorem.

\paragraph{The {\tt Thread} class.} 
Now we are ready to modify class {\tt Thread} of Section~\ref{sec:forkjoin}'s verification system to handle reentrant locks.
To handle reentrant locks, we modify
class {\tt Thread}'s method contracts as shown in Figure~\ref{fig:locks:class:thread}.
Intuitively, we forbid {\tt fork} and {\tt join}'s contracts (\ie~{\tt preFork} and {\tt postJoin})
to depend on the caller's lockset. This would not make sense since {\tt Lockset} predicates
are interpreted \wrt to the current thread. Obviously, 
a thread calling {\tt fork} (or {\tt join}) differs from the newly created (or the joined) thread.
We forbid {\tt fork} and {\tt join}'s contracts to depend on the caller's lockset by
(1) adding {\tt Lockset(S)} in {\tt fork}'s precondition: because
callers of {\tt fork} have to establish {\tt fork}'s precondition, this forbids
{\tt preFork} to depend on a {\tt Lockset} predicate (recall
that two {\tt Lockset} predicates cannot be $\LLAnd$-combined) and
(2) by adding {\tt Lockset(S)} in {\tt run}'s postcondition: this forbids
{\tt postJoin} to depend on a {\tt Lockset} predicate:
\begin{figure}[hbtp]
  \begin{tabbing}\tt
  cl\=\tt ass Thread ext Object\lbr 
  \\[\jot]
  \>\tt\xy \hlspec{pred  preFork = true;} \\
  \>\tt\xy \hlspec{group postJoin<perm p> = true;}
  \\[\jot]
  \>\xy \hlspec{requires \changed{\tt Lockset\lpa S\rpa} * preFork; ensures \changed{\tt Lockset\lpa S\rpa};} \\
  \>\tt \hlspec{final} void fork(); 
  \\[\jot]
  \>\xy \hlspec{requires Join(this,p); ensures postJoin<p>;} \\
  \>\tt \hlspec{final} void join(); 
  \\[\jot]
  \>\xy \hlspec{final} \hlspec{requires \changed{\tt Lockset\lpa nil\rpa} * preFork;} \\
  \>\xy\phantom{final} \hlspec{ensures ~\changed{\tt \lpa ex~Lockset~S\rpa\lpa Lockset\lpa S\rpa\rpa} * postJoin<1>;} \\
  \>\tt void run() \lbr\ null \rbr 
  \\[\jot]\tt
  \rbr 
  \end{tabbing}
  \caption{Class {\tt Thread}}
  \label{fig:locks:class:thread}
\end{figure}

\subsection{Verified Programs}
\label{sec:verified:programs:locks}


We need to update Section~\ref{subsec:verified:programs:fj}'s rules for runtime states to account
for reentrant locks. 
There are two changes to rule~\ref{rule:thread:fj}:
(1) premise $\dom \ahplock = \set o$ is added to ensure that a thread's
augmented heap only tracks the locks held by this thread and
(2) the thread's postcondition is modified to reflect
the change in {\tt join}'s postcondition in class {\tt Thread}. 
\begin{center}
\staterule{(Thread)}
  {
   \begin{array}{c} 
   \ahpg(o) \leq \sem \binfr
   \quad
   \Gamma \vdash \hastype \sigma {\Gamma'}
   \quad
   \goodstore {\Gamma,\Gamma'} s
   \\
   \cfv c \cap \dom{\Gamma'} = \emptyset
   \quad
   \changed{\dom \ahplock = \set o}
   \quad
   \holds {\Gamma[\sigma]} \pe \ahp s {F[\sigma]}    
   \\
   \hoare r {\Gamma,\Gamma'} F c \Void {\changed{\Ex S \LocksetTy {\Lockset S}} \LAnd \scalar \binfr {\Pred o \postJoin 1}}
   \end{array}
  }
  {\goodthread \ahp {\Thread o c s}}
\label{rule:thread:locks}
\end{center}

We define the set $\ready\ahp$ of
all initialized objects whose locks are not held, and the function $\kw{conc}$
that maps abstract lock tables to concrete lock tables:
\begin{displaymath}
\begin{array}{c}
    \ready \ahp \deq 
    \ahpinit \setminus
    \setcomp {\,o\,} {\,(\exists p)(o \in \lt(p))\,}
\\[\jot]
\\[-1ex]
\concapp \lt o \deq
\left\{
\begin{array}{l l}
  (p,\lt(p)(o)) & \mbox{iff } o \in \lt(p) \\
  \free & \mbox{otherwise}
\end{array}
\right.
\end{array}
\end{displaymath}

In $\kw{conc}$'s definition, we let $\lt(p)(o)$ stand for the
multiplicity of $o$ in $\lt(p)$. Note that $\kw{conc}$ is well-defined, 
by axiom~\ref{rsc-locks} for augmented heaps (see page~\pageref{rsc-locks}). The new rule for states ensures that there exists a
augmented heap $\ahp$ to satisfy the thread pool $\tpool$ and an augmented heap 
$\ahp'$ to satisfy the resource invariants of the locks that are ready to be
acquired. In addition, function $\kw{conc}$ relates the program's lock table
to the top level augmented heap's abstract lock table:
\begin{center}
  \staterulelabelbis
  {(State)}
  {
    \begin{array}{l}
    h = (\ahp \LAnd \changed{\ahp'})_{\kw{hp}}
    \\
    \changed{\lm = \conc {\ahplock}}
    \qquad
    \goodpool {\ahp} \tpool
    \quad
    \changed{
      \compatible \ahp {\ahp'}
      \quad
      \ahplock' = \emptyset
    }
    \\
    \changed{
      \kw{fst}(\ahph') \subseteq \kw{fst}(h) = \Gamma
      \quad
      \holds \Gamma \pe {\ahp'} \emptyset {\varoast_{o \in \ready {\ahp}} {\MonInv o}}
    }
  \end{array} }
  {
    \goodstate {\State \tpool {h,\changed l}}
  }
\label{rule:state:lock}
\end{center}

As in Section~\ref{subsec:verified:programs:fj},
using case analysis on the shape of the ($\goodstate \stt$)'s proof
tree, we have shown that the preservation
Theorem~\ref{thm:preservation} also holds for the language with
reentrant locks.
As corollaries we have shown that  verified programs satisfy the following properties:
null error freeness, partial correctness, and data race freeness (details in~\cite[Chap. 6]{HurlinPhd}).

Finally, we can also show that an {\tt IllegalMonitorException} cannot
occur in a verified program. Suppose we model an {\tt
  IllegalMonitorException} as a \emph{monitor error} in our
language. A head command $\hc$ is called a \emph{monitor error} iff it
tries to call {\tt wait}, {\tt notify} or {\tt notifyAll} on a lock
that is not held.  Now we can state the \emph{Monitor Error Free
  theorem} as a corollary of the preservation theorem for the language
with reentrant locks (see page 118 in~\cite[Chap. 6]{HurlinPhd}).
\begin{thm}[Verified Programs are Monitor Error Free] 
\label{theorem:monitor-error}
If $\hastype {(\cls,c)} \ok$ 
and 
$\initcmd c \ssteps \cls
	{\stt = 
    \State 
     { \tpool 
       \parpop 
       \IdThread {o} {(\BareThread {\hc;c} {s})} 
     }
     { \hp }}$
then $\hc$ is not a monitor error.
\end{thm}

\proof Similar to Theorem~\ref{theorem:null-error}, the theorem can be
proved by contradiction.  By $\hastype {\initcmd c} \ok$ and the
preservation theorem, we know that $\goodstate \stt$.  Inspecting the
derivation of the cases \ref{rule:red-wait},
\ref{rule:red-notify-one}, \ref{rule:red-notify-all} (see page 125
of~\cite[Chapter 6]{HurlinPhd}) shows that it is impossible for a
thread to invoke {\tt wait()}, {\tt notify()} or {\tt notifyAll()} if
the object is not locked. As these are the only statements that could
result in a monitor error, this concludes the proof.

\qed

\subsection{Examples of Reasoning with Reentrant Locks}
\label{sec:examples:lock}

In this section, we show examples of reasoning with reentrant locks.
We provide two examples: first we show a specification of a typical
container class, using reentrant locks, and the wait-notify mechanism;
next we discuss an advanced lock coupling example.

\paragraph{A Typical Container: class {\tt Set}}
\label{subsec:examples:lock:set}

For container classes of the Java library, lock reentrancy is crucial
to avoid duplication of method implementations, as they typically
contain methods that can be called by clients, and by other methods in
the container. We illustrate this by discussing a class {\tt Set},
containing a public method {\tt has} that is also called by other
methods in the class. Additionally, the container is developed for a
concurrent setting, using the wait-notify mechanism when a thread
tries to retrieve an object that is not in the container yet.

Class {\tt Set} contains a method {\tt has} that is used to check if
some element belongs to the receiver set. In addition, there is a
method {\tt add}, which adds an element to the receiver set if not
already present.  Moreover, it defines a method {\tt visit}, which
blocks until a particular element is present in the set.  All methods
lock the receiver set. Hence, as methods {\tt add} and {\tt visit}
call {\tt has}, reentrant locks are crucial for class {\tt Set}'s
implementation.

Internally, class {\tt Set} is backed up by a list, shown in
Figure~\ref{fig:list-reentrant-lock}. Class {\tt List} is a
\emph{shallow} container: lists do not have permission to access their
values.  Instead, values must be accessed by synchronizing on
them. That is why the {\tt state} predicate ensures that a list only
contains initialized values. Additionally, predicate {\tt state} gives
access to field {\tt next} of the list's first node and to all {\tt
  next} fields of subsequent nodes, and it provides references to the
values stored in the list.

{\small
\begin{figure}
\begin{tabbing}
~\=\tt class List extends Object\lbr \\
\>~~\=\tt Object element; List next; \\
\>~~\=\tt \\
\>\> \tt\xy \hlspec{pred  state = PointsTo(element,1,v) * PointsTo(next,1,n) * } \\
\>\>~~\=\tt\xy \phantom{pred state =} \hlspec{v.initialized * n.state;} \\
\>\>\\
\>~~\xy \hlspec{requires  init * o.initialized; ensures state@List;} \\
\>\>\tt void init(Object o, List n)\lbr~element = o; next = n; \rbr \\
\>\> \tt \\
\>~~\=\xy \hlspec{requires  state; ensures state;} \\
\>\>\tt bool has(Object o)\lbr \\
\>\>~~\=\tt bool result; \\
\>\>\>\tt if(element == o)\lbr~result = true; \rbr \\
\>\>\>\tt else\lbr~  if(next != null)\lbr~result = next.has(o); \rbr~ \rbr \\
\>\>\>\tt result; \\
\>\>\tt \rbr \\
\>\>\\
\>\>\xy \hlspec{requires  state * o.initialized; ensures state;} \\
\>\>\tt void add(Object o)\lbr~List l = new List; l.init(o,this); ~\rbr\\
\>\rbr
\end{tabbing}
\caption{Class {\tt List} } \label{fig:list-reentrant-lock}
\end{figure}
}

Figure~\ref{fig:set-reentrant-lock} presents the implementation of
class {\tt Set}, which ensures that an object cannot appear
twice in the underlying list. For simplicity, we identify two objects
if they have the same address in the heap (\ie we use Java's {\tt
  ==})\footnote{Alternatively, we could put Java's {\tt equals} in
  class {\tt Object} and use it here.}.

{\small
\begin{figure}
\begin{tabbing}
\=\tt class Set extends Object\lbr \\
\>~~\=\tt List rep;  \\
\>\>\\
\>\>\tt\xy \hlspec{pred inv = PointsTo(rep,1,r) * r.state;} \\
\>\>\\
\>\> \xy \hlspec{requires Lockset(S) * init * fresh *} \\
\>\>~~\xy~~~~~~~~~\hlspec{Set classof this * o.initialized;} \\
\>\> \xy \hlspec{ensures~ Lockset(S) * !(S contains this) * initialized;} \\
\>\> \tt void init(Object o)\lbr~ rep = new List; rep.init(o,null); \hlspec{commit};  \rbr \\
\>\>\\
\>\>\xy \hlspec{requires Lockset(S) * (S contains this -* inv) * initialized;} \\
\>\>\xy \hlspec{ensures~ Lockset(S) * (S contains this -* inv);} \\
\>\>\tt bool has(Object o)\lbr \\
\>\>~~\=\tt lock(); List result = rep.has(o); unlock(); result; \\
\>\>\tt \rbr \\
\>\> \\
\>\>\xy \hlspec{requires Lockset(S) * !(S contains this) *}\\
\>\>\xy~~~~~~~~~\hlspec{initialized * o.initialized;} \\
\>\>\xy \hlspec{ensures~ Lockset(S) * !(S contains this);} \\
\>\>\tt void add(Object o)\lbr \\
\>\>\>\tt lock(); if(!has(o))\lbr~ rep.add(o); notifyAll(); \rbr \  unlock(); \\
\>\>\tt \rbr \\
\>\>\\
\>\> \xy \hlspec{requires Lockset(S) * !(S contains this) * initialized; }\\
\>\> \xy \hlspec{ensures Lockset(S) * !(S contains this) * initialized;} \\
\>\> \tt void visit(Object o)\lbr \\
\>\>~~\=\tt lock(); while(!has(o))\lbr~  wait(); \rbr \  unlock(); \\
\>\>\tt \rbr \\
\>\rbr
\end{tabbing}
\caption{Class {\tt Set} in presence of reentrant locks} \label{fig:set-reentrant-lock}
\end{figure}
}

The resource invariant of a {\tt Set} consists of (1) the field {\tt
rep} and (2) the list pointed to by the field {\tt rep}. This is
specified in predicate {\tt inv}'s implementation.  
A {\tt Set} \emph{owns} its underlying list {\tt rep}: while the
receiver set is locked when clients call {\tt has}, {\tt add} or {\tt visit}, the
underlying list is never locked. Access rights to the underlying list
are packed into the resource invariant of the set (see {\tt inv}'s
definition). 

Elements of sets should be accessed by synchronizing on them. Although
there is no {\tt get} method in class {\tt Set}'s implementation, we
make sure that elements of sets are {\tt initialized} (see {\tt
state}'s implementation in class {\tt List} and {\tt o.initialized} in
various contracts). Hence, a {\tt get} method would have {\tt
result.initialized} as a postcondition, allowing clients to lock
returned elements. 

Method {\tt init} both (1) initializes field {\tt rep} and (2)
initializes the set's resource invariant (with the {\tt commit}
command). Point (2) is formalized by having {\tt fresh} in {\tt
init}'s precondition and having {\tt initialized} in {\tt init}'s
postcondition. In addition, {\tt init}'s precondition includes {\tt
Set classof this}. This is required to verify that {\tt commit} is
sound \ie that the monitor invariant is established before {\tt
commit}.

The contract of method {\tt has} in class {\tt Set} allows lock-reentrant
calls. If a lock-reentrant call is performed, however, {\tt inv} is
required (as expressed by {\tt (S contains this -* inv)}). Methods {\tt
add} and {\tt visit} in class {\tt Set} could be specified similarly
allowing lock reentrant calls. Notice that verification of all these methods is straightforward.

A simpler implementation of methods {\tt add} and {\tt visit} in class {\tt Set} would
call {\tt has} on the underlying list. In this way, the lock-reentrant
call would be avoided. However, our implementation is safer: if method
{\tt has} is overridden in subclasses of {\tt Set} (but not method
{\tt add} or method {\tt visit}), our implementation is still correct; while the simpler
implementation could exhibit unexpected behaviors.

Class {\tt Set} exemplifies a typical use of lock reentrancy and the
wait-notify mechanism in the Java library. We believe that our
verification system fits well to verify such classes. In addition,
this example shows how our system supports programs that include
objects that must be locked before access and objects that are
accessed without synchronization. Importantly, the addition of locks
does not force programmers to indicate {\tt Lockset} predicates
everywhere in contracts: class {\tt List}, which backs up class {\tt
  Set}, does not mention any {\tt Lockset} predicates.

Finally, suppose class {\tt Set} would be extended by a subclass {\tt
  BoundedSet} containing a field {\tt count} to keep track of the
number of elements stored in the set. The resource invariant of {\tt
  BoundedSet} would be defined as {\tt pred inv = PointsTo(count,1,\_)}, which implicitly would be conjoined with the inherited resource
invariant from {\tt Set}. Thus, locking any instance of the
{\tt BoundedSet} would provide access to the field {\tt count},
\emph{and} to the underlying list representation of the set.

\paragraph{Lock Coupling}
\label{subsec:examples:lock:lc}

Next we illustrate how our verification system handles lock coupling.
We use the following convenient abbreviations: 
\begin{displaymath}
\begin{array}{c}
\locked \specval {\specval'} \deq \Lockset {\specval\cdot \specval'}
\\
\unlocked \specval {\specval'} \deq 
  \Lockset {\specval'} \LAnd \jnot {\lpa \contains {\specval'} \specval \rpa} 
\end{array}
\end{displaymath}

Suppose we want to implement a
sorted linked list with repetitions. For simplicity, assume that the
list has only two methods: \texttt{insert()} and \texttt{size()}. The former 
inserts an integer into the list, and the latter returns the
current size of the list. To support a constant-time \texttt{size()}-method,
each node stores the size of its tail in a \texttt{count}-field. 
Each node $n$
maintains the invariant $\GGet n {\java{count}} \jdeq \GGet n {\java{next}}.{\java{count}} \jplus 1$.  

In order to allow multiple threads inserting simultaneously, we want to 
avoid using a single lock for the whole list.  
We have to be careful, though: a naive locking policy that simply 
locks one node at a time would be unsafe, because several threads trying to
simultaneously insert the same integer can cause a semantic data race, so that
some integers get lost and the \texttt{count}-fields get out of sync with the
list size. The lock coupling technique avoids this by simultaneously holding
locks of two neighboring nodes at critical times.  

Lock coupling has been used as an example by 
Gotsman \etal~\cite{GotsmanBCRS07} for
single-entrant locks. The additional problem with reentrant locks is that
\texttt{insert()}'s precondition must require that none of the list nodes is
in the lockset of the current thread. This is necessary to ensure that
on method entry the current thread is capable of acquiring all nodes's resource
invariants: 

\begin{tabbing}
\xy\hlspec{requires this.unlocked(S) * \emph{no list node is in S};}\\
\xy\hlspec{ensures~ Lockset(S);}\\
\tt void insert(int x); 
\end{tabbing}

The question is how to formally represent the informal condition written in italic.
Our solution makes use of class parameters. We require that nodes of a
lock-coupled list are \emph{statically owned} by the list object, i.e., they
have type $\java{Node}\ANGLE o$, where $o$ is the list object. Then we can
approximate the above contract as follows:  

\begin{tabbing}
\xy\hlspec{requires this.unlocked(S) * \emph{no $\This$-owned object is in S};} \\
\xy\hlspec{ensures~ Lockset(S);} \\
\tt void insert(int x); 
\end{tabbing}

To express this formally, we define a marker interface, \ie
an interface with no content, for owned objects:
\enlargethispage{\baselineskip}

\begin{tabbing}
  \tt interface Owned\hlspec{<Object owner>} \lbr\ /* a marker interface */ \rbr
\end{tabbing}

Next we define an auxiliary predicate $\OwnedUnlocked \specval {\specval'}$
(read as ``if the current thread's lockset is $\specval'$, then the aggregate
owned by object $\specval$ is traversable''). Concretely, this predicate says
that no object owned by $\specval$ is contained in $\specval'$:
\begin{displaymath}
\begin{array}{l}
\OwnedUnlocked \specval {\specval'} \Deq
\\ \qquad
\java{
\lpa 
     fa\ \Object\ owner,\, 
     Owned\ANGLE{owner}\ x
\rpa\lpa
    \jnot \lpa \contains {\meta \specval'} x \rpa
    \, \COr\, 
    owner \jneq {\meta \specval}
\rpa}
\end{array}
\end{displaymath}

Note that in our definition of $\OwnedUnlocked \specval {\specval'}$, we
quantify over a type parameter (namely the \texttt{owner}-parameter of the
\texttt{Owned}-type). Here we are taking advantage of the fact that program
logic and type system are inter-dependent.

Now, we can formally define an interface for sorted integer lists:

\begin{tabbing}
\tt interface\ SortedIntList\ \lbr 
\\[\jot]
~~\tt\xy \hlspec{pred inv<int c>;} // c is the number of list nodes 
\\[\jot]
~~\xy \hlspec{requires this.inv<c>;\ ensures this.inv<c> * result==c;}
\\
~~\tt int size(); 
\\[\jot]
~~\xy\hlspec{requires this.unlocked(S) * this.traversable(S);}\\
~~\xy\hlspec{ensures Lockset(S);} \\
~~\tt void insert(int x);
\\[\jot]
\rbr
\end{tabbing}

\begin{figure}[hbtp]{\small
\begin{tabbing}\tt
\ \ \ \ \ \=\ \ \ \ \ \=\ \ \ \ \ \=\ \ \ \ \ \= \kill \\ \tt
class LockCouplingList implements SortedIntList\lbr
\\[\jot]
\>\tt Node\hlspec{<this>} head;
\\[\jot]
\>\tt\xy \hlspec{pred inv<int c> = (ex Node<this> n)(}  \\
\>\>\tt\xy \hlspec{ PointsTo(head, 1, n) * n.initialized * PointsTo(n.count, 1/2, c)  );}
\\[\jot] 
\>\xy\hlspec{requires this.inv<c>; ensures this.inv<c> * result==c;} \\
\>\tt int size() \lbr\ return head.count;\ \rbr
\\[\jot]
\>\xy\hlspec{requires Lockset(S) * !(S contains this) * this.traversable(S);} \\
\>\xy\hlspec{ensures~ Lockset(S);} \\
\>\tt void insert(int x) \lbr \\
\>\>\tt lock(); Node\hlspec{<this>} n = head; \\
\>\>\tt if (n!=null) \lbr \\
\>\>\>\tt  n.lock(); \\
\>\>\>\tt  if (x <= n.val) \lbr \\
\>\>\>\>\tt n.unlock(); head = new Node\hlspec{<this>}(x,head); \hlspec{head.commit}; 
            unlock(); \\
\>\>\>\tt \rbr\ else \lbr\ unlock(); n.count++; n.insert(x); \rbr \\
\>\>\tt \rbr\ else \lbr\ head = new Node\hlspec{<this>}(x,null); unlock();  \rbr\ 
\rbr\ 
\rbr
\\[3ex]
\tt class Node\hlspec{<Object owner>} implements Owned\hlspec{<owner>}\lbr 
\\[\jot]
\>\tt int count; int val; Node\hlspec{<owner>} next;
\\[\jot]
\>\tt\xy \hlspec{public pred couple<int count\_this, int count\_next> =} \\
\>\>\tt\xy \hlspec{(ex Node<owner> n)(} \\
\>\>\>\>\tt\xy \hlspec{PointsTo(this.count, 1/2, count\_this) * PointsTo(this.val, 1,int)} \\ 
\>\>\>\tt\xy \hlspec{* \,PointsTo(this.next, 1, n) * n!=this * n.initialized} \\
\>\>\>\tt\xy \hlspec{* ( n!=null -* PointsTo(n.count, 1/2, count\_next) )}  \\
\>\>\>\tt\xy \hlspec{* ( n==null -* count\_this==1 ) )}; 
\\[\jot]
\>\tt\xy \hlspec{public pred inv<int c> = couple<c,c-1>;} 
\\[\jot]
\>\xy \hlspec{requires PointsTo(next.count, 1/2, c);} \\
\>\xy \hlspec{ensures PointsTo(next.count, 1/2, c)} \\
\>\tt\xy \hlspec{\phantom{en}* ( next!=null -* PointsTo(this.count, 1, c+1) )} \\
\>\tt\xy \hlspec{\phantom{en}* ( next==null -* PointsTo(this.count, 1, 1) )} \\
\>\tt\xy \hlspec{\phantom{en}* PointsTo(this.val, 1, val) * PointsTo(this.next, 1, next);}
\\ 
\>\tt Node(int val, Node\hlspec{<owner>} next) \lbr \\
\>\>\tt if (next!=null) \lbr\ this.count = next.count+1; \rbr\ else \lbr\ this.count = 1; \rbr \\
\>\>\tt this.val = val; this.next = next; 
\rbr
\\[\jot]
\>\xy \hlspec{requires Lockset(this$\,\cdot\,$S) * owner.traversable(S) * this.couple<c+1,c-1>;} \\ 
\>\xy \hlspec{ensures Lockset(S);} \\ 
\>\tt void insert(int x) \lbr \\
\>\>\tt Node\hlspec{<owner>} n = next; \\
\>\>\tt if (n!=null) \lbr \\
\>\>\>\tt n.lock(); \\
\>\>\>\tt if (x <= n.val) \lbr \\
\>\>\>\>\tt n.unlock(); next = new Node\hlspec{<owner>}(x,n); \hlspec{next.commit}; unlock(); \\
\>\>\>\tt \rbr\ else \lbr\ unlock(); n.count++; n.insert(x); \rbr \\
\>\>\tt \rbr\ else \lbr\ next = new Node\hlspec{<owner>}(x, null); unlock(); \rbr\ 
\rbr\  
\rbr
\end{tabbing}}
\caption{A lock-coupling list}
\label{fig:lock-coupling}
\end{figure}

Figure~\ref{fig:lock-coupling} shows 
a tail-recursive lock-coupling implementation of \texttt{SortedIntList}. 
The auxiliary
predicate $\Pred n {\java{couple}} {c,c'}$, as defined in the \texttt{Node}
class, holds in states where $\GGet n {\java{count}} \jdeq c$ and $\GGet
{\GGet n {\java{next}}} {\java{count}} \jdeq c'$. 
Figure~\ref{fig:lock-coupling}'s
implementation has been verified in our system.

But how can clients of lock-coupling lists establish \texttt{insert()}'s
precondition? The answer is that client code needs to track the types of 
locks held by the current thread. For instance, if $C$ is
not a subclass of \texttt{Owned}, then \texttt{list.insert()}'s precondition
is implied by the following assertion, which is satisfied when the current
thread has locked only objects of types $C$ and \texttt{Owned<$\lvar$>}.

\begin{small}
\begin{center}
\tt list.unlocked(S) * $\lvar$!=list *
\\
\tt (fa Object z)(!(S contains z) | z instanceof $C$ | z instanceof Owned<$\lvar$>)  
\end{center}
\end{small}

This example demonstrates that we can handle fine-grained concurrency
despite the technical difficulties raised by lock reentrancy
(\ie\xspace {\tt lock}'s precondition is harder to prove). However, we
have to fall back on the type system to verify this
example. Consequently, ownership becomes \emph{static}; however based
on the design decision of the data structure this is acceptable.
Usually when it is necessary that nodes can be transferred from one
container to another, all nodes have to come from a dedicated node
pool. In that case, our approach would still work, but with the node
pool as the owner.  


\section{Related Work}
\label{sec:relatedwork}

The work that is closest related to our work is Parkinson's
thesis~\cite{Parkinson05b} (recently represented
in~\cite{ParkinsonB13}). This formalizes a subset of singlethreaded
Java to specify and verify such programs with separation logic. There
are, however, a few differences: we feature value-parameterized
classes, we do not include casts (but it would be straightforward to
add them, as we did in our earlier work~\cite{HaackH08b}), we do not
model constructors, we do not provide block scoping, and, contrary to
Parkinson, programs written in our model language are not valid Java
programs. While Parkinson introduced abstract predicates and
permissions, he does not combine them as we do.  Later, both Parkinson
and Bierman~\cite{ParkinsonB08} and Chin
\etal~\cite{ChinDNQ08} provided a flexible way to handle
subclassing. 



Separation-logic-based approaches for
parallel programs~\cite{OHearn07,BerdineCO05} focused on a theoretically elegant, but unrealistic,
parallel operator. Notable exceptions are Hobor \etal~\cite{HoborAZ08} and Gotsman \etal~\cite{GotsmanBCRS07}
who studied (concurrently to us) Posix threads for C-like programs.
Contrary to us, Hobor \etal do not model join as a native method, instead they require
programmers to model join with locks. For verification purposes, this means
that Hobor \etal would need extra facilities to make reasoning about fork/join
as simple as we do. Gotsman \etal's work is very similar to Hobor \etal's work.

There are a number of similarities between our work and
Gotsman \etal~\cite{GotsmanBCRS07}'s work. For instance in the treatment of
initialization of dynamically created locks, our \texttt{initialized}
predicate corresponds to what Gotsman calls lock handles (with his lock handle
parameters corresponding to our class parameters). Since Gotsman's language
supports deallocation of locks, he scales lock handles by fractional
permissions in order to keep track of sharing. This is not necessary in a
garbage-collected language. In addition to single-entrant locks, Gotsman also
treats thread joining. We cover thread joining in a simpler and more powerful way,
because we allow multiple read-only joining.
The essential differences between Gotsman's and our
paper are \un that we treat reentrant locks, which are a different
synchronization primitive than single-entrant locks, and \deux that 
we treat subclassing and extension of resource invariants in
subclasses. Hobor \etal's work~\cite{HoborAZ08} is very similar to 
\cite{GotsmanBCRS07}.

Zhao in his thesis~\cite{Zhao} developed a permission-based type
system for a concurrent Java-like language to detect data races and
deadlocks. 
His permission system is an extension of Boyland's original permission system~\cite{Boyland03}. 
Nested permission are used to model
protected objects, while guards can be passed as class parameters. The
type syste handles reentrant locks, but without counting the
reentrancy level. Moreover, joins are not supported in his work, and the system
can only verified a fixed set of properties, \emph{i.e.}, it has no
support for user-specified contracts. 

A different approach is
taken by Vafeiadis, Parkinson \etal~\cite{VafeiadisP07,WickersonDP10},
combining rely/guarantee reasoning with separation logic. On one hand,
this is both powerful and flexible: fine-grained concurrent algorithms
can be specified and verified.  On the other hand, their verification
system is more complex than ours.  This line of research has been
extended by Dodds \etal, proposing deny-guarantee
reasoning~\cite{DoddsFPV09} to tame dynamically scoped threads. 
The idea of deny-guarantee reasoning is
to lift separation logic to assert about the possible interferences between
threads. 
Recently, Concurrent Abstract Predicates (CAP)~\cite{dinsdaleyoung10cap} have
been proposed by Dinsdale-Young \etal as a further follow-up on
deny-guarantee reasoning. 
They proposed a logic by which interferences can be asserted with actions instrumented by permissions. 
Permission-based actions can describe how a thread can treat the state.
Using CAP to specify a mutable data structure, one can
distinguish between the \emph{internal} shared states and local states
of the data structure.  
Abstract predicates in CAP encapsulate both resources and interferences
which allows one to reason about the client program without
having to deal with all the underlying interferences and resources. 
Initially, it
was not possible to use CAP to reason about synchronizer object,
because they protect \emph{external} shared resources. 
However, inspiring from~\cite{Jacobs:2011:EMF} and ~\cite{DoddsJP11} Svendsen extended CAP for
higher-order separation logic to specify library usage protocols.
The development of CAP and HOCAP seem to be an
important progress to reason about concurrent programs. However, there
is no well-developed tool support for them yet, the approach does not
consider reentrant locks, and it results in a highly-complicated
verification technique, especially when it should be applied to a
realistic programming language such as Java.

Another related line of work is by Jacobs \etal~\cite{JacobsSPS06} who
extend the Boogie methodology for reasoning about object
invariants~\cite{BarnettDFLS04} to a multithreaded Java-like
language. While their system is based on classical logic (without
operators like $\LLAnd$ and $\LLImplies$), it includes built-in
notions of ownership and access control. Their system deliberately
enforces a certain programming discipline (like concurrent separation
logic and our variant of it also do) rather than aiming for a complete
program logic. In this approach, objects can be in two states:
unshared or shared.  Unshared objects can only be accessed by the
thread that created them; while shared objects can be accessed by all
threads, provided these threads synchronize on this object.  This
partially correspond to our method: Jacobs \etal's \texttt{shared}
objects (objects that are shared between threads) directly correspond
to our
\texttt{initialized} objects (objects whose resource invariants are
initialized). While Jacobs \etal's policy is simple, it is too
restrictive: an object cannot be passed by one thread to another
thread without requiring the latter thread to synchronize on this
object.  Jacobs \etal's system prevents deadlocks, by imposing a
partial order on locks. As a consequence of their order-based deadlock
prevention, their programming discipline statically prevents
reentrancy, although it may not be too hard to relax this at the cost
of additional complexity. 

Smans \etal~\cite{SmansJPS08,SmansJP09} automatically verify sequential programs
using \emph{implicit dynamic frames}. While their approach uses first-order logic,
it is close to separation logic, because their verification algorithm
approximates the set of locations accessed by methods (like specifications in separation logic).
On the upside, Smans \etal's approach alleviates the burden
of specifying the set of locations accessed by methods, because such sets
are inferred from functional specifications. Furthermore, \un like other first-order logic
based approaches; they can use off-the-shelf theorem provers and \deux
they implemented their approach. On the downside,
solving the verification conditions generated by Smans \etal's tool is much slower
than using symbolic execution and separation logic (like~\cite{DiStefanoP08}).
Another drawback is that they cannot write
specifications that mirrors separation logic's magic wand $\LLImplies$. The
magic wand is crucial to specify data structures that temporarily ``lend''
a part of their representation to clients, like iterators~\cite{HaackH09}. 

Like Smans \etal, Leino and M\"uller~\cite{LeinoM09} presented
a verification system for multithreaded programs that uses implicit dynamic
frames and SMT solvers. Contrary to their previous work~\cite{JacobsSPS06}
they do not impose a programming model: they use
fractional permissions to handle concurrency. They do not
support multiple readonly joiner threads but they
prevent deadlock. Consequently, even if they do not handle reentrant locks,
these locks could be handled without a major effort.
 
Finally, in a more traditional approach, De Boer~\cite{Boer07} extends
the results of Abraham \etal~\cite{AbrahamBRS02} with a sound and
complete proof system based on the Owicki/Gries method, to generate
interference freedom tests for dynamically created threads in Java.
In his approach, interferences between threads are annotated as
\emph{global} assertions and \emph{local} properties are proved in
sequential Hoare logic.  Java's {\tt synchronized} methods are
considered as the programs' synchronization mechanisms and static
auxiliary variables are defined to control the owner and reentrancy
level of the lock.  While this work covers dynamic thread creation, it
lacks support for \emph{reentrant locks} and object-oriented features
of Java.

\section{Conclusion}
\label{sec:conclusion}

In this paper, we have presented a variant of permission-based
separation logic that allows reasoning about object-oriented
concurrent programs with dynamic threads and reentrant locks. The
main selling point of this logic is that it combines several
existing specification techniques, \emph{and} that it is not developed for
an idealized programming language. Together this makes it powerful and
practical enough to reason about real-life concurrent Java programs,
as has been demonstrated on several examples, both in a sequential and
in a concurrent setting. 

An essential ingredient of the logic is the use of permissions. These
ensure that in a verified program, data races cannot occur, while
shared readings are allowed. Thus concurrent execution of
the program is restricted as little as possible. Further, the logic
also contains abstract predicates, as proposed by Parkinson, which are
suitable to reason about inheritance, and class parameters. This paper
is the first to combine these three different features in a single
specification language for a realistic programming language.

Currently, a tool~\cite{vct-www, AmighiBHZ12} is being developed for
this logic in the context of the VerCors
project\footnote{http://fmt.cs.utwente.nl/research/projects/VerCors/}. Throughout,
the tool is developed with practical usability in mind: eventually it
should provide sufficient support for a programmer to prove
correctness of his or her applications.

To ensure practical usability involves several topics: \un improving
readability of the specification language, for example by merging it
with an an existing specification language such as
JML~\cite{LeavensPCCRCMKC07}; \deux development of appropriate proof
theories to automatically discharge proof obligations; and \trois
development of techniques to reason about the absence of aliasing in
the context of lock-reentrancy. The first topic has also been
investigated both by Tuerk~\cite{Tuerk09} and Smans
\etal~\cite{SmansJP10}, while the second topic has been investigated
by Parkinson \etal~\cite{DiStefanoP08}. However, in both cases the
results have to be further extended to fit in our framework, in
particular because they do not consider the magic wand.  Concerning
the third topic, the lock-coupling example
(Section~\ref{subsec:examples:lock:lc}), uses class parameters to
model ownership. We will investigate how this can be done more
systematically. At present, (simplified versions of) the examples in this
paper can be verified by the tool.

Further we are also extending the application domain of the
logic, to be able to reason about a larger class of concurrent Java
programs, and to verify also functional properties of these
applications. We mention in particular the following recent results
and plans for future work:
\begin{itemize}
\item We specified the
\texttt{BlockingQueue} hierarchy from the
\texttt{java.util.concurrent} library using a history-based
specification~\cite{abs-1209-2239}.  The specifications can be used to
derive funcional properties about queues, for example to show that in
a concurrent environment the order of elements is always preserved.
\item 
We also developed formal specifications for several 
synchronization classes, such as the (reentrant and
read-write) locks, semaphores and latches from the Java API.  
\item We are developing techniques to reason about functional
  properties that have to hold throughout an execution, so-called
  \emph{strong invariants}.
\item We are formally specify classes from \texttt{atomic} package
  from the Java API to support reasoning about lock-free data
  structures. As a first step we have
specified \texttt{AtomicInteger} as a primitive synchronizer and
proved the correctness of several synchronization patterns on top of
this. 
\item We also plan to investigate whether permission
annotations can be generated, instead of being written by the
programmer.  
\item We have been adapting the current logic to reason about GPU
  kernels~\cite{HuismanM13}.
\end{itemize}

\section*{Acknowledgments}
We thank Ronald Burgman for working out a first version of the
specification of the sequential and parallel mergesort algorithms.

\bibliographystyle{alpha}
\bibliography{bibli}

\appendix


\section{Auxiliary Definitions}
\label{sec:add-defs}
\label{chap:ad}

\subsection{Definitions of lookup functions}
\label{sec:lookup}

\begin{display}
{Field Lookup, $\fields {\TyApp C \specvals} = \Fld \Ts \fs$:}
\staterule{(Fields Base)}
  { }
  {\fields \Object = \emptyset}
\quad
\staterulecond{(Fields Ind)}
  {
    \fields {\TyApp D {\subst \perms \alphas {\perms'}}} 
    = 
    \Fld {\Ts'} {\fs'}
  }
  {
     \CClass {C} \Ts \alphas {\TyApp D {\perms'}} \Us
     {\Fld \Ts \fs\ \pds\ \axs\ \mds}
  }
  {
    \fields {\TyApp C \perms} 
    = 
    \subst \perms \alphas {(\Fld \Ts \fs)},\,\Fld {\Ts'} {\fs'}
  }
\label{rule:fields-ind-lock}
\end{display}

\begin{display}{Axiom Lookup, $\axioms {\TyApp t \specvals} = F$:}
\clause{
\begin{array}{rcl}
\alkup \axs & \deq & 
  \left\{\begin{array}{l@{\quad\mbox{if }}l}
  \True & \axs = ()
  \\
  F \LAnd \alkup {\SEQ {\ax}} & \axs = (\AX F, \SEQ {\ax})
  \end{array}\right.
\\[2ex]
\alkup \Ts & \deq & 
  \left\{\begin{array}{l@{\quad\mbox{if }}l}
  \True & \Ts = () \mbox{ or } \Ts = (\Object)
  \\
  \alkup U \LAnd \alkup {\Vs} & \Ts = (U,\Vs)
  \end{array}\right.
\end{array}
}
\\[\GAP]
\staterule{(Ax Class)}
  { 
    \CClass {C} \Ts \logvars U \Vs 
       {\fds\;\pds\;\axs\;\mds}
  }
  { 
    \alkup {\TyApp C \specvals} = 
    \alkup {\subst \specvals \logvars \axs} 
    \LAnd\, 
    \alkup {\subst \specvals \logvars {(U,\Vs)}}
  }
\\[\GAP]
\staterule{(Ax Interface)}
  { 
   \Interface I \Ts \logvars \Us 
      {
     \ptys\;\axs\;\mts
      }
  } 
  { 
    \alkup {\TyApp I \specvals}= 
    \alkup {\subst \specvals \logvars \axs} 
    \LAnd\, 
    \alkup {\subst \specvals \logvars {\Us}}
  }
\end{display}


\noindent Remarks on method lookup (defined below):
\begin{itemize}
\item
  In $\kw{mbody}$ and $\kw{mtype}$, 
  we replace the implicit self-parameter $\This$ 
  by an explicit method parameter
  (separated from the other method parameters by a semicolon).
  This is technically convenient for the theory.
\end{itemize}

\begin{display}{Method Lookup,
$\mtype m {\TyApp t \perms} = \mt$
and $\mbody m {\TyApp C \perms} = \Mbody \logvars \rvars c$:}
\staterule{(Mlkup \texttt{Object})}
   { 
      \java{class\ Object\ } 
      \lbr 
        \dots\, \MD \Ts \alphas \mspec U m \Vs \rvars c\, \dots
      \rbr
   }
   {
     \kw{mlkup}(m,\Object) = 
     \MD \Ts \alphas \mspec U m \Vs \rvars c

   }
\label{rule:mlkup-object}
\\[\GAP]
\staterule{(Mlkup Defn)}
  { 
    \CClass {C} {\Ts'} {\alphas'} {U'} {\Vs'}
       {\dots\, \MD \Ts \alphas \mspec U m \Vs \rvars c\, \dots }
  }
  { 
    \mlkup {m, \TyApp C {\perms}} = 
    \subst \perms {\alphas'} 
                  {(\MD \Ts \alphas \mspec U m \Vs \rvars c)}
  }  
\\[\GAP]
\staterulecond{(Mlkup Inherit)}
  { m \not\in\dom\mds }
  { 
    \CClass {C} \Ts \alphas 
            {\TyApp {D} {\perms'}} \Us {\fds\;\pds\;\mds}
    \quad
    \mlkup  {m, \TyApp D {\subst \perms \alphas {\perms'}}} = \md'
  }
  { \mlkup  {m, \TyApp C \perms} = \md' }
\\[\GAP]
If $\mlkup {m, \TyApp C \perms} = 
    \MD \Ts \logvars {\Mspec F G} U m \Vs \rvars c$, then:
\\ \quad
\clause{\begin{array}{rcl}
  \mbody m {\TyApp C \perms}  
  & \deq &
  \Mbody \logvars {\This;\rvars} c 
  \\
  \mtype m {\TyApp C \perms}  
  & \deq &
  \MMMT \Ts \alphas {\Mspec F { \lpa \result \rpa }} 
        U m {\TyApp C \perms} {\,\This}  \Vs \rvars
\end{array}
}
\\[\GAP]
\staterule{(Mtype Interface)}
  { 
   \Interface I \Ts \logvars \Us 
      { 
        \dots\,
	\MMT {\Ts'} {\logvars'} {\Mspec F G} {U'} m {\Vs'} {\rvars};
	\,\dots
      }
  } 
  { 
    \mtype m {\TyApp I \specvals} = 
    \subst \specvals \logvars 
    {(\SMT {\TypedVar {\Ts'} {\logvars'}}
           F {\lpa \result \rpa } {U'} m 
           {\TypedVar {\TyApp I \specvals} \This;\,
	    \TypedVar {\Vs'} \rvars})}
  }
\label{rule:mtype-interface}
\\[\GAP]
\staterulecond{(Mtype Interface Inherit)}
  { 
       \Interface I \Ts \logvars {\Us,V,\Us'} {\ptys\;\axs\;\mts}
  }
  { 
     m \not\in\dom\mts 
     \quad 
     (\forall U \in \Us,\Us')
     (\mtype m {\subst \specvals \logvars U} = \kw{undef})
     \quad
     \mtype m {\subst \specvals \logvars V} = \mt
  } 
  { 
    \mtype m {\TyApp I \specvals} = \mt
  }
\label{rule:mtype-interface-inherit}
\\[\GAP]
\staterulecond{(Mtype Interface Inherit \texttt{Object})}
  { 
       \Interface I \Ts \logvars {\Us} {\ptys\;\axs\;\mts}
  }
  { 
     m \not\in\dom\mts 
     \quad
     (\forall U \in \Us)
     (\mtype m {\subst \specvals \logvars U} = \kw{undef})
     \quad
     \kw{mtype}(m,\Object) = \mt
  } 
  { 
    \mtype m {\TyApp I \specvals} = \mt
  }
\label{rule:mtype-interface-object}
\end{display}


\noindent Remarks on predicate lookup:
\begin{itemize}
\item
  The ``$\kw{ext}\ \Object$'' in 
  $\plkup {\initpred,\Object}$ and~\ref{rule:plkup-object}
  is included to match the format of the relation.
  There is nothing more to this.
\item
  Each class implicitly defines the \texttt{init}-predicate,
  which gives write permission to all fields of the class frame.
  In \ref{rule:plkup-init}, $\df T$ is the default value of type $T$ ($\kw{df}$ is
  formally defined in Section~\ref{subsec:semantics:jll}).
\end{itemize}

\begin{display}{Predicate Lookup, $\ptype P t \specvals = \spty$
and $\pbody {\GGet \specval P} {\specvals'} C {\specvals''} = F\;\kw{ext}\;
    T$:}
\clause{
  \plkup{\initpred,\Object} =
  \NewPDNoArg {} \initpred \True \;\kw{ext}\;\Object
}
\\[\GAP]
\staterule{(Plkup \texttt{Object})}
   { 
      \java{class\ Object\ } 
      \lbr \dots\,\NewPD P {\Ts} {\logvars} F;\,\dots \rbr
   }
   {
     \kw{plkup}(P,\Object) = 
     \NewPD P {\Ts} {\logvars} F \;\kw{ext}\;\Object
   }
\label{rule:plkup-object}
\\[\GAP]
\staterule{(Plkup Defn)}
  {
    \CClass {C} {\Ts'} {\logvars'} U \Vs
       {\dots\,\NewPD P {\Ts} {\logvars} F; \,\dots}
  }
  {
    \plkup {P,\TyApp C \specvals} =
    \subst \specvals {\logvars'}
    {(\NewPD P {\Ts} {\logvars} F \;\kw{ext}\;\Object)}
  }
\label{rule:plkup-defn}
\\[\GAP]
\staterule{(Plkup $\initpred$)}
  {
    \CClass {C} {\Ts'} {\logvars'} U \Vs
       {\fds\;\pds\;\mds}
    \quad
     F = \varoast_{\Fld T f \in \fds} \PointsTo \This f \one {\df T}
  }
  {
    \plkup {\initpred,\TyApp C {\specvals}}
    =
    \subst \specvals {\logvars'}
    { ( \PDNoArg {} \initpred F \;\kw{ext}\;U \,) }
  }
\label{rule:plkup-init}
\\[\GAP]
\staterulecond{(Plkup Inherit)}
  {
    P \not \in \dom \pds 
  }
  {
    \CClass {C} {\Ts'} {\logvars'} U \Vs
       {\fds\;\pds\;\mds}
    \quad
    \plkup{P,U} = \NewPD P {\Ts} {\logvars} F \;\kw{ext}\;U'
  }
  {
    \plkup {P, \TyApp C {\specvals}} =
      \subst \specvals {\logvars'}
        {(\NewPD P {\Ts} {\logvars} \True \;\kw{ext}\;U)}
  }
\label{rule:plkup-inherit}
\\[\GAP]
If 
$\plkup {P, \TyApp C \specvals} =
 \NewPD P \Ts \logvars F \kw{ext}\;V$,
then:
\\
\clause{\quad
\begin{array}{rcl}
  \ppbody {\GGet \specval P} {\specvals'} {\TyApp C \specvals}
  & \deq &
  {(F \;\kw{ext}\;V)}[\specval/\This,\specvals'/\logvars]
\\
  \ptype P C \specvals
  & \deq &
  \PTNew P \Ts \logvars
\end{array}
}
\\[\GAP]
\staterule{(Ptype Interface)}
  {
   \Interface I \Ts \logvars \Us
      {\dots\,\PTNew P {\Ts'} {\logvars'};\,\dots}
  }
  {
    \ptype P I \specvals =
    \subst \specvals \logvars {(\PTNew P {\Ts'}  {\logvars'})}
  }
\\[\GAP]
\staterulecond{(Ptype Interface Inherit)}
  { 
       \Interface I \Ts \logvars {\Us,V,\Us'} {\ptys\;\axs\;\mts}
  }
  { 
     P \not\in\dom\pts 
     \quad 
     (\forall U \in \Us,\Us')
     (\kw{ptype}(P, \subst \specvals \logvars U) = \kw{undef})
     \quad
     \kw{ptype}(P, \subst \specvals \logvars V) = \spty
  } 
  { 
    \ptype P I \specvals = \spty
  }
\label{rule:ptype-interface-inherit}
\\[\GAP]
\staterulecond{(Ptype Interface Inherit \texttt{Object})}
  { 
       \Interface I \Ts \logvars {\Us} {\ptys\;\axs\;\mts}
  }
  { 
     P \not\in\dom\ptys 
     \quad
     (\forall U \in \Us)
     (\kw{ptype}(P, \subst \specvals \logvars U) = \kw{undef})
     \quad
     \kw{ptype}(P,\Object) = \spty
  } 
  { 
    \ptype P I \specvals = \spty
  }
\label{rule:ptype-interface-inherit-object}
\end{display}

\noindent The partial function $\ptype P t \specvals$ is extended to 
predicate selectors $\PAt P C$ as follows:
\begin{displaymath}
\ptype {\PAt P C} t \specvals \Deq
\left\{\begin{array}{l@{\quad}l}
\ptype P t \specvals & \mbox{if } t = C 
\\
\undef & \mbox{otherwise}
\end{array}\right.
\end{displaymath}

\subsection{Semantics of operators}
\label{sec:sem:op}
To define the semantics
of the command assigning the result of an operation (case $\HdOp \lvar \op \vs$ of our
command language), we define the semantics of operators. 

Let $\arityFun$ be a function that assigns to each operator its arity.
We define: 
\begin{displaymath}
\begin{array}{l l l}
  \arity {\tjkw{==}} \deq 2 & \arity{\tjkw{\&}} \deq 2 & \arity{\tjkw{|}} \deq 2
  \\
  \arity {\tjkw{!}} \deq 1 & \arity {\isclassof C} \deq 1 & \arity {\iinstanceof T} \deq 1
\end{array} 
\end{displaymath}

Let $\optypeFun$ be a function that maps each operator $\op$ to a 
partial function $\optype\op$ of type 
$\ParFun {\set{\IntTy,\BoolTy,\Object,\PermTy}^{\arity\op}}
  {\set{\IntTy,\BoolTy,\PermTy}}$.
We define: 
\begin{displaymath}
\begin{array}{l l}
  \multicolumn{2}{l}{
    \optype {\tjkw{==}} \deq \setcomp {\ ((T,T),\BoolTy)\ } 
    {\ T \in \set{\IntTy,\BoolTy,\Object,\PermTy,\LocksetTy}\ }
  }
  \\
  \optype {\tjkw{!}} \deq \set{\ (\BoolTy,\BoolTy)\ }
  &
  \quad
  \optype {\tjkw{\&}} \deq 
  \optype {\tjkw{|}} \deq \set{\ ((\BoolTy,\BoolTy),\BoolTy)\ }
  \\
  \multicolumn{2}{c}{
	  \optype {\isclassof C}  \deq  \set {\ (\Object,\BoolTy)\ } 
  }
  \\
  \multicolumn{2}{c}{
	  \optype {\iinstanceof T}  \deq  \set {\ (\Object,\BoolTy)\ }
  }
\end{array}
\end{displaymath}

We assume that each operator $\op$ is interpreted by a function of the
following type: 
%
\begin{displaymath}
\sem \op \ \ \in\ \ 
\Fun \HeapSet {\bigcup_{(\Ts,U)\,\in\,\optype\op} \Fun {\sem \Ts} {\sem U}}
\end{displaymath}

For the logical operators \texttt{!}, \texttt{|} and
\texttt{\&}, we assume the usual interpretations. 
Operator $\texttt{==}$ is interpreted as the identity relation.
The semantics of {\tt isclassof} and {\tt instanceof} is as follows:
\begin{displaymath}
  \begin{array}{r c l}
    \sem {\isclassof C} ^h (o) & \deq &
      \left\{
        \begin{array}{l l}
	  \True & \text{if $o \neq \Null$ and $\Fst {h(o)} = \TyApp C \specvals$ for some $\specvals$} \\
	  \False & \text{if $o \neq \Null$, $\Fst {h(o)} = \TyApp D \specvals$, and $D \neq C$} \\
	  \False & \text{if $o = \Null$} \\
	  \kw{undef} & o \not\in\dom h \\
	\end{array}
      \right.
  \end{array}
\end{displaymath}
\begin{displaymath}
  \begin{array}{r c l}
    \sem {o\;{\tt instanceof}\;T}^h & \deq &
      \left\{
        \begin{array}{l l}
	  \True & \text{if $o \neq \Null$ and $\Fst {h(o)} \subty T$} \\
	  \False & \text{if $o \neq \Null$ and $\Fst {h(o)} \not\subty T$} \\
	  \False & \text{if $o = \Null$} \\
	  \kw{undef} & \text{if $o \not\in\dom h$} 
        \end{array}
      \right.
  \end{array}
\end{displaymath}

\noindent Formally, the semantics of operators is expressed as follows:

\begin{display}{Semantics of Operators: $\hastype {\sem {\op(\vs)}} {\Fun \HeapSet {\ParFun \StoreSet \ClValSet}}$:}
\staterule{(Sem Op)}
  {
    \esem {w_1} h s =  v_1
    \quad \cdots\quad 
    \esem {w_n} h s = v_n
    \quad
    \osem \op h {v_1,\dots,v_n} =  v
  }
  {
   \esem {\op(w_1,\dots,w_n)} h s =  v
  }
\label{rule:sem-exp-op(b)}
\end{display}

\subsection{Semantics of specification values and expressions.}
\label{sec:sem:specs}

Expressions contain specification values, read-write variables, and
operators. Therefore, we give the semantics of specification values
and the semantics of expressions together.  Let $\SemValSet$ be the
semantic domain of specification values.  For the moment, $\SemValSet$
is simply $\ClValSet$; but it is extended in
Sec.~\ref{subsubsec:sl:semantics} as we extend specification
values. We range over $\SemValSet$ with meta-variable $\semval$.

{
\begin{display}{{\footnotesize Semantics of Specification Values and Expressions}, 
$\hastype {\sem e} {\Fun \HeapSet {\ParFun \StoreSet \SemValSet}}$:}
\staterule{(Sem SpecVal)}
  { \sem \specval = \semval }
  {\esem { \specval} h s =  \semval}
\label{rule:sem-exp-val}
\;
\staterule{(Sem Var)}
  {s(\lvar) = v }
  {\esem \lvar h s = v}
\label{rule:sem-exp-var}
\;
\staterule{(Sem Op)}
  {
    \esem {w_1} h s =  v_1
    \;\cdots\;
    \esem {w_n} h s = v_n
    \quad
    \osem \op h {v_1,\dots,v_n} =  v
  }
  {
   \esem {\op(w_1,\dots,w_n)} h s =  v
  }
\label{rule:sem-exp-op}
\end{display}
}

{Note that, we do not have to define a semantics of logical variables
$\alpha$, because we deal with them by substitution.}

\subsection{Small-step reduction}
\label{subsec:apx:reductions}

The state reduction relation $\rightarrow_\cls$ is given with respect
to a class table $\cls$ in Section~\ref{sec:smallstep-reduction}. In the reduction rules, we use the following
abbreviation for field updates: $ \updtheap h o f v = \updtfun h o
{(\Fst{h(o)},\updtfun {\Snd{h(o)}} f v)} $.

\begin{display}{State Reductions, $\stt \sstep \cls \stt'$:}
\RuleRedDcl {\stateaxiomcond{(Red Dcl)}} p c T
\label{rule:red-dcl}
\\[\jot]
\RuleRedFinDcl {\stateaxiomcond{(Red Fin Dcl)}} p c {c'}
\label{rule:red-fin-dcl}
\\[\jot]
\RuleRedVarSet {\stateaxiomcond{(Red Var Set)}} p c v
\label{rule:red-var-set}
\\[\jot]
\RuleRedOp {\stateaxiomcond{(Red Op)}} p c
\label{rule:red-op}
\\[\jot]
\RuleRedGet {\stateaxiomcond{(Red Get)}} p c o
\label{rule:red-get}
\\[\jot]
\RuleRedSet{\stateaxiomcond{(Red Set)}} p c o v
\label{rule:red-set}
\\[\jot]
\RuleRedNew {\stateaxiomcond{(Red New)}} p \cmd o
\label{rule:red-new}
\\[\jot]
\RuleRedCallSeq{\stateaxiomcond{(Red Call)}} p o c {c_m} {c'}
\label{rule:red-call}
\\[\jot]
\RuleRedReturn {\stateaxiom{(Red Return)}} p \cmd
\label{rule:red-return}
\\[\jot]
\RuleRedIfTrue {\stateaxiom{(Red If True)}} p \cmd {\cmd'} {\cmd''}
\label{rule:red-if-true}
\\[\jot]
\RuleRedIfFalse {\stateaxiom{(Red If False)}} p \cmd {\cmd'} {\cmd''}
\label{rule:red-if-false}
\\[\jot]
{
\RuleRedWhileTrue {\stateaxiomcond{(Red While True)}} p e \cmd {\cmd'} 
\label{rule:red-while-true}
}
\\[\jot]
{
\RuleRedWhileFalse {\stateaxiomcond{(Red While False)}} p e \cmd {\cmd'} 
\label{rule:red-while-false}
}
\label{idx:op:sem}
\end{display}

\subsection{Typing rules} Here we define typing rules needed for Section~\ref{subsec:semantics:jll},  Section~\ref{sec:forkjoin} and Section~\ref{sec:locks}.
\label{subsec:apx:types}
\label{subsubsec:types}

\paragraph{Rules for Section~\ref{subsec:semantics:jll}}

Because the semantics of formulas depends on a typing judgment,
we need to define typing rules before giving the formulas' semantics.

A \emph{type environment} is a partial function of type
$\ParFun {\ObjIdSet \cup \VarSet} \TySet$. We use the meta-variable $\Gamma$
to range over type environments. $\hpenv \Gamma$ denotes the \emph{restriction
  of $\Gamma$ to $\ObjIdSet$}: 
\begin{displaymath}
\hpenv\Gamma \Deq \setcomp {\ (o,T) \in \Gamma\ } 
                           {\ o\in \ObjIdSet\ }
\end{displaymath}
\label{idx:gamma}

A type environment is \emph{good} when objects within its domain are well-typed:

\begin{display}{Good Environments, $\goodenv \Gamma$:}
\staterule{(Env)}
  {
   (\forall x \in \dom\Gamma)(\goodtype \Gamma {\Gamma(x)})
   \quad
    (\forall o \in \dom\Gamma)(
      \Gamma(o) \subty \Object
      \mbox{ and }
      \goodtype {\hpenv\Gamma} {\Gamma(o)}
    )
  }
  { \goodenv \Gamma }
\label{rule:env}
\end{display}

We define a sanity condition on types: primitive types are always sane,
while user-defined types must be such that (1) type identifiers are in the class table and
(2) type parameters are well-typed. Below, the existential quantification
in~\ref{rule:ty:ref}'s second premise enforces typing derivations to be finite.

\begin{display}{Good Types, $\goodtype \Gamma T$:}
\staterule{(Ty Primitive)}
  {
    T \in \set {\Void,\IntTy,\BoolTy,\PermTy}
  }
  {\goodtype \Gamma T}
\qquad
\staterulecond{(Ty Ref)}
  {
   \TyApp t {\TypedVar \Ts \logvars} \in \cls
  }
  {
   (\exists \Gamma' \subset \Gamma)(
   \goodenv {\Gamma'}
   \quad
   \Gamma' \vdash \hastype \specvals {\subst \specvals \logvars \Ts}) 
  }
  {
   \goodtype \Gamma {\TyApp t \specvals} 
  }
\label{rule:ty:ref}
\end{display}

%
%
We define a \emph{heap extension order} 
on well-formed type environments:
\begin{displaymath}
\Gamma' \hpsup \Gamma
\quad\mbox{iff}\quad
\goodenv {\Gamma'},\ 
\goodenv {\Gamma},\
\Gamma' \supseteq \Gamma
\mbox{ and } \restrict {\Gamma'} \VarSet = \restrict \Gamma \VarSet 
\end{displaymath}

As models of formulas are tuples that contain a heap and a stack (see Section~\ref{subsubsec:sl:augmented-heaps}),
we define a well-typedness judgment for objects, heaps, and stacks:

\begin{display}{Well-typed Objects, $\goodobj \Gamma \obj$:}
\staterulecond{(Obj)}
   { \dom\ostr \subseteq \dom{\fields {\TyApp C \perms}}  }
   {
     \goodtype \Gamma {\TyApp C \perms}
     \quad
     (\forall f \in \dom\ostr)
     (\Fld T f \in \fields {\TyApp C \perms}
      \ \Rightarrow\ 
      \Gamma \vdash \hastype {\ostr(f)} T)
   }
   { \goodobj \Gamma {(\TyApp C \perms,\ostr)}}
\label{rule:obj}
\end{display}

Note~~~that~~~we~~~~require~~~~$\dom\ostr~~\subseteq~~\dom{\fields {\TyApp C \perms}}$,~~~~not~~~~$\dom\ostr~~~=\\
\dom{\fields {\TyApp C \perms}}$.
Thus, we allow partial objects.
This is needed, because $\LLAnd$ joins heaps on a per-field basis.

Below, we use function $\hastype {\kw{fst}} {\Fun \HeapSet {(\ParFun \ObjIdSet \TySet)}}$ to extract the function
that maps object identifiers to their dynamic types from a heap:
\begin{displaymath}
  \hp(o) = \Cpl T \whatever \Rightarrow \kw{fst}(\hp)(o) = T
  \label{idx:fst}
\end{displaymath}

We now define well-typed heaps and stacks:

\begin{display}{Well-typed Heaps and Stacks, $\goodheap \Gamma h$ and
   $\goodstore \Gamma s$ :}
\staterule{(Heap)}
   {
    \goodenv \Gamma
    \quad 
    \Gamma \subseteq \kw{fst}(h)
    \quad
    (\forall o \in \dom h)(\goodobj \Gamma {h(o)})
   }
   {\goodheap \Gamma h}
\label{rule:heap}
\\[\GAP] 
\staterule{(Stack)}
   {
     \goodenv \Gamma
     \quad
     (\forall x \in \dom s) ( \Gamma \vdash \hastype {s(x)} {\Gamma(x)})
   }
   {\goodstore \Gamma s}
\label{rule:store}
\end{display}

Because formulas include expressions, we define a well-typedness
judgment for values, specification values, and expressions (recall that expressions
include specification values of type {\tt bool}).
 
\begin{display}{Well-typed Values and Specification Values, 
    $\Gamma \vdash \hastype v T$ and
    $\Gamma \vdash \hastype \specval T$:}
\staterule{(Val Var)}
  {
    \goodenv \Gamma
    \quad
    \Gamma(x) = T
  } 
  {
    \Gamma \vdash \hastype x T
  }
  \label{rule:val-var}

\staterule{(Val Oid)}
  {
    \goodenv \Gamma
    \quad
    \Gamma(o) = T
  } 
  {
    \Gamma \vdash \hastype o T
  }
  \label{rule:val-oid}

  \staterule{(Val Sub)}
  {\Gamma \vdash \hastype \specval T 
    \quad 
    T \subty U}
  {\Gamma \vdash \hastype \specval U}   
  \label{rule:val-exp-sub}

\staterule{(Val Null)}
  {\goodtype \Gamma {\TyApp t \specvals}}
  {\Gamma \vdash \hastype \Null {\TyApp t \specvals}}
  \label{rule:val-null}
\\[\GAP]
\staterule{(Val Int)}
  { \goodenv \Gamma }
  {\Gamma \vdash \hastype \enn \IntTy}
  \label{rule:val-int}
\quad
  \staterule{(Val Bool)}
  { \goodenv \Gamma }
  {\Gamma \vdash \hastype \bee \BoolTy}
  \label{rule:val-bool}
\quad
\staterule{(Val Full)}
  {
    \goodenv \Gamma
  }
  { 
    \Gamma \vdash \hastype \full \PermTy
  }
  \label{rule:val:full}
\quad 
  \staterule{(Val Split)}
  {
    \Gamma \vdash \hastype \specval \PermTy
  }
  {
    \Gamma \vdash \hastype {\Split \specval} \PermTy
  } 
  \label{rule:val:split}
\end{display}

\newcommand{\ClSpecValSet}{\kw{ClSpecValSet}}

\begin{display}{Well-typed Expressions, $\Gamma \vdash \hastype e T$:} 
  \staterule{(Exp Sub)}
  {\Gamma \vdash \hastype e T 
    \quad 
    T \subty U}
  {\Gamma \vdash \hastype e U}   
\qquad
  \staterule{(Exp Var)}
  {
    \goodenv \Gamma
    \quad
    \Gamma(\lvar) = T
  }
  {
    \Gamma \vdash \hastype \lvar T
  }
\qquad
\staterule{(Exp Op)} 
   {
    \Gamma \vdash \hastype {\es} {\Us}
    \quad
    \optype \op (\Us) = T
   }
   {\Gamma \vdash \hastype {\op(\es)} T}
\end{display}

We now have all the machinery to define well-typed formulas.
Below, the partial function $\ptype P C
\specvals$ (formally defined in~\ref{sec:add-defs})
looks up the type of predicate $P$ in the least supertype of
$\TyApp C \specvals$ that defines or extends $P$. 

\begin{display}{Well-typed Formulas, $\goodform \Gamma F$:}
  \staterule{(Form Bool)}
  {\Gamma \vdash \hastype e \BoolTy}
  {\goodform \Gamma e}
  \label{rule:good-bool}
  \quad
  \staterule{(Form Points To)}
  { 
    \Gamma \vdash \hastype {e} U
    \quad
    \Gamma \vdash \hastype {\specval} {\PermTy}
    \quad
    \Fld T f \in \fields {U} 
    \quad
    \Gamma \vdash \hastype {e'} T
  }
  { 
    \goodform \Gamma {\PointsTo e f {\specval} {e'} } 
  }
  \label{rule:form-points-to}
  \\[\GAP]
  \staterule{(Form Log Op)}
  {
    \goodform \Gamma {F,F'}
  }
  {
    \goodform \Gamma {F \;\bop\; F'}
  }
  \label{rule:form:op}
  \qquad
  \staterule{(Form Pred)}
  {
    \Gamma \vdash \hastype \specval U
    \quad
    \pptype \kappa U = \PTNew P \Ts \logvars
    \quad
    \Gamma \vdash \hastype {\specvals'} \Ts
  }
  {
    \goodform \Gamma {\Pred \specval \kappa {\specvals'}}
  }
  \label{rule:form:pred}
  \\[\GAP]
  \staterule{(Form Quant)}
  {
    \goodtype \Gamma T
    \quad
    \goodform {\Gamma,\hastype \logvar T} F
  }
  {
    \goodform \Gamma {\Qt \logvar T F}
  }
\end{display}

\paragraph{Rules for Section~\ref{sec:forkjoin}}

To cover Section~\ref{sec:forkjoin}'s {\tt Join} formula,
we extend 
the judgment for well-typed formulas as follows:

\begin{display}{Well-typed Formulas, $\goodform \Gamma F$:}
  \dots
  \quad
  \staterule{(Form Join)}
  { 
    \Gamma \vdash \hastype {e} \ThreadTy
    \quad
    \Gamma \vdash \hastype {\specval} {\PermTy}
  }
  { 
    \goodform \Gamma {\Join e \specval} 
  }
  \label{rule:form-join}
  \quad
  \dots
\end{display}

\paragraph{Rules for Section~\ref{sec:locks}}

To accommodate Section~\ref{sec:sl:locks}'s $\LocksetTy$'s type, we
update the previous typing rule for good types:

\begin{display}{Good Types, $\goodtype \Gamma T$:}
  \dots
  \quad
\staterule{(Ty Primitive)}
  {
    T \in \set {\Void,\IntTy,\BoolTy,\PermTy,\changed{\LocksetTy}}
  }
  {\goodtype \Gamma T}
  \quad
  \dots
\end{display}

The following typing rule extends typing to values representing locksets: 
\begin{displaymath}
\recallmath
  { \semval \in \Bag \ObjIdSet }
  { \Gamma \vdash \hastype \semval \LocksetTy }
\end{displaymath}

To cover formulas about locksets and the state of locks,
we extend the 
judgment for well-typed formulas:
\begin{display}{Well-typed Formulas, $\goodform \Gamma F$:}
  \dots
  \quad
  \staterule{(Form Lockset)}
    { \Gamma \vdash \hastype \specval \LocksetTy}
    { \goodform \Gamma {\Lockset \specval} }
  \quad 
  \staterule{(Form Contains)}
  {
    \Gamma \vdash \hastype {\specval,e} {\LocksetTy, \Object}
  }
  {
    \goodform \Gamma {\contains \specval {e}}
  }
  \quad
  \staterule{(Form Fresh)}
    { \Gamma \vdash \hastype e \Object }
    { \goodform \Gamma {\Fresh e}}
  \label{rule:form-fresh}    
  \\[\GAP]
  \staterule{(Form Initialized)}
    { \Gamma \vdash \hastype e \Object }
    { \goodform \Gamma {\Initialized e}}
  \label{rule:form-initialized}    
  \quad
  \dots
\end{display}

\subsection{Verification} 
\label{subsec:apx:verifications}
In this section we present a complete list of natural deduction rules (Section~\ref{subsec:natrualdeductionrules}) and Hoare rules (Section~\ref{subsec:sl:hoare}). 
In addition, the formal definitions of the sanity conditions required for the verification of the interfaces and classes (see Section~\ref{subsec:verifiedinterfacesclasses}) will be given.

\paragraph{Logical Consequences}

\begin{display}{Logical Consequence, $\entails v \Gamma \Fs G$:}
\staterule{(Id)}
  {\Gamma \vdash \hastype  {v,\Fd,G} {\Object,\ok} }
  {\entails v \Gamma {\Fd,G} G}
\label{rule:con-id}
\qquad 
\staterule{(Ax)}
  {
    \axiom v \Gamma G 
    \quad
    \Gamma \vdash \hastype {v,\Fd,G} {\Object,\ok}
  }
  {\entails v \Gamma \Fd G}
\label{rule:con-ax}
\\[\GAP]
\staterule{($\LLAnd$ Intro)}
  {\entails v \Gamma \Fd {H_1} \quad \entails v \Gamma \Gd {H_2}}
  {\entails v \Gamma {\Fd,\Gd} {H_1 \LAnd H_2}}
\label{rule:con-and-intro}
\qquad
\staterule{($\LLAnd$ Elim)}
  {\entails v \Gamma \Fd {G_1\LAnd G_2}
   \quad
   \entails v \Gamma {\Es,G_1,G_2} H}
  {\entails v \Gamma {\Fd,\Es} H}
\label{rule:con-and-elim}
\\[\GAP]
\staterule{($\LLImplies$ Intro)}
  {\entails v \Gamma {\Fd,G_1} {G_2}}
  {\entails v \Gamma \Fd {G_1 \LImplies G_2}}
\label{rule:con-imp-intro}
\qquad 
\staterule{($\LLImplies$ Elim)}   
  {\entails v \Gamma \Fd {H_1 \LImplies H_2}
   \quad
   \entails v \Gamma \Gd {H_1}}
  {\entails v \Gamma {\Fd,\Gd} {H_2}}
\label{rule:con-imp-elim}
\\[\GAP]
\staterule{($\CCAnd$ Intro)}
  {\entails v \Gamma \Fd {G_1} \quad \entails v \Gamma \Fd {G_2}}
  {\entails v \Gamma {\Fd} {G_1 \CAnd G_2}}
\label{rule:con-add-and-intro}
\qquad
\staterule{($\CCAnd$ Elim 1)}
  {\entails v \Gamma \Fd {G_1 \CAnd G_2}}
  {\entails v \Gamma {\Fd} {G_1}}
\label{rule:con-add-and-elim-one}
\qquad
\staterule{($\CCAnd$ Elim 2)}
  {\entails v \Gamma \Fd {G_1 \CAnd G_2}}
  {\entails v \Gamma {\Fd} {G_2}}
\label{rule:con-add-and-elim-two}
\\[\GAP]
\staterule{($\CCOr$ Intro 1)}
  {\entails v \Gamma \Fd {G_1}}
  {\entails v \Gamma \Fd {G_1 \COr G_2}}
\label{rule:con-or-intro-one}
\quad
\staterule{($\CCOr$ Intro 2)}
  {\entails v \Gamma \Fd {G_2}}
  {\entails v \Gamma \Fd {G_1 \COr G_2}}
\label{rule:con-or-intro-two}
\quad
\staterulecond{($\CCOr$ Elim)}
  {\entails v \Gamma \Fd {G_1 \COr G_2} } 
  {
    \entails v \Gamma {\Ed,G_1} H 
    \quad
    \entails v \Gamma {\Ed,G_2} H
  } 
  {\entails v \Gamma {\Fd,\Ed} H}
\label{rule:con-or-elim}
\\[\GAP]
\staterulecond{(Ex Intro)}
  {
   \goodform {\Gamma,\hastype \logvar T} G
  }
  {
   \Gamma \vdash \hastype {\specval} T
   \quad
   \entails v \Gamma \Fd {\subst {\specval} \logvar G}
  }
  {\entails v \Gamma \Fd {\Ex \logvar T G}}
\label{rule:con-ex-intro}
\qquad
\staterulecond{(Ex Elim)}
  { \logvar \not\in \Fd,H }
  {
   \entails v \Gamma \Ed {\Ex \logvar T G}
   \quad
   \entails v {\Gamma,\hastype \logvar T} {\Fd,G} H
  }
  {
   \entails v \Gamma {\Ed,\Fd} H
  }
\label{rule:con-ex-elim}
\\[\GAP] 
\staterule{(Fa Intro)}
  {\logvar \not\in \Fd \quad \entails v {\Gamma,\hastype \logvar T} \Fd G}
  {\entails v \Gamma \Fd {\Fa \logvar T G}}
\label{rule:con-fa-intro}
\quad
\staterule{(Fa Elim)}
  {
   \entails v \Gamma \Fd {\Fa \logvar T G}
   \quad 
   \Gamma \vdash \hastype {\specval} T    
  }
  {
    \entails v \Gamma \Fd {\subst {\specval} \logvar G}
  }
\label{rule:con-fa-elim}
\end{display}

\paragraph{Hoare Rules}

\begin{display}{Hoare Rules}
\staterulelabel{(Fld Set)}
  {
    \Gamma \vdash \hastype {u,w} {U,W}
    \quad
    \Fld W f \in \fields U
  }
  {
    \hhoare v \Gamma 
       {\Contains u f \one W} 
       {\HdFSetNoFin u f w}
       {\Contains u f \one w}
  }
  \\[\GAP]

\staterulelabel{(Get)}
   {  
     \Gamma \vdash \hastype {u,\specval,w} {U,\PermTy,W}
     \quad
     \Fld W f \in \fields U
     \quad
     W \subty \Gamma(\lvar)
    }
   {
     \hhoare v \Gamma {\Contains u f \perm w} 
                      {\HdGet \lvar u f} 
		      {
                        \Contains u f \perm w 
                        \LAnd 
			\lvar \jdeq w
                      }
   } 
%
  \\[\GAP]

\staterulelabel{(New)}
  {
    \TyApp C {\TypedVar \Ts \logvars} \in \cls
    \quad
    \Gamma \vdash \hastype \specvals {\subst \specvals\logvar \Ts}
    \quad
    \TyApp C \perms \subty \Gamma(\lvar)
  }
  {
      \hhoare v \Gamma 
         \True
         {\HdNew \lvar C \perms}
        {
        \GGet \lvar \initpred
        \,\LAnd\, \iisclassof C \lvar
        }
  }
%
  \\[\GAP]

\staterulelabelbis{(Call)}
  {
  \begin{array}{c}
    \mtype m {\TyApp t \perms} 
    = \TyAbs {\TypedVar \Ts \alphas}\; {\Mspec {G} 
					{ \lpa {\logvar'} \rpa \lpa {G'} \rpa} }
   \\ \hspace{1cm}U\;m\;\lpa{\TyApp t \perms}\;{\rvar_0} , \Ws \ \rvars \rpa
   \\
    \quad	
    \sigma = (u/\rvar_0,\perms'/\logvars,\ws/\rvars)
    \quad
    \Gamma \vdash \hastype {u,\specvals',\ws} 
        {\TyApp t \perms,\Ts[\sigma],\Ws[\sigma]}
    \quad
    U[\sigma] \subty \Gamma(\lvar) 
  \end{array}}
  {
     \hhoare v \Gamma {u \jneq \Null \LAnd\, G[\sigma]}
        {\HdCall {\lvar} u m \ws} 
     {
       \Ex {\logvar'} {U[\sigma]\;} {
         \logvar' \jdeq \lvar 
         \,\LAnd\;
         G'[\sigma] }
     }
  }
  \\[\GAP]
\staterulelabel{(Val)}
  {
   \entails v \Gamma F {\subst w \logvar G}
   \quad
   \Gamma \vdash \hastype w {U \subty T} 
   \quad
   \goodform{\Gamma,\hastype \logvar U} G
  }
  {
    \hoare v \Gamma F w T {\lpa U \ \logvar \rpa \lpa G \rpa } 
  }
%
  \\[\GAP]

\staterulelabel{(Dcl)}
  { 
    \begin{array}{l}
      \lvar \not\in F,G 
      \quad
      \hoare v {\Gamma,\hastype \lvar T} {F \LAnd\, \lvar \jdeq \df T} c U {G}
    \end{array}
  }
  {\hoare v \Gamma F {\Dcl T \lvar c} U {G} }
%
  \\[\GAP]

\staterulelabel{(Fin Dcl)}
  { 
    \begin{array}{l}
    \rvar \not\in F, G, v
    \quad
    \Gamma \vdash \hastype \lvar T
    \quad
    \hoare v {\Gamma,\hastype {\rvar} T} {F \LAnd\, \rvar \jdeq \lvar} c U {G} 
    \end{array}
  }
  { \hoare v \Gamma F {\FSetVarNoFin T {\rvar} \lvar c} U {G}}
%
  \\[\GAP]

\staterulelabel{(Seq)}
  {
   \hhoare v \Gamma {F} \hc {F'} 
   \quad
   \hoare v \Gamma {F'} c T {G}
  }
  {
    \hoare v \Gamma F {\hc;c} T G
  }

  \\[\GAP]

\staterulelabel{(Frame)}
  { 
    \hhoare v \Gamma F \hc G
    \quad
    \goodform \Gamma H
    \quad
    \fv H \cap \writes \hc = \emptyset
  } 
  {
    \hhoare v \Gamma {F \LAnd H} \hc {G \LAnd H}
  }
%
  \\[\GAP]

\staterulelabelbis{(Consequence)}
  { 
    \begin{array}{l}
    \hhoare v \Gamma {F'} \hc {G'}
    \\
    \entails v \Gamma F {F'}
    \quad
    \entails v \Gamma {G'} G
    \end{array}
  }
  {
    \hhoare v \Gamma F \hc G
  }
%
  \\[\GAP]

\staterulelabel{(Exists)}
  {
    \hhoare v {\Gamma,\hastype \logvar T} F \hc G
  }
  {
    \hhoare v \Gamma {\Ex \logvar T F} \hc {\Ex \logvar T G}
  }
 \\[\GAP]
\staterulelabel{(Var Set)}
  {
   \Gamma \vdash \hastype {w} {\Gamma(\lvar})
  } 
  {\hhoare v \Gamma \True {\HdSetVar \lvar w} {\lvar \jdeq w}}
\label{rule:hoa-var-set}
\\
\staterulelabel{(Op)}
  {
    \Gamma \vdash \hastype {\op(\ws)} {\Gamma(\lvar)}
  }
  {
    \hhoare v \Gamma \True {\HdOp \lvar \op \ws} {\lvar \jdeq \op(\ws)}
  }
\label{rule:hoa-op}
%
\\[\GAP]
\staterulelabel{(If)}
  {
    \begin{array}{l}
    \Gamma \vdash \hastype w \BoolTy
    \quad
    \hoare v \Gamma {F \LAnd w} c \Void G
    \quad
    \hoare v \Gamma {F \LAnd \jnot w} {c'} \Void G
    \end{array}
  }
  {
    \hhoare v \Gamma F {\HdCond w c {c'}} G
  }
\\[\GAP]
{
    \staterulelabel{(While)}
    {
    \Gamma \vdash \hastype {e,F} {\BoolTy,\diamond}
    \quad
    \hoare v \Gamma {F \CAnd e} \cmd \Void F
    }
    {
    \hoare v \Gamma F {\HdInvWhile e F \cmd} \Void {F \CAnd \jnot e}
    }
    \label{rule:while}
}
\\[\GAP]
\staterulelabel{(Assert)}
  {
    \entails v \Gamma F G
  }
  {
    \hhoare v \Gamma F {\HdAssert G} F 
  }
\label{rule:hoa-assert}
%
\\[\GAP]
\staterulelabel{(Return)} 
  {
    \begin{array}{l}
	    \Gamma \vdash \hastype v T
	    \quad 
	    \entails o \Gamma F {\subst v \logvar G}
	    \quad
	    T \subty U 
	    \\
	    \hoare o {\Gamma, \hastype \lvar U} 
		     {\Ex \logvar T {\logvar \jdeq \lvar\; \LAnd\; G}} c V H
    \end{array}
  } 
  {
    \hoare o {\Gamma,\hastype \lvar U} {F} {\returnand \lvar v c} V H 
  }
\label{rule:hoa-return}

\end{display}

\paragraph{Remarks}
The rule~\ref{rule:hoa-assert} is defined for a specification-only
\texttt{assert} statement, that is formally defined on
page~\pageref{idx:spec:command}. Intuitively, \texttt{assert($G$)}
expresses that $G$ should hold at that point in the execution. It
is used to express a corollary about partial correctness of a
verified program. 
The rule~\ref{rule:hoa-return} is for the auxiliary \texttt{return}
statement, defined in Section~\ref{subsec:semantics:jll}. As
explained, source code programs do not contain this statement, but we
need the rule to prove soundness of the proof system.

\paragraph{Well-formedness}

The following presents formal definitions of well-formed predicate types, method types and verified interfaces
\begin{display}{Well-formed Predicate Types, 
Method Types
, Verified Interfaces
:}
\staterulelabel{(Pred Type)}
  {
    \goodtype \Gamma \Ts
  }
  {
    \good \Gamma {\PTNew P \Ts \alphas}
  }
	\quad

\staterulelabel{(Ax)}
  { \goodform \Gamma F }
  { \good \Gamma {\AX F}}
\\[\GAP]
\staterulelabel{(Mth Type)}
  { 
      \good {\Gamma,\hastype \logvars \Ts, \hastype \rvars \Vs} {\Ts,F,U,\Vs}
      \quad
      \goodform {\Gamma,\hastype \logvars \Ts, \hastype \rvars \Vs,\hastype \result U} G
  }  
  { 
    \good \Gamma {\MMT \Ts \logvars {\Mspec F G} U m \Vs \rvars }
  }
\\[\GAP]
\staterulelabelbis{(Int)}
  {      
  \begin{array}{c}
     \TypeExtends I \Ts \alphas \Us
     \quad
     \initpred \not\in \dom\pdtys 
  \\
     \goodtype {\hastype \alphas \Ts} {\Ts,\Us,\pdtys}
     \quad
     \good {\hastype \alphas \Ts, \hastype \This {\TyApp I \alphas}}
           {\ax,\mts}  
	\end{array}  
  }
  {
    \hastype 
      { \Interface I \Ts \alphas \Us {\pdtys\;\axs\;\mts} } \ok
  }
\end{display}

We write $\cfv c$ for the set of variables that occur freely in an object creation command in $c$.
Rule~\ref{rule:cls} below is the main judgment for verifying classes.
Premises
$\Extends C \Ts \alphas U$ and $\Implements C \Ts \alphas \Vs$ enforce class $C$ to be sane. Premise $\Sound C \Ts \alphas$ enforces
$C$'s axioms to be sound. Premise
$\good {\hastype \logvars \Ts,\hastype \This {\TyApp C \logvars}} {\fds,\axs,\mds}$
enforces $C$'s methods ($\mds$) to be verified.

Rule~\ref{rule:mth} below verifies methods.
In this rule, we
prohibit object creation commands to contain logical method parameters
because our operational semantics does not keep track of
logical method parameters (while it does keep track of class parameters).

\begin{display}{Verified Classes, $\hastype \cl \ok$:}
\staterulelabelbis{(Cls)}
  {
  \begin{array}{c}
    \Extends C \Ts \alphas U
    \quad
    \Implements C \Ts \alphas \Vs
    \quad
    \Sound C \Ts \alphas
    \quad
    \initpred \not\in \dom\pds 
\\
    \goodtype {\hastype \logvars \Ts} {\Ts,U,\Vs}
    \quad
    \goodin {\hastype \logvars \Ts} \pds {\TyApp C \logvars}
    \quad
    \good {\hastype \logvars \Ts,\hastype \This {\TyApp C \logvars}}
          {\fds,\axs,\mds} 
	\end{array}  
  }
  {  
      \hastype 
        {\CClass C \Ts \alphas U \Vs {\fds\;\pds\;\axs\;\mds} }
	 {\ok}
  }
  \label{rule:cls}
\\[\GAP]
\staterulelabel{(Fld)}
  { \goodtype \Gamma T }
  { \good \Gamma {\Fld T f}}
\label{rule:fld}
\qquad
\staterulelabel{(Pred)}
  {
    \good \Gamma {\PTNew P \Ts \alphas}
    \quad
    \Gamma,\hastype \This U,\hastype \logvars \Ts \vdash \hastype F \ok
  }
  { \goodin \Gamma {\NewPD P \Ts \alphas F} U }
\label{rule:pred-def}
\\[\GAP]
\staterulelabelbis{(Mth)}
  {
    \begin{array}{c}
     \good \Gamma {\MMT \Ts \logvars {\Mspec F G} U m \Vs \rvars }    
     \quad
     \cfv c \cap \logvars = \emptyset
     \\
     \Gamma' = {\Gamma,\hastype \logvars \Ts,\hastype \rvars \Vs}
     \quad
     \hoare \This {\Gamma'}
         {F \LAnd\,\This\neq\Null} c U { \lpa \result \rpa \lpa G \rpa } 
    \end{array}
  }
  {  
   \good \Gamma 
       {\MD \Ts \logvars {\Mspec F G} U m \Vs \rvars c}
  }
\label{rule:mth}
\end{display}

\paragraph{Sanity Conditions for Class Extensions and Interface Implementations}

In the definition below, we treat the partial functions $\kw{mtype}$
and $\kw{ptype}$ (formally defined in~\ref{sec:lookup}) 
as total functions that map elements outside their domains to
the special element $\undef$. Furthermore, we extend the subtyping relation:
\begin{displaymath}
\subty \ = \setcomp {(T,U)} {T \subty U} \,\cup\,\set{(\undef,\undef)} 
\end{displaymath}

\begin{displaymath}
\begin{array}{c}
\Extends C \Ts \logvars U 
\deq
\left\{\begin{array}{l}
U \mbox{ is a parameterized class} 
\\
f \in \dom{\fields U} \ \Rightarrow \ f \not\in \declared C
\\
\left.\begin{array}{l}(\forall m,\mt)(\mtype m U = \mt
            \ \Rightarrow\  \\\phantom{(\forall m,\mt)(}
	    \hastype \logvars \Ts
	    \vdash
	    \mtype m {\TyApp C \logvars} \subty \mt)\end{array}\right.
\\
(\forall P,\pdty)(\pptype P U = \pdty
            \ \Rightarrow\  
	    \ptype P C \logvars \subty \pdty)
\end{array}\right.
\end{array}
\end{displaymath}

\begin{displaymath}
\begin{array}{c}
\TypeExtends I \Ts \logvars U 
\deq
\left\{\begin{array}{l}
U \mbox{ is a (parameterized) interface}
\\
(\forall m,\mt)(\mtype m U = \mt
            \ \Rightarrow\  \\
\phantom{(\forall m,\mt)(}
	    \hastype \logvars \Ts
	    \vdash
	    \mtype m {\TyApp I \logvars} \subty \mt)
\\
(\forall P,\pdty)(\pptype P U = \pdty
            \ \Rightarrow\ \\ 
\phantom{(\forall P,\pdty)(}\ptype P I \logvars \subty \pdty)
\end{array}\right.
\end{array}
\end{displaymath}
\begin{displaymath}
\TypeExtends I \Ts \logvars \Us
\Deq
(\forall U \in \Us)(\TypeExtends I \Ts \logvars U)
\end{displaymath}

\begin{displaymath}
\begin{array}{c}
\Implements C \Ts \logvars U 
\deq
\left\{\begin{array}{l}
U \mbox{ is a (parameterized) interface}
\\
(\forall m,\mt)(\mtype m U = \mt
            \ \Rightarrow\ \\ 
\phantom{(\forall m,\mt)(} \mtype m {\TyApp C \logvars} \neq \undef)
\\
(\forall P,\pdty)(\pptype P U = \pdty
            \ \Rightarrow\  
	    \ptype P C \logvars \neq \undef)
\\
(\forall m,\mt)(\mtype m U = \mt
            \ \Rightarrow\ \\
\phantom{(\forall m,\mt)(}
	    \hastype \logvars \Ts
	    \vdash
	    \mtype m {\TyApp C \logvars} \subty \mt)
\\
(\forall P,\pdty)(\pptype P U = \pdty
            \ \Rightarrow\  
	    \ptype P C \logvars \subty \pdty)
\end{array}\right.
\end{array}
\end{displaymath}
\begin{displaymath}
\Implements C \Ts \logvars \Us
\Deq
(\forall U \in \Us)(\Implements C \Ts \logvars U)
\end{displaymath}

\end{document}